\documentclass[a4paper,fleqn,usenatbib,useAMS]{mnras}

\usepackage{graphicx}   
\graphicspath{{plots/}}
\usepackage{amsmath}    
\usepackage{braket}     
\usepackage{color}      
\usepackage{hyperref}   
\usepackage{geometry}   
\usepackage{lscape}     
\usepackage{tabularx}   
\usepackage{adjustbox}  

\newcommand{\pd}{\partial}
\newcommand{\ji}{\varphi}

\newcommand{\SuperNu}{{\tt SuperNu}}
\newcommand{\WinNET}{{\tt WinNET}}
\newcommand{\gcc}{{\textrm{g}\ \textrm{cm}^{-3}}}
\newcommand{\cmg}{\textrm{cm}^2\ \textrm{g}^{-1}}
\newcommand{\ergs}{\textrm{erg}\ \textrm{s}^{-1}}

\usepackage[T1]{fontenc}
\usepackage{ae,aecompl}

\usepackage{newtxtext,newtxmath}

\title[Electromagnetic signatures of neutron star mergers]
      {Impact of ejecta morphology and composition on the electromagnetic
      signatures of neutron star mergers}

  \author[R. T. Wollaeger et al.]
         {Ryan T. Wollaeger$^1$,
         Oleg Korobkin$^1$,\thanks{email: korobkin@lanl.gov}
         Christopher J. Fontes$^1$,
         Stephan K. Rosswog$^2$,
         \newauthor
         Wesley P. Even$^{1,3}$,
         Christopher L. Fryer$^{1,4,5}$,
         Jesper Sollerman$^2$,
         Aimee L. Hungerford$^1$,
         \newauthor
         Daniel R. van Rossum$^6$,
         and Allan B. Wollaber$^{1}$
         \\
         \\
         $^1$Center for Theoretical Astrophysics, Los Alamos National Laboratory, Los Alamos, NM 87545, USA\\
         $^2$The Oskar Klein Centre, Department of Astronomy, AlbaNova,
             Stockholm University, SE-106 91 Stockholm, Sweden\\
         $^3$Southern Utah University, Cedar City, UT 84720, USA\\
         $^4$University of New Mexico, Albuquerque, NM 87131, USA\\
         $^5$University of Arizona, Tucson, AZ 85721, USA\\
         $^6$Radboud Radio Lab, Radboud University, Nijmegen, Netherlands
  } 

\date{Last updated \today}

\pubyear{2017}

\begin{document}
\label{firstpage}
\pagerange{\pageref{firstpage}--\pageref{lastpage}}
\maketitle

\begin{abstract}
The electromagnetic transients accompanying compact binary mergers
($\gamma$-ray bursts, afterglows and 'macronovae')
are crucial to pinpoint the sky location of gravitational wave sources.
Macronovae are caused by the radioactivity from freshly synthesised
heavy elements, e.g. from dynamic ejecta and various types of winds.
We study macronova signatures by using multi-dimensional radiative transfer
calculations. We employ the radiative transfer code \SuperNu\ and state-of-the
art LTE opacities for a few representative elements from the wind and
dynamical ejecta (Cr, Pd, Se, Te, Br, Zr, Sm, Ce, Nd, U) to calculate synthetic
light curves and spectra for a range of ejecta morphologies.
The radioactive power of the resulting macronova is
calculated with the detailed input of decay products. We assess the detection
prospects for our
most complex models, based on the portion of viewing angles that are sufficiently
bright, at different cosmological redshifts ($z$).
The brighter emission from the wind is unobscured by the lanthanides (or actinides)
in some of the models, permitting non-zero detection probabilities
for redshifts up to $z=0.07$.
We also find the nuclear mass model and the resulting radioactive heating
rate are crucial for the detectability. While for the most pessimistic
heating rate (from the FRDM model) no reasonable increase in the ejecta mass
or velocity, or wind mass or velocity, can possibly make the light curves
agree with the observed nIR excess after GRB130603B, a more optimistic
heating rate (from the Duflo-Zuker model) leads to good agreement.
We conclude that future reliable macronova observations would constrain
nuclear heating rates, and consequently help constrain nuclear mass
models.
\end{abstract}

\begin{keywords}
neutron stars, mergers, electromagnetic counterparts, gravitational waves
\end{keywords}

\section{Introduction}

Neutron star mergers (NSMs) realize extreme conditions, probing the limits of
fundamental theories.
The matter evolves in a curved space-time at several times nuclear density
and at temperatures in excess of $10^{11}$ K.
Moreover, the high density and curved space-time have the potential to
generate magnetic fields beyond magnetar strength.
These events announce themselves through a variety of channels: electromagnetic,
gravitational, nucleosynthetic signatures and even neutrinos in the (un-)lucky
occurrence of a nearby event.
The most conspicuous signatures of NSMs are thought to be short $\gamma$-ray
bursts \citep[GRBs, see][]{popham99,fryer99,bloom99,piran05,lee07,nakar07,
fong13,berger14a}, and it is very likely that advanced detector facilities such as
AdLIGO \citep{aasi15a}, Advanced VIRGO \citep{acernese15a} and KAGRA
\citep{akutsu15}, will also detect the long-awaited NSM gravitational wave
signals.

In a NSM several physical mechanisms conspire to unbind
material from the merging stars, releasing neutron-rich outflows into the
surrounding galactic environment.
The NSM outflows can be subdivided into several classes according to their
ejecta amounts, neutron richness, morphologies, and expansion
velocities. First, the \emph{dynamical ejecta} are expelled by gravity,
centrifugal and pressure forces at the moment of the merger itself.
Numerical simulations indicate that this type of outflow has velocities in
the subrelativistic regime $\sim0.1-0.3\ c$, is very neutron-rich
($Y_e\sim0.03-0.2$), and has masses
in the range $\sim10^{-4}-0.05\ M_\odot$
\citep{rosswog13a,bauswein13a,hotokezaka13,sekiguchi16a,lehner16a,rosswog17a,endrizzi16}.
If the collapse to a black hole (BH) is delayed, an intense \emph{neutrino-} and
\emph{accretion-driven wind} is launched from the hot surface of the resulting
hypermassive neutron star~\citep[HMNS,][]{dessart09a,perego14a,just15a,martin15}.
This wind has a higher electron fraction $Y_e\sim0.3-0.5$, but
lower velocity and mass.
Additional outflow can be launched from the HMNS by the strong magnetic
fields \citep{siegel14b,ciolfi15a}.
Finally, nuclear recombination assisted by viscous magnetic forces unbinds
outer layers of the post-merger accretion disk and launches \emph{disk wind}
outflows \citep{chenwx07,metzger08e,fernandez13b,just15a}.
The disk winds have estimated velocities $v\sim0.05-0.1\ c$, moderately
neutron rich composition with electron fraction $Y_e\sim0.2-0.4$ and
a mass comparable to that of dynamical ejecta.
A fair pictorial representation of NSM outflows can be found e.g. in \cite{rosswog13a}.
If scientists can observationally distinguish the wind mass loss for systems
that spend more than $100\,ms$ as HMNSs versus those that collapse quickly to a
BH, they can probe the equation of state of dense matter\citep{fryer15}.

It has been suggested that the neutron-rich outflows from NSMs can be
important sites for the "strong" $r$-process nucleosynthesis
\citep{lattimer74,lattimer77,eichler89,freiburghaus99}\footnote{However, see
\cite{cote17b} for a literature review and a recent critical discussion of
$r$-process sites from the perspective of chemical evolution and population
synthesis modelling.}.
This hypothesis has attracted much attention recently, after attempts to
robustly produce heavy $r$-process elements in core-collapse supernova simulations encountered
significant difficulties
\citep{arcones07a,fischer10a,roberts10a,thielemann11}.
Other indirect observational evidence also points to a rare, robust event
(such as a NSM) as the main "strong" $r$-process producer: the robust pattern of
abundances in old metal-poor $r$-process stars \citep{sneden08}, the absence of any
traces of recent $^{244}$Pu in deep sea reservoirs
\citep{turner04,wallner15,hotokezaka15c},
and the newly discovered "$r$-process galaxy" in the family of ultra-faint dwarf
galaxies \citep{ji15,hirai15}.

Residual $r$-process radioactivity can potentially power an electromagnetic
transient, a so-called ``macronova'' \citep[or ``kilonova'', see][for
discussion of the naming conventions]{metzger16c}.
This idea was originally proposed in \cite{li98}, revived in \cite{kulkarni05}
and further developed in \cite{metzger10a,metzger12a}.
The opacities adopted for dynamical ejecta in these early works were seriously
underestimated and led to overly optimistic prediction for detectability. Macronova detection
prospects became dimmer after it was realized that the opacities in the optical
and near infrared are a few orders of magnitude higher due to heavy
line blanketing by lanthanides \citep{kasen13,barnes13,fontes15a,fontes17a}.
Subsequent studies \citep{grossman14,rosswog14a,kyutoku13,metzger14a} including detailed
radiative transfer simulations \citep{tanaka13,tanaka14,kasen15a} with updated opacities
predicted dimmer light curves that would peak after a few days in the infrared
part of the spectrum, implying more pessimistic prospects for macronova detection.

Despite these difficulties, as of now, several candidate kilonova/macronova
events have been identified
\citep{tanvir13,deUgartePostigo13,berger13a,yang15,jin15}, but their nature is
still very ambiguous due to sparse observational data and uncertainties in
theoretical models.  These uncertainties include the partition of radioactive
energy between different decay products,
which then have different capacities for thermalization \citep{hotokezaka15d,barnes16a}.
The influence of the radioactive heating rates was studied by
\cite{lippuner15}, who found that the heating profile remains quite featureless
\citep[see also][]{barnes16a,rosswog17a,wu16}.
\cite{barnes16a} explored four different nuclear mass models and analysed
thermalization in detail. Their results show that the uncertainty in thermalization
has a sub-dominant effect on light curves relative to the theoretical uncertainty in
the nuclear mass model (see their Fig.17).

Detection of electromagnetic counterparts would provide crucial information to
localize the astrophysical environments of gravitational wave signals
\citep{metzger12a,nissanke13,piran13a,singer14,chu15,ghosh15,bartos16,abbott16a}.
Preliminary searches for the electromagnetic macronova-like transients following
gravitational wave candidate triggers \citep[e.g.][]{aasi14a,copperwheat16}
were not successful in finding plausible candidates, and neither was a recent
search in the dark energy surveys \citep{doctor16}. Additionally, nearby short
GRBs (GRB160314A, GRB160821B) did not exhibit clear signs of bright
macronovae (Kasliwal et al. in prep, Troja et al. in prep).
These non-detections indicate that many macronovae are indeed as
faint as predicted, possibly due to the high opacity of lanthanides and the low
ejected mass (relative to supernovae).
Recent comprehensive reviews of electromagnetic counterparts can be found in
\cite{rosswog15a}, \cite{fernandez15a} and \cite{metzger16c}.

Thus, accurate and reliable macronova light curve predictions are needed
to constrain the detection prospects of NSMs.
Previous studies with detailed multidimensional radiative transfer
\citep{kasen13,barnes13,tanaka13,tanaka14,kasen15a,fernandez16,barnes16a}
used the Sobolev expansion opacity formalism to treat the substantial
number of lines that can occur in the spectra of lanthanide and actinide
elements.  In the present work, we consider an alternative line-smeared
approach that conserves the integral of the opacity over frequency
\citep{fontes15a,fontes17a}. The latter method can produce significantly higher
opacities compared to the expansion opacity formalism.
In this study, we extend the work of \cite{fontes15a,fontes17a} with a state-of-the art
open source radiative transfer code, \SuperNu\footnote{
\url{https://bitbucket.org/drrossum/supernu/wiki/Home}}\ \citep{wollaeger14a},
which implements a 3D semi-implicit multigroup Monte Carlo solver.
With \SuperNu\ and the line-smeared opacities, we explore the effects of
varying NSM ejecta morphology, composition (or opacity), and r-process
decay heating on macronova light curves and spectra.

The morphology of the outflow from a NSM depends on the binary mass ratio and
the nuclear equation of state. Tidal dynamical ejecta, which are expelled from the
system on a dynamical merger timescale, tend to preserve a quasi-toroidal
configuration. On the other hand, general relativistic
simulations with soft equations of state show highly irregular hot outflows
from the shocked interface,
which become almost isotropic.
Here, we explore both a sequence of toroidal dynamical ejecta configurations
from binary neutron star merger simulations \citep{rosswog14a}
and a sequence of spherically symmetric ejecta configurations from an
analytic hydrodynamical model.

The composition of the NSM outflow determines both the nuclear heating rates,
which power the macronova, and the opacity of the ejecta.
In our models, the nuclear heating rates are taken directly from the output of
the $r$-process network \WinNET\ \citep[similar to ][]{rosswog17a}.
Time-dependent detailed compositions of decaying isotopes allow accurate
calculation of nuclear energy partitioning between different decay products
($\alpha$-, $\beta$-, $\gamma$- radiation and fission products).
We then apply analytic fits from \cite{barnes16a} to compute energy
thermalization for each of the decay products.
For the $\gamma$-ray thermalization efficiency, in multiple dimensions,
we either ray-trace from the origin to obtain optical depths or perform
Monte Carlo.
In either case, we use a grey, pure-absorption $\gamma$-ray opacity,
calibrated to accurately reproduce energy deposition from Compton scattering
and photoionization \citep[see Fig.5 in][]{barnes16a}.

Our opacity treatment is limited to detailed multifrequency
opacities for a few selected representative elements, with an assumption of
local thermodynamic equilibrium (LTE opacities). The opacities are
calculated with the Los Alamos suite of atomic physics codes \citep{fontes15b}.
The elements are selected either due to their higher abundance in dynamical
ejecta or wind, or due to an open $f$-shell in their atomic structure. We also
explore simple density-weighted mixtures of representative elements.

Because of heavy line blanketing in lanthanides, and even actinides,
\citep{mendozatemis15} abundantly present in the dynamical ejecta
\citep[][find mass fractions $>20\%$]{rosswog17a}, the detection of
electromagnetic counterparts directly from the heavy $r$-process ejecta is
very difficult. However, if the distribution of lanthanides has a quasi-toroidal
morphology due to preferentially equatorial ejection or neutrino irradiation
in polar regions \citep{wanajo14}, there is a possibility of detecting an
additional blue component from the lanthanide-free "polar caps". In this
study, we consider a range of configurations of dynamical ejecta and wind
outflows, and investigate the "opening angle" of visibility for these
configurations in optical bands.

The uncertainties and interdepencies of the morphology, composition,
opacity, and nuclear heating in NSM outflow make characterizing the macronova
signal a challenging problem.
In this work, we attempt to isolate and examine the impact of each of these
aspects on the macronova signal; we first summarize the methods and
approximations for the simulations.
Specifically, in Sect.~\ref{sec:exp_models}, we describe the origin and
hydrodynamics of various types of NSM outflows, give typical estimates of their
parameters, and derive an analytic spherically-symmetric homologously expanding
solution.
In Sect.~\ref{sec:composition}, we provide motivation for the
composition and r-process heating rates that dictate the opacity
and provide the power source for the luminosity.
In Sect.~\ref{sec:opacities}, we discuss the radiative transfer
and opacity methods employed to obtain light curves and spectra.
Here we also discuss some past and current code verifications.
In Sect.~\ref{sec:sensitivity}, we study various aspects of macronovae
for a range of models with increasing level of sophistication, starting from
simple spherically-symmetric models with grey opacity, and ending with
complex combined 2D axisymmetric models with dynamical ejecta and wind, having
detailed elemental opacities.
In Sect.~\ref{sec:realistic_models}, we synthesize light curves
and spectra for our most realistic models, which include mixed
compositions for wind and dynamical ejecta and detailed $r$-process
radioactive energy source.
In Sect.~\ref{sec:detectability} we assess the detection prospects of our
most realistic models using limiting magnitudes from VISTA and LSST.
We consider these theoretical detection prospects in the context of
recent estimates for macronova detection rates.
Finally, in Sect.~\ref{sec:discussion}, we discuss and summarize our
findings.

\section{Methodology}

\subsection{Expansion dynamics}
\label{sec:exp_models}

NSM outflows can be divided into two main classes: dynamical ejecta
and "winds".  The "winds" are assumed to be ejected by
the sum of all other processes, such as powerful neutrino emission, viscous
and magnetic stresses, and energy which is released in the post-merger
accretion disk due to nuclear recombination \citep{dessart09a,perego14a,
martin15,chenwx07,siegel14b,ciolfi15a,metzger08e,fernandez13b,just15a,wu16}.

Dynamical ejecta have been studied extensively \citep{rosswog99,rosswog13a,
bauswein13a,hotokezaka13,sekiguchi16a,lehner16a,rosswog17a,endrizzi16},
and the consensus on the value of total ejected mass is the range between
$10^{-4}$ and $\sim0.05\ M_\odot$.
These are also the ranges used in recent population synthesis and chemical
evolution studies~\citep{fryer15,cote17b}.
Eccentric binaries or parabolic encounters can
unbind an order of magnitude more mass
\citep{rosswog13a,gold12,east12a,radice16a},
but such events are expected to be very rare. Mergers of neutron stars with
black holes can release up to $0.2\ M_\odot$ of material
\citep{rosswog05a,foucart15,kyutoku15,kawaguchi15,foucart16a}.

Dynamic ejecta become undbound at the moment of contact. They fall in two
categories, "tidal ejecta" unbound by gravitational torques and "interaction
ejecta" that become unbound due to hydrodynamic processes (see Fig.2 in
\cite{korobkin12}). The first component is cold and
extremely neutron-rich (with electron fractions $Y_e<0.04$), while the second
component can potentially have higher $Y_e$ resulting from the copious
production of $e^-e^+$-pairs, which rapidly drives matter
to a more symmetric state \citep{wanajo14,radice16a}. Further irradiation of
the rapidly receding dynamical ejecta by neutrinos from the surface of the hot
transient hypermassive neutron star, however, does not alter its
composition very much, because the ejecta are sufficiently far away when
neutrino emission becomes significant \citep{radice16a,foucart15}. Most recent
studies agree that the combined electron fraction of the dynamical ejecta is
in the range $Y_e\sim0.04-0.25$, which allows for one or more
nuclear fission cycles and a robust main $r$-process nucleosynthesis
\citep{korobkin12}. As a consequence, dynamical ejecta will have high abundances
of elements with an open $f$-shell -- lanthanides and actinides
\citep[see Table 1 in][]{rosswog17a}.
The open $f$-shell of these elements furnish extremely high opacity in visible
bands and lead to dimmer and slower evolving transients peaking in the infrared
\citep{barnes13,kasen15a,fontes15a,fontes17a}.

The morphology of the dynamical ejecta depends on the compactness of the merging stars
and the binary mass ratio. Higher mass ratios produce more massive tidal ejecta
\citep{rosswog13a} which tend to have a toroidal shape \citep{rosswog14a},
while softer equations of state and inclusion of general relativistic gravity
enhances shocks which lead to more irregular and isotropic outflow shapes,
dominated by the interaction component \citep{bauswein13a}.

Increased interest due to the possibility of an additional, bluer component
from secondary outflows has resulted in a number of recent wind studies
\citep{dessart09a,grossman14,perego14a,martin15,metzger14a,just15a,fernandez15c}.
The general consensus here is that the wind component has higher $Y_e$, which
prevents formation of lanthanides. The morphology of the wind outflow is very
sensitive to a variety of factors, but the studies converge on the fact
that wind outflows are generally slower than the dynamical ejecta
($0.01$-$0.15$~c vs. $\sim0.1$-$0.3$~c). The mass of the wind component is
also highly uncertain -- estimates vary from $10^{-4}\ M_\odot$ up to a few
$10^{-1}\ M_\odot$, depending on the assumptions about the lifetime of the
hypermassive neutron star \citep{perego14a} or mass of the accretion disk
\citep{metzger14a}.
In asymmetric mergers the disk masses can easily reach several
${0.1\ M_\odot}$ \citep{giacomazzo13} and as much as $20\ \%$ of these masses
can become unbound at late times \citep{fernandez13b,just15a}.

\begin{figure}
  \begin{center}
  \begin{tabular}{c}
    \includegraphics[width=0.23\textwidth]{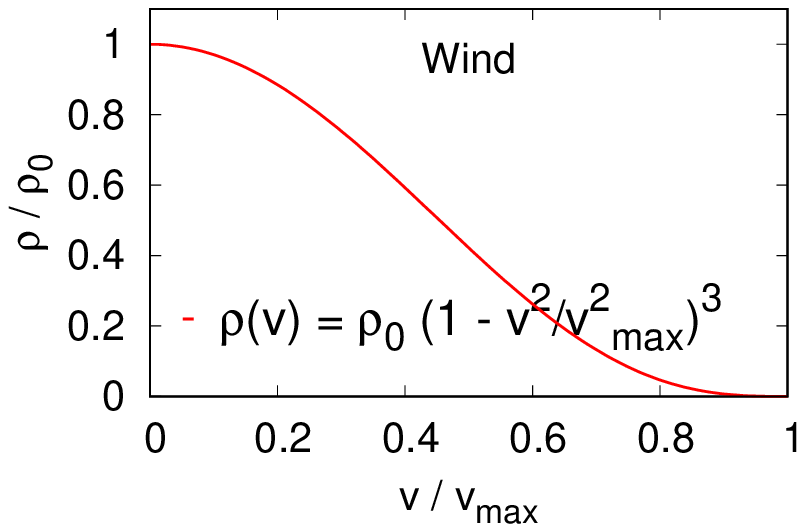}
    \includegraphics[width=0.23\textwidth]{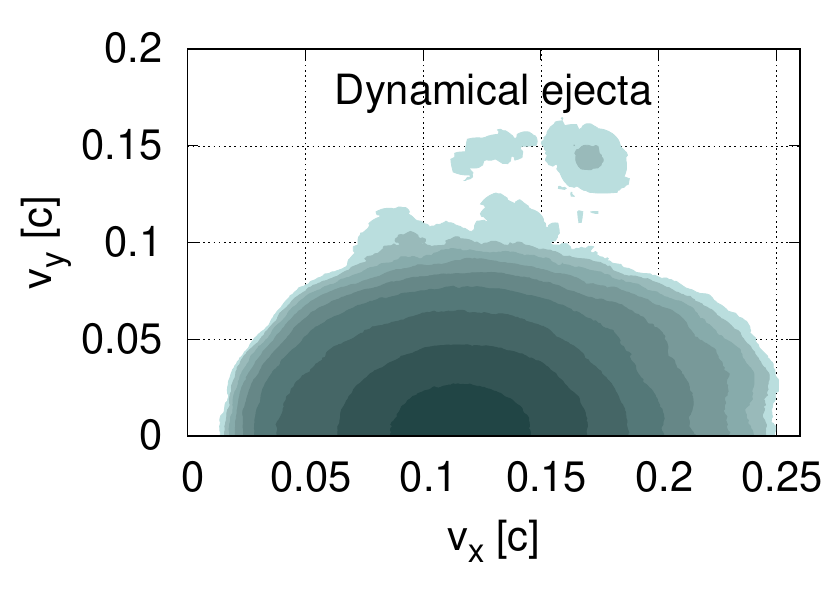}
    \\
    \includegraphics[width=0.49\textwidth]{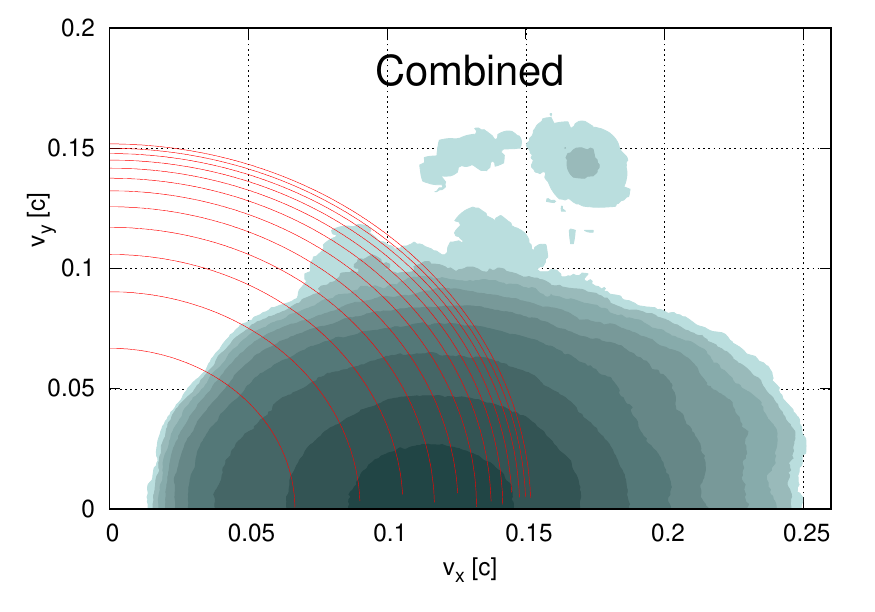}
  \end{tabular}
  \end{center}
  \caption{
     Density profiles, illustrating three types of morphology of the outflows
     explored in this study.
     Top left: radial density profile of the spherically-symmetric analytic
     models. Top right: axisymmetric averaged profile of dynamical ejecta from
     NSM simulations.
     Bottom: density profiles of combined models, with the wind and
     dynamical ejecta density and composition superimposed.
     Density contours of the two latter axisymmetric plots are in log space,
     separated by 0.25~dex.
  } 
  \label{fig:superimposed}
\end{figure}

The morphology of the outflow is crucial for the visibility of the blue
transient. Here we explore three types of morphologies: spherically-symmetric
analytic density profiles (Sect.~\ref{sec:selfsim}), axisymmetric dynamical
ejecta from NSM simulations (Sect.~\ref{sec:dyn_ejecta}), and combined models
where we superimpose the first two models (Sect.~\ref{sec:dyn_wind}), as
illustrated in Fig.~\ref{fig:superimposed}.
We explain the naming conventions of our models in Table~\ref{tab:naming} with
more detailed parameters for each model listed in Table~\ref{tab:allmodels}.

\begin{table}
  \caption{
     Naming conventions for the models used in this paper.
  } 
\begin{adjustbox}{width=0.5\textwidth}
\begin{tabular}{lrll}
\hline
  & Notation & & Comments \\
\hline
spherically    &            & {\tt m1,m2,m3}    & varying mass \\
symmetric      & {\tt SA} + & {\tt v1,v2,v3}    & expansion velocity \\
analytic models&            & {\tt k0,k1,k2,k3} & grey opacity \\
\hline
sph.-symmetric &            & {\tt Se,Br,Te,Pd,Cr,Zr}  & lighter elements \\
models with    & {\tt SA} + & {\tt Sm,Ce,Nd,U}         & lanthanides/actinide \\
multigroup     &            & {\tt d}                  & mixture for dynamical ejcta \\
opacity        &            & {\tt w1,w2}              & two types of wind \\
\hline
\begin{tabular}{@{}l@{}}dynamical\\ejecta\\morphologies\end{tabular}
               & $\left.\begin{array}{c}{\tt A}\\{\tt B}\\{\tt C}\\{\tt D}\end{array}\right\}$ +
               & $\left\{\begin{array}{cc}{\tt 1d}\\{\tt 2d}\end{array}\right\}$ + {\tt Sm}
               & \begin{tabular}{@{}l@{}}spherically (1d) or\\axisymmetrically (2d)\\
                 averaged morphologies\\with opacity of Sm\end{tabular}\\
\hline
\begin{tabular}{@{}l@{}}axisymmetric (2d)\\ejecta + wind (W)\\models\end{tabular}
               & {\tt W2} +
               & \begin{tabular}{@{}l@{}}
                       {\tt A,B,C,D}\\
                       {\tt Se,Br,Te,Pd,Cr,Zr}\\
                       {\tt light}/{\tt heavy}\\
                       {\tt slow}/{\tt fast}
                 \end{tabular}
               & \begin{tabular}{@{}l@{}}ejecta morphologies\\
                                         wind opacity\\
                                         wind mass\\
                                         wind velocity
                 \end{tabular}\\
\hline
\begin{tabular}{@{}l@{}}detailed\\
    composition\\
\end{tabular}  & & $X_1$, $X_2$
               & \begin{tabular}{@{}l@{}}opacity mixtures\\
                     for dynamical ejecta \\
                     + two types of wind  \\
                 \end{tabular}\\
and nuclear    & & $DZ_1$, $DZ_2$
               & enhanced heating rates \\
heating        & & {\tt Xnh1, Xnh2}
               & $X_{1,2}$ + detailed heating \\
\hline
most realistic & & $\gamma A_1$, $\gamma B_1$, $\gamma C_1$, $\gamma D_1$, & {\tt Xnh1, Xnh2} + $\gamma$-transfer\\
models         & & $\gamma A_2$, $\gamma B_2$, $\gamma C_2$, $\gamma D_2$  & + morphologies {\tt A-D}\\
\hline
\end{tabular}
\end{adjustbox}
\label{tab:naming}
\end{table}

\subsubsection{Analytic models in spherical symmetry}
\label{sec:selfsim}

Consider a spherically-symmetric outflow expanding in vacuum. The motion of
the fluid can be described by the Euler equations
of ideal hydrodynamics in spherical coordinates:
\begin{align}
&\frac{\pd\rho}{\pd t} +
\frac{\pd}{\pd r}(\rho v) = -\frac{2}{r}\rho v,
\\
&\frac{\pd v}{\pd t} + v\frac{\pd v}{\pd r}
+\frac{1}{\rho}\frac{\pd p}{\pd r} = 0,
\end{align}
where $v$ is the radial velocity, $\rho$ and $p$ are density and pressure, and
$r$ and $t$ are the radial coordinate and time.
These equations represent conservation of mass and momentum, and if the flow
is adiabatic then the conservation of energy follows.
For a self-similar homologous solution there exist functions
$R(t)$ (scale parameter) and $\ji(x)$ (shape function with
the dimensionless radius coordinate $x=r/R(t)$) such that the density
and the velocity can be expressed as:
\begin{align}
\label{eq:ansatz}
\rho(t,r) &= R(t)^{-3}\ji(r/R(t)),\\
v(t,r)    &= r\dot{R}(t)/R(t).
\end{align}
This ansatz automatically satisfies the continuity equation.
The momentum conservation equation becomes:
\begin{align}
r\frac{\ddot{R}}{R}
+\frac{1}{\rho}\frac{\pd p}{\pd\rho}R^{-4}\ji'
= 0,
\end{align}
(where the prime superscript and over-dot indicate the derivative with respect
to $x$ and $t$, respectively).
Using a polytropic equation of state above, the momentum equation can be
rewritten as a sum with one term containing the time dependence while the
other depends on the dimensionless radius $x$:
\begin{align}
\ddot{R}R^{3\Gamma-2}+K\Gamma\ji^{\Gamma-2}\ji'\cdot\frac{1}{x} = 0.
\end{align}
But this is only possible if both terms are constant:
\begin{align}
 \ddot{R}R^{3\Gamma-2} = -K \Gamma\ji^{\Gamma-2}\ji'/x = C.
\end{align}

Both ODEs admit closed-form solutions for special choices of $\Gamma$.
For radiation-dominated flows with $\Gamma=4/3$ it is convenient to express
the solution in the following closed form:
\begin{align}
  &\ji(x) = \rho_0 R_0^3\left(1-x^2\right)^3,
  \label{eq:jit}
\\
  &(t - t_0) = \frac{R(t)}{V}\sqrt{1-\frac{R_0}{R(t)}}
               + \frac{R_0}{V}
               \log{\left[\frac{R(t)}{R_0}
               \left(1-\sqrt{1-\frac{R_0}{R(t)}}\right)^2
               \right]},
  \label{eq:Rt}
\end{align}
where $R_0$ is the initial characteristic radius of the outflow, $\rho_0$ is
the initial central density, and $V$ is the expansion velocity.

Notice that for $t\gg t_0$ equation (\ref{eq:Rt}) reduces to a trivial linear
dependence: $R(t) \approx V t$. Because the condition ${t\gg t_0}$ is
certainly valid during the time when electromagnetic signals are expected, we
can safely ignore any nonlinearity in (\ref{eq:Rt}) and arrive at the
following expansion profile:
\begin{align}
  \rho(t,r)=
  \rho_0
  \left(\frac{t}{t_0}\right)^{-3}
  \left(1-\frac{r^2}{v^2_{\rm max}\;t^2}\right)^3.
\label{eq:rho_sphsym}
\end{align}
Here, $\rho_0$ is initial central density at time $t_0$ and $v_{\rm max}$ is
the velocity of the expansion front (see Fig.\ref{fig:superimposed}, top left
panel for an illustration). These parameters can be easily related to the
total mass $m_{\rm ej}$ and average velocity $\bar{v}$ of the outflow:
\begin{align}
 &m_{\rm ej}= 4\pi\;\rho_0\;t_0^3\;v_{\rm max}^3 \int_0^1(1-x^2)^3\;x^2\;dx
            = \frac{64\pi}{315}\rho_0\;t_0^3\;v_{\rm max}^3,
 \\
 &\bar{v} = \frac1{m_{\rm ej}} \int 4\pi r^2 \rho(r)\;v\;dr
          = \frac{63}{128} v_{\rm max}
          \approx \frac12 v_{\rm max}.
\label{eq:mas_vave}
\end{align}

The analytic solution is based on the assumptions that:
(a)~the internal energy of the outflow is negligible compared to its kinetic
energy; and (b)~the outflow is radiation-dominated and thus can be
described by a polytropic equation of state $p=K\rho^\Gamma$ with
$\Gamma=4/3$. For dynamical ejecta, these assumptions have been shown to be
accurately fulfilled \citep{rosswog14a}.
Although the second assumption breaks down at later times
when radiation can freely escape, by then it has already established a
homologous expansion pattern, with shells at different radii being out of
sonic contact. Finally, we adopt a non-relativistic approach, consistent with
the expansion velocities $\ll c$ (but note that our radiative transfer solver
\SuperNu\ takes into account relativistic corrections up to $O(v/c)$ in the
treatment of Monte Carlo photon particles; see Sect.~\ref{sec:opacities}).

An ideal gas equation of state also suggests the following profile for the radial
shape of the temperature:
\begin{align}
T(r, t) = T_0 \left(\frac{\rho(r,t)}{\rho_0}\right)^{1/3}
        = T_0 \left(\frac{t}{t_0}\right)^{-1}
                    \cdot\left(1-\frac{r^2}{v^2_{\rm max}\;t^2}\right).
\label{eq:simpmodel-T}
\end{align}
Here, $T_0$ is the temperature at the center at initial time $t=t_0$. However,
the temperature is much more sensitive to the details of the equation of state
and interaction between matter and radiation and nuclear energy
input, so this temperature dependence has to be regarded only as a very
simple estimate. In our radiative transfer simulations, the
temperature is recomputed inside \SuperNu\ based on detailed composition,
radiative losses and local energy input from the radioactive source (see
Sect.~\ref{sec:opacities}). Consequently, we only use equation
(\ref{eq:simpmodel-T}) to initialize our radiative transfer simulations.

\subsubsection{Dynamical ejecta models}
\label{sec:dyn_ejecta}

Spherically-symmetric models are often used as an approximation for isotropic
dynamical ejecta or for the case when the dynamical ejecta completely obscure
the blue transient from the wind.
To verify this approximation and test the impact of ejecta asphericity on the
light curves, we explore axisymmetric dynamical ejecta based on morphologies from
\cite{rosswog14a} \citep[the same as used in][]{kasen15a,fontes15a,fontes17a}.
The latter were computed by long-term hydrodynamic evolution (up to 100 years
after the merger) with radioactive heating source \citep{rosswog14a},
following simulations of NSMs \citep{rosswog13a}. NSM simulations were
performed with the smoothed particle hydrodynamics (SPH) method in Newtonian
gravity \citep{rosswog00,rosswog05a,rosswog07c,rosswog15b}, with a nuclear
equation of state \citep{shen98a,shen98b} and an opacity-dependent multiflavour
neutrino leakage scheme \citep{rosswog03a} to take care of the changes in the
neutron to proton ratio and the cooling by neutrino emission.

Relevant parameters of the models of dynamical ejecta are given in
Table~\ref{tab:dyn_models} and notation (A--D) is the same as in
\cite{rosswog14a} and \cite{grossman14}.
For each of these four 3D morphologies we compute three different effective 1D
and 2D density distributions, distinguished by three different types of averaging.
Models {\tt A1dSm}--{\tt D1dSm} are computed by spherical averaging of the
density:
\begin{align}
  \rho(r) = \frac{1}{4\pi}\int_{4\pi} \rho(r,\theta,\ji)\;d\Omega.
\end{align}
Models {\tt A2dSm}--{\tt D2dSm} are computed by azimuthal averaging:
\begin{align}
  \rho(R,z) = \frac{1}{2\pi}\int_0^{2\pi} \rho(R,z,\ji)\;d\ji.
\end{align}
Finally, in models {\tt A1dmSm}--{\tt D1dmSm} the abbreviation "{\tt m}" stands
for "density maximum": we first find the radius $R_{\rm max}$ of the circle at
which the density in the equatorial plane reaches its maximum, and then
average the density distribution with respect to the distance to that circle:
\begin{align}
  \rho(\xi) = \frac1{4\pi^2}\int_0^{2\pi}\int_0^{2\pi} \rho(\xi,\alpha,\ji)\;d\alpha\;d\ji.
\end{align}
Here ${\{\xi,\alpha,\ji\}}$ are toroidal coordinates around the circle of
maximal density. This type of averaging is designed to test how much the light
curve is affected by the optical depth of the emitting layer versus geometry
of the outflow.

\begin{figure}
  \begin{center}
  \begin{tabular}{c}
    \includegraphics[width=0.49\textwidth]{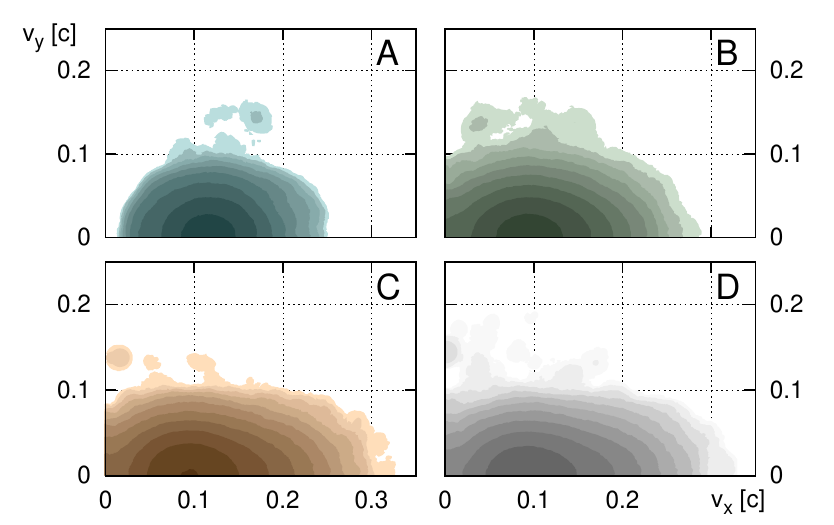} \\
    \includegraphics[width=0.49\textwidth]{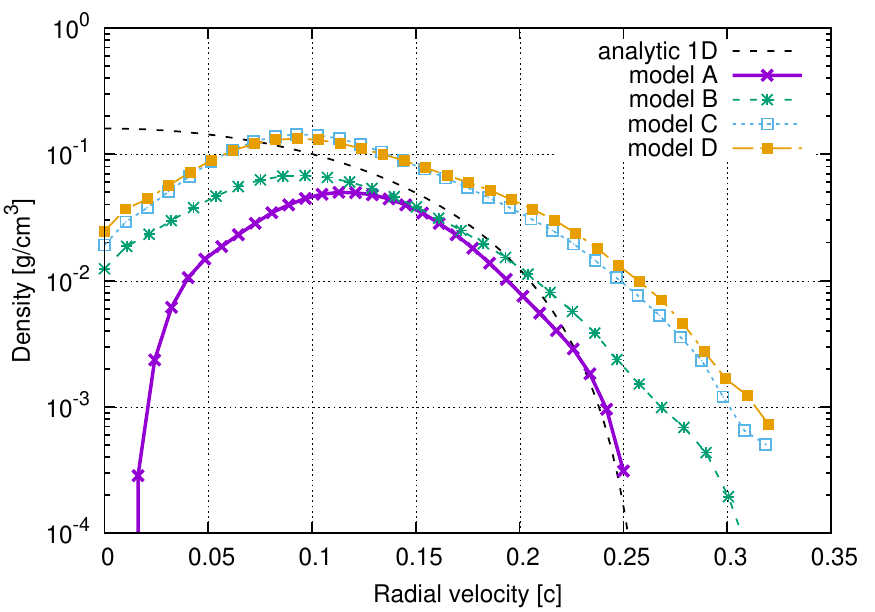}
  \end{tabular}
  \end{center}
  \caption{
     Morphology of the dynamical ejecta models A-D from binary neutron star
     merger simulations, plotted in velocity space. Top: azimutally averaged
     density; the contours are in log space, separated by 0.25~dex. Bottom:
     spherically averaged density profiles.  Thin dashed line represents an
     analytic fit with $v_{\rm max}=0.3\ c$, $m_{\rm ej}=0.013\ M_\odot$.
  } 
  \label{fig:SKR_models_2d}
\end{figure}

Figure~\ref{fig:SKR_models_2d} displays the resulting azimuthally and spherically
averaged density distributions. Since we assume a simple homologous expansion
that is attained within about one hour after the merger \citep[as demonstrated
in][]{rosswog14a}, the density profiles are shown in velocity space. Given the
initial density profile $\rho_0(\mathbf{v})$ at time $t_0$, dynamical ejecta
density at a later time $t$ at a point $\mathbf{r}$ is calculated as:
\begin{align}
  \rho(t,\mathbf{r}) =
  \left(\frac{t}{t_0}\right)^{-3}\rho_0\left(\frac{\mathbf{r}}{t}\right).
\label{eq:rho_num2d}
\end{align}

Figure~\ref{fig:SKR_models_2d} (bottom panel, thin dashed line), shows a
fit of the analytic density profile of type described by
Eq.(\ref{eq:rho_sphsym}) in comparison to a spherically averaged numerical
density profile. It agrees with the density profile for model A for large
velocities and deviates from it significantly near the origin where model A
has a hole.

\subsubsection{Combined models of dynamical ejecta and wind}
\label{sec:dyn_wind}

In combined models of dynamical ejecta and wind, we take axisymmetric models
of the dynamical ejecta and amend them with various parameterized
spherically-symmetric density distributions for the wind, as illustrated in
Fig.~\ref{fig:superimposed} (bottom panel).
Because morphology, mass and composition of the wind are rather uncertain, we
explore a range of parameters listed in Table~\ref{tab:models_2d}.
The added density profile for the wind is modelled with the analytic
spherically-symmetric distribution (\ref{eq:rho_sphsym}).
When combining two outflows, we simply add the corresponding densities and
weighted compositions at every point and ignore potential hydrodynamical
interaction between the wind and dynamical ejecta.
This is certainly a strong simplification, we leave the exploration
of this hydrodynamic interaction to future work.

\subsection{Matter composition}
\label{sec:composition}

We compute the compositional evolution within the ejecta with the network code
\WinNET\ \citep{winteler12a,winteler12b} that is derived from the {\tt BasNet} network
\citep{thielemann11}. The network includes 5831 isotopes reaching up to
$Z=111$ between the neutron drip line and stability. The reaction rates are
from the compilation of \cite{rauscher00a} for the finite range droplet model
\citep[FRDM;][]{moeller95a} and the weak interaction rates ($e^-e^+$-captures
and $\beta$-decays) are the same as used in \cite{arcones11a}. For fission and
neutron capture, we use fission rates of \cite{panov10a} and $\beta$-delayed
fission probabilities as described in \cite{panov05}.

We use FRDM as our baseline model, but it needs to be stressed that the
nuclear heating rates for ejecta that contain matter beyond the platinum peak
is strongly impacted by the used mass formula, see \cite{barnes16a} and
\cite{rosswog17a}. From the four mass formulae explored by \cite{barnes16a},
FRDM yielded the smallest and Duflo-Zucker (DZ) the largest nuclear heating
rates. At the times most relevant for macronovae, the heating from DZ can be
an order of magnitude larger than the one from FRDM, see Fig.~7, left panel,
in \cite{rosswog17a}.

Dynamical ejecta composition and nuclear heating is approximated using
a single particle trajectory from model B in \cite{rosswog14a}, and two
representative tracers from previous studies on neutrino-driven winds: tracers
H1 and H5 from \cite{perego14a}. A single trajectory from dynamical ejecta may
be sufficient to represent nuclear heating, since the heating contribution is
relatively robust \citep{metzger10a,goriely11a,lippuner15}, as is the final
nucleosynthetic pattern \citep{korobkin12,lippuner15}.

\begin{figure}
  \begin{center}
  \includegraphics[width=0.49\textwidth]{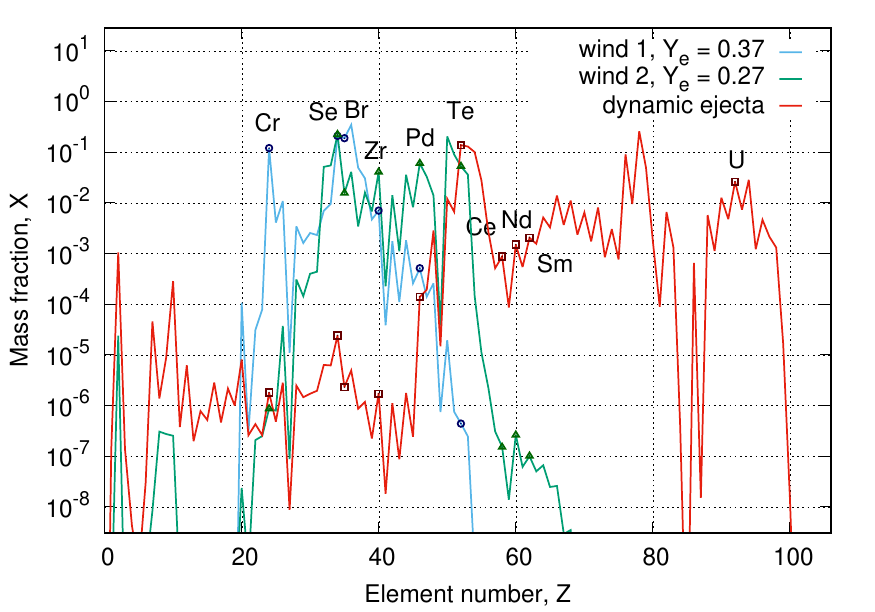}
  \end{center}
  \caption{
     Mass fractions of the elements in different parts of the wind and
     in the ejecta, as a function of atomic number Z.
  } 
  \label{fig:mass_fractions}
\end{figure}

For the wind, we pick
two representative tracers with initial electron fractions $Y_e=0.37$ and
$Y_e=0.27$. Fig.~\ref{fig:mass_fractions} displays the computed composition of
dynamical ejecta and wind tracers for $t=1\ $day. Tracer H5 from
\cite{perego14a} is our "wind 1" model, it has a peak in abundances around
iron group elements and around $r$-process first peak (Br). The other tracer,
H1, is the "wind 2" model, and it produces the $r$-process pattern between first
(Br) and second (Xe) peaks. Broader ranges of potential nucleosynthetic paths
will be explored elsewhere.

For radiative transfer, a few representative elements are mixed in the same
proportion as the one encountered in the composition of each type of
outflow (dynamical ejecta, wind 1 and wind 2). Fig.~\ref{fig:mass_fractions}
marks the mass fractions of the elements that we picked for detailed opacity
calculation, and Table~\ref{tab:mfs_repr} lists their numerical values in each
of the model outflows. These mass fractions are then used to mix approximate
opacities in the mixed-composition models {\tt X1}, {\tt X2} and all
$\gamma$-models (see Table~\ref{tab:models_2d} and Sect.~\ref{sec:combined} for
details on these models).

As can be seen from the plot, winds contain a negligible fraction of elements
with open $f$-shell (lanthanides or actinides), and as such are expected to
be more transparent. Notice that the "wind 2" composition additionally
contains very little elements with open $d$-shell. This
makes "wind 2" more transparent than "wind 1", which is polluted by iron-group
elements.

\begin{table}
  \caption{
     Mass fractions of representative elements in the two types of
     wind outflow and in the dynamical ejecta.
  } 
\begin{tabular}{lccc}
\hline
Elem. & Wind 1 & Wind 2 & Dynamical ejecta \\
\hline
${}_{24}$Cr & $0.120$     & $8.6\times10^{-7}$ & $1.8\times10^{-6}$ \\
${}_{34}$Se & $0.208$     & $0.222$            & $2.4\times10^{-5}$ \\
${}_{35}$Br & $0.188$     & $0.0156$           & $2.3\times10^{-6}$ \\
${}_{40}$Zr & $0.007$     & $0.0405$           & $1.7\times10^{-6}$ \\
${}_{46}$Pd & $5.1\times10^{-4}$ & $0.0598$    & $1.4\times10^{-4}$ \\
${}_{52}$Te & $4.4\times10^{-7}$ & $0.0523$    & $0.137$ \\
${}_{58}$Ce & $<10^{-20}$ & $1.5\times10^{-7}$ & $0.00087$ \\
${}_{60}$Nd & $<10^{-20}$ & $2.6\times10^{-7}$ & $0.00149$ \\
${}_{62}$Sm & $<10^{-20}$ & $1.0\times10^{-7}$ & $0.00203$ \\
${}_{92}$U  & $<10^{-20}$ & $<10^{-20}$        & $0.026  $ \\
\hline
\end{tabular}
\label{tab:mfs_repr}
\end{table}

We first explore nuclear heating in the ejecta with the analytic power law fit
\citep[cf.][]{korobkin12}:
\begin{align}
  \dot{\epsilon}(t) = \epsilon_{\rm th}\cdot 2\times10^{10}\;t_d^{-1.3}\;
                      {\rm erg}\;{\rm g}^{-1}\;{\rm s}^{-1},
\label{eq:analytic_nuclear_heating}
\end{align}
where $t_d$ is time in days and $\epsilon_{\rm th}$ is a fraction of energy
that is left for thermalization (after all neutrinos and a certain fraction of
gammas escaped). This fraction is normally taken to be
$\epsilon\sim0.2-0.5$; see \cite{metzger10a} for details. We adopt a value of
$\epsilon=0.25$ in our models.

Models {\tt DZ}$_{1}$ and {\tt DZ}$_{2}$ explore the impact of increased
nuclear heating. The rates of nuclear heating depend on the properties of the nuclei at the
$r$-process path, which are currently unknown experimentally and highly
uncertain theoretically. In particular, compared to other nuclear mass models,
the FRDM nuclear mass model adopted in this work tends to underestimate heating
rates for the time scales of macronovae \citep[as demonstrated in][]{wu16}.
Fig.~7 in \cite{rosswog17a} shows one order of magnitude higher heating rates
for the Duflo-Zucker DZ31 nuclear mass model~\citep{duflo95}, computed with
the network of \cite{mendozatemis15}.
In models {\tt DZ}$_{1}$ and {\tt DZ}$_{2}$, we use the expression
(\ref{eq:analytic_nuclear_heating}) for heating, but increase the heating rate
in the dynamical ejecta by a factor of 10.
Otherwise, these models are identical to {\tt X}$_1$ and {\tt X}$_2$.

\begin{figure}
  \begin{center}
  \includegraphics[width=0.49\textwidth]{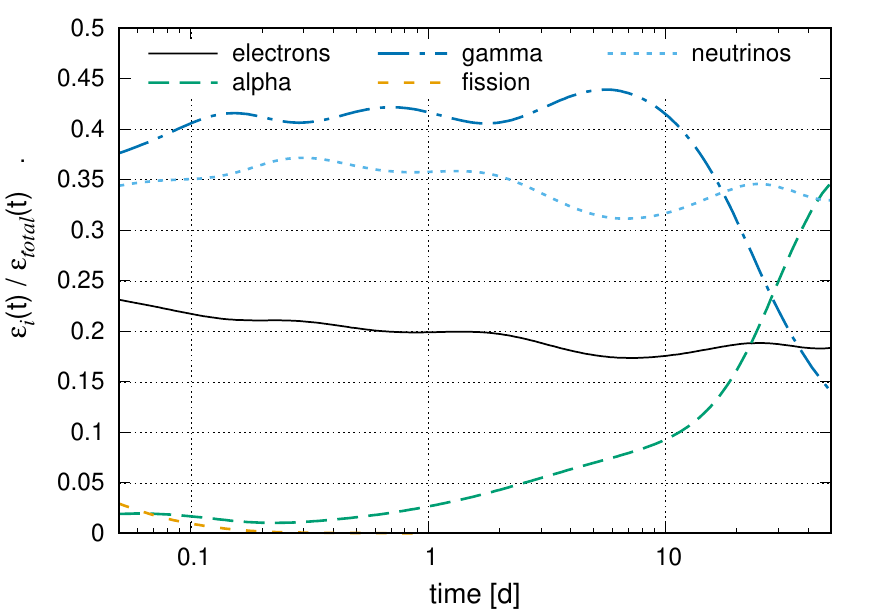}
  \\
  \includegraphics[width=0.49\textwidth]{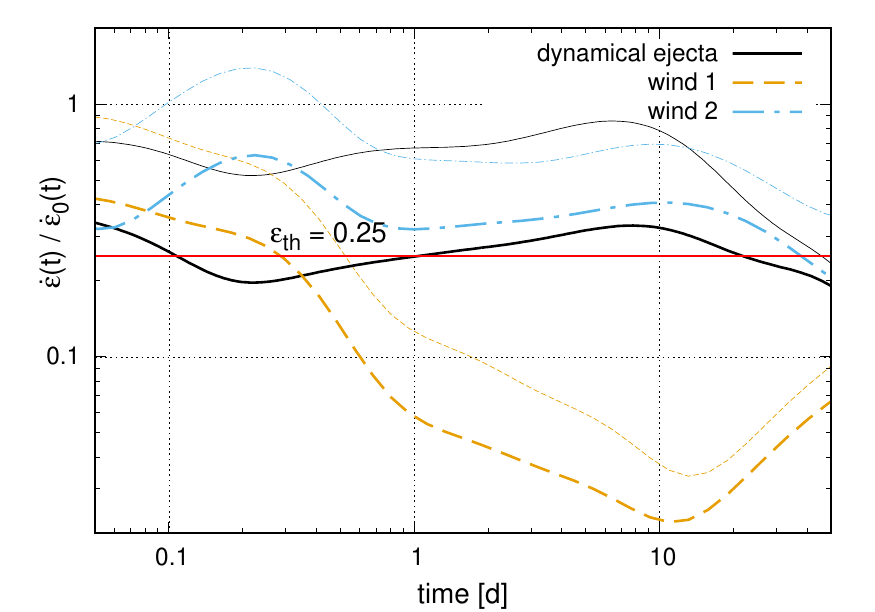}
  \end{center}
  \caption{
     Top: partitioning of nuclear energy release between different radioactive
     species, as a function of time, calculated by \WinNET\ for dynamical
     ejecta outflow.
     Bottom: peculiarities of the nuclear heating with (thin lines) and
     without (thick lines) contribution of $\gamma$-radiation, in dynamical
     ejecta and two different models of wind. Here, neutrinos are excluded,
     and the heating rates are normalized to the analytic power law
     (\ref{eq:analytic_nuclear_heating}). The rate with
     $\epsilon_{\rm th}=0.25$ used in simpler models is also shown for
     reference.
  } 
  \label{fig:hrate_dyn}
\end{figure}

For our advanced models {\tt X1nh}, {\tt X2nh} and for all $\gamma$-models (as
listed in Table~\ref{tab:models_2d}) we use instead detailed time-dependent
nuclear heating output from nucleosynthesis network which distinguishes
different radiation species.
The top panel of Fig.~\ref{fig:hrate_dyn} shows the evolution of the fractions of
nuclear heating rates which are carried away by different species, as computed
by \WinNET, and the bottom panel demonstrates total heating rates (without
neutrinos), normalized to the energy generation given in
Eq.~(\ref{eq:analytic_nuclear_heating}) for comparison.
Following the methodology developed by \cite{barnes16a}, we apply pointwise
density-dependent analytic prescription for thermalization efficiencies in the
wind and in the dynamical ejecta.

For a particle species "$i$" ($\alpha$-, $\beta$- or fission fragments),
the thermalization efficiency is calculated as follows:
\begin{align}
f_i(t,\mathbf{r})= \frac{\log{(1+2\eta_i^2)}}{2\eta_i^2},
\end{align}
where the coordinate- and time-dependent quantity $2\eta_i^2$
\citep[c.f.][]{rosswog17a} is defined as:
\begin{align}
2\eta^2_i(t,\mathbf{r})= \frac{2A_i}{t \rho(t,\mathbf{r})},
\end{align}
and the constants $A_i$ determine thermalization times:
$\{A_{\alpha},\;A_\beta,\;A_{\rm ff}\} =
\{1.2,\,1.3,\,0.2\}\times10^{-11} {\rm g}\,{\rm cm}^{-3}\,{\rm s}$.
These constants correspond to the choice of average particle energies
$E_{\alpha,0}=6\;{\rm MeV}$, $E_{\beta,0}=0.5\;{\rm MeV}$,
$E_{{\rm ff},0}=100\;{\rm MeV}$, and the same values of energy-loss rates for
different species as originally computed in \cite{barnes16a}
(cf. their Eqs. 19 and 25).

In models {\tt X1nh} and {\tt X2nh} we adopt coordinate-dependent
thermalization for $\gamma$-particles as well. Specifically, we use
a thermalization efficiency of $f_\gamma=1-e^{-\tau}$, where
\begin{align}
  &\tau(t, r) = \int_{vt}^{v_{\rm max}t} \kappa_{\gamma}\rho(t,r)dr,
\end{align}
and $\kappa_\gamma$ is an average opacity in the $\gamma$-ray band.
Instead of adopting an approximate expression for $\tau$, we calculate
the radial optical depth directly from our ejecta morphology and spatial
grid, assuming density is piecewise constant over the spatial cells.
The piecewise-constant treatment for density in the calculation of $\tau$
is consistent with the treatment of opacity and energy deposition
for Monte Carlo.
For all models with detailed r-process heating, we use a grey
$\gamma$-ray opacity value of $\kappa_\gamma = 0.1 \cmg$
\citep[same as used in][]{barnes16a}.

Total nuclear input at a position $\mathbf{r}$ and time $t$ is calculated as
a weighted average:
\begin{align}
\dot\epsilon(t,\mathbf{r})=
\frac{\rho_{\rm wind}(t,\mathbf{r}) \dot\epsilon_{\rm wind}(t,\mathbf{r})
     +\rho_{\rm dyn} (t,\mathbf{r}) \dot\epsilon_{\rm dyn} (t,\mathbf{r})}
     {\rho_{\rm wind}(t,\mathbf{r}) + \rho_{\rm dyn} (t,\mathbf{r})},
\end{align}
where heating contributions from the wind (model 1 or 2) and dynamical ejecta are
calculated separately according to the detailed nucleosynthesis and
composition in each of the components:
\begin{align}
\dot\epsilon_{\rm wind}= \sum_i f_i^{\rm wind}(t,\mathbf{r}) \dot\epsilon_i^{\rm wind}, \;\;
\dot\epsilon_{\rm dyn} = \sum_j f_j^{\rm dyn}(t,\mathbf{r})  \dot\epsilon_j^{\rm dyn}.
\end{align}
Subscripts $i$ and $j$ indicate the radioactive species.
The sums give the total heating rate available for local heating in
a parcel of the wind or dynamical ejecta~\citep{barnes16a,rosswog17a}.
In taking the average weighted by partial density of the sums, we are
assuming the dynamical ejecta and winds are uniformly mixed in the spatial
cells where they overlap.

\subsection{Radiative transfer and opacity}
\label{sec:opacities}

We compute our light curves and spectra with the radiative
transfer software \SuperNu~\citep{wollaeger14a}, with opacity
from the state-of-the-art Los Alamos suite of atomic physics
codes~\citep{fontes15a,fontes15b}.
Here, we describe some aspects of the radiative transfer and
opacity that make them viable for macronova simulations.

\subsubsection{Radiative transfer}
\label{sec:radiative_transfer}

\SuperNu\ is a multidimensional Monte Carlo radiative transfer code
specialized for synthesizing light curves and spectra of supernovae
\citep[see][]{wollaeger14a,wollaeger13,vanrossum16}.
More generally, the code is designed for modeling thermal radiative
transfer in expanding, partially ionized plasma with radioactive
sources \citep{wollaeger14a}.
\SuperNu\ has an implementation of Implicit Monte Carlo (IMC) and
Discrete Diffusion Monte Carlo (DDMC); DDMC accelerates simulations
with optically thick regions
\citep[see][]{fleck71,densmore07,densmore12,abdikamalov12}.
The Monte Carlo particles are tracked through a velocity grid, which
relates to the spatial grid through the homologous approximation
\citep{kasen06},
\begin{equation}
  \label{eq:homologous}
  \vec{v} = \frac{\vec{r}}{t} \;\;,
\end{equation}
where $\vec{v}$, $\vec{r}$, and $t$ are the velocity,
radial coordinate and time.

Relativistic corrections are accounted for in the radiative
transfer to order O($v/c$), and the effect of the radiation on the
ejecta momentum is assumed to be negligible.
These are often reasonable approximations for supernovae and
macronovae~\citep[see][]{kasen06,barnes13}.

The resulting tally of energy absorbed by the ejecta is used to update
the temperature in each spatial cell, using the standard IMC approach
\citep{fleck71}.
The IMC equation for temperature is~\citep{fleck71,wollaeger13},
\begin{equation}
  \label{eq:imc_temp}
  C_{v,n}\frac{DT}{Dt} = \mathcal{E} - f_{n}\sigma_{P,n}acT_{n}^{4}
  + f_{n}\rho_{n}\dot{\epsilon} \;\;,
\end{equation}
where $C_{v,n}$, $\sigma_{P,n}$, $T_{n}$, and $\rho_{n}$ are
the heat capacity, Planck opacity, temperature, and density at
time step $n$.
The Fleck factor,
\begin{equation}
  \label{eq:fleck_factor}
  f_{n} = \frac{1}{1+4aT_{n}^{3}\sigma_{P,n}c\Delta t_{n}/C_{v,n}} \;\;,
\end{equation}
is a result of semi-implicitly discretizing the temperature
(or internal energy) equation in time~\citep{fleck71}.
The value of $\mathcal{E}$ is the rate of energy effectively
absorbed in the comoving frame during time step $n$, per unit volume.
Approximations made to obtain equation (\ref{eq:imc_temp}) are consistent
with those typically made for radiative transfer in supernovae
\citep[see][]{kasen06}, and are also valid for macronovae
\citep{barnes13,kasen13,hotokezaka13}.

\begin{figure}
  \begin{center}
    \begin{tabular}{c}
    \includegraphics[width=0.49\textwidth]{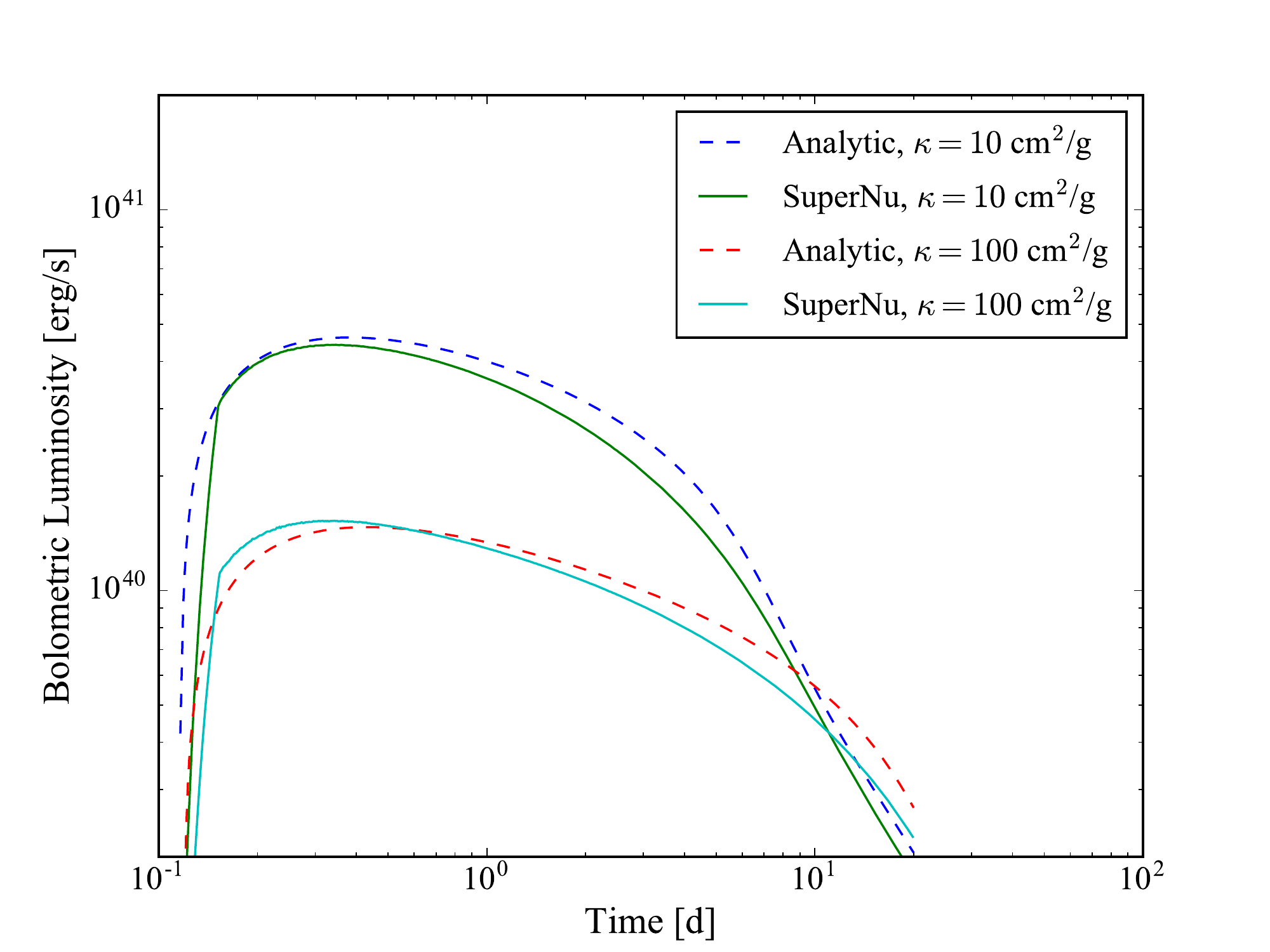}
    \end{tabular}
  \end{center}
  \caption{
    Bolometric light curves for a simple macronova-type
    spherical outflow from \SuperNu\ and an analytic model.
  }
\label{fig:supernu_vs_analytic}
\end{figure}

We supply a simple analytic verification here to ensure the radiative
transfer produces accurate luminosities for macronova-type problems.
The test problem has a uniform density with a total mass of 0.01 $M_{\odot}$,
a maximum outflow speed of 0.25$c$, and a uniform grey absorption opacity
of either 10 or 100~$\cmg$.
The r-process heating rate is the analytic model in equation
(\ref{eq:analytic_nuclear_heating}), with $\epsilon_{\rm th}=0.25$.
The problem is started at 10000 seconds, with an initial uniform temperature
of $1.5\times10^4$ K.
To derive the luminosity benchmark, we employ the normalizations and
Fourier series expansion technique described by~\cite{pinto00}.
For this problem, Fig.~\ref{fig:supernu_vs_analytic} has bolometric light
curves from the analytic model and \SuperNu.
The analytic solution is of the equilibrium comoving radiation diffusion
equation with a simple outer-boundary condition.
Further details of this solution can be found in the Appendix~\ref{app:analytic}.

Apart from analytic radiative transfer solutions
\citep{wollaeger13,wollaeger14a}, \SuperNu\ has been tested against
other supernova light curve codes.
Benchmarks include the deterministic code {\tt PHOENIX}~\citep{vanrossum12}
for the W7 model of Type Ia supernovae (SN Ia)~\citep{wollaeger14a}, and the codes
{\tt STELLA}~\citep{blinnikov06}, {\tt RHMC}~\citep{noebauer12}, and
{\tt V1D}~\citep{livne93} for a grey pair-instability supernova
model~\citep{kozyreva17}.
For the SN Ia W7 comparison with the {\tt PHOENIX} code, the peak bolometric
luminosities differ by $\sim10-15$~\% (with subsequent more controlled comparisons
bringing this to $\sim5-10$ \%), and very close spectral profiles
\citep[see Figs. 7 and 8 of][]{wollaeger14a}.
Similarly close agreement was found with the other codes for the grey pair-instability
supernova model \citep[see Fig. 9 of][]{kozyreva17}.
For a double-degenerate white dwarf merger model, light curves and spectra from
\SuperNu\ have been compared to observations of the slowly declining SN Ia, SN 2001ay,
producing similar broadband magnitudes and spectra
\citep[see Figs. 8 and 12 of][]{vanrossum16}.

For realistic opacities, \SuperNu\ calculates bound-bound contributions
with line lists, and tabulated data for bound-free and free-free contributions
\citep{verner96,sutherland98}.
These contributions are added into a 100-1000 group wavelength grid, which
is defined in the ejecta's comoving frame, typically spanning UV (.01 $\mu$m)
to IR (3.2 $\mu$m) for supernovae~\citep{wollaeger14a}.
Lines are treated like Dirac delta functions when they are grouped, so they
each only contribute to one group.
This also spreads the contribution of the line over the group.
During the transport phase, Monte Carlo particles sample collision
distances from only the resulting grouped opacity structure.
Thus, unlike the typical Sobolev expansion opacity formalism in Monte Carlo
codes (see, for instance,~\cite{kasen06,kromer09}), line transfer is not
directly treated by~\SuperNu.

\subsubsection{Opacity}
\label{sec:opacity_subsec}

We use the Los Alamos suite of atomic
physics codes \citep{fontes15b} to calculate the detailed multifrequency LTE
opacities for the few representative elements listed in
Table~\ref{tab:mfs_repr}. The elements are selected to
represent the variety of compositions in the dynamical ejecta and in different
types of winds (see Fig.~\ref{fig:mass_fractions}): Lanthanides (Sm, Ce, Nd), an
Actinide (U), lighter wind (Cr, Se, Br) and heavier wind (Zr, Pd, Te).
The opacities are calculated on a 27-point temperature grid
$0.01\ {\rm eV} \le k_BT \le 5\ {\rm eV}$ for density values sampled for every
decade from $\rho_{\rm min}=10^{-20}\ \gcc$ to $\rho_{\rm max}=10^{-4}\ \gcc$.
These temperature and density ranges suffice to cover the typical
thermodynamic conditions encountered in expanding dynamical ejecta around the
epoch when macronovae peak.

\begin{figure}
  \begin{tabular}{c}
    \includegraphics[width=0.49\textwidth]{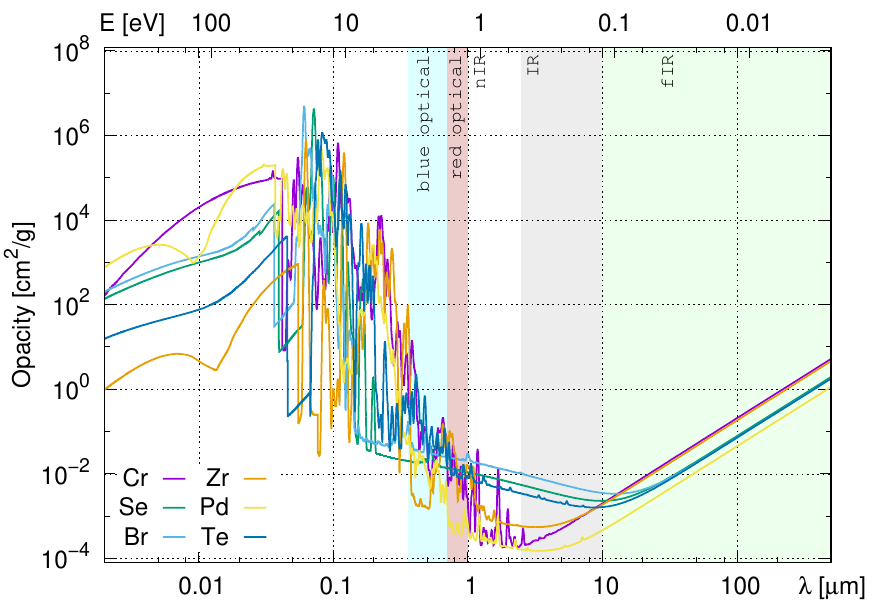}
    \\
    \includegraphics[width=0.49\textwidth]{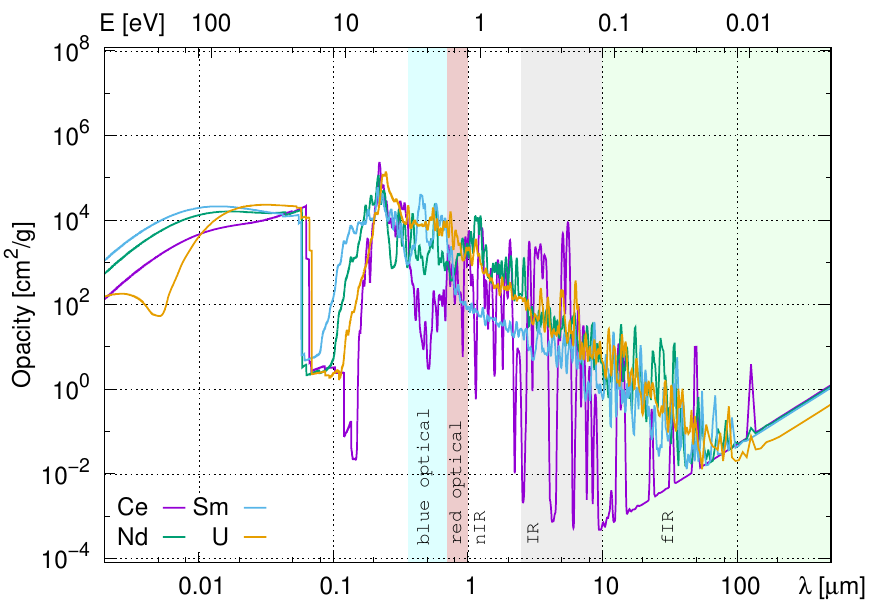}
  \end{tabular}
  \caption{
     Profiles of opacity for pure elemental plasma at LTE for temperature
     $T=5800$~K ($k_BT=0.5$~eV) and density $\rho=10^{-13}\;\gcc$. Top: elements representing
     the wind; bottom: heavy $r$-process lanthanides and actinides from  dynamical ejecta.
     The dips in opacity around 0.1~$\mu m$ are artificial, caused by the
     limited choice of transitions included in the atomic physics models.
  } 
  \label{fig:opac_sm_loglog}
\end{figure}

Figure~\ref{fig:opac_sm_loglog} illustrates typical opacity profiles for a number
of representative elements for plasma density $\rho=10^{-13}\;\gcc$ and temperature
$T=0.5$~eV.
As can be seen in the lower panel of Fig.~\ref{fig:opac_sm_loglog}, the opacities contain an
artificial window from $\sim0.06-0.17$ $\mu$m, due to the limited choice of transitions
that were included in the atomic physics model in that range.
For some of our simulations, this window in the opacity causes artificially
enhanced emission in that wavelength range.

In our simulations, we do not apply the expansion opacity formalism
\citep{karp77,eastman93}, which employs the Sobolev approximation
\citep{sobolev60} and is adopted in previous detailed macronova radiative
transfer calculations~\citep{barnes13,kasen13,tanaka13}.
The method for calculating the opacities employed in this work has been
previously described by \cite{fontes15a,fontes17a}. Briefly, the lines are
broadened using an effective Doppler width in the Voigt profile by
$\Delta\lambda/\lambda\sim\Delta v/c\sim0.01$.
For the deterministic calculations of~\cite{fontes17a}, this approach is
used to take into account the shifting of lines due to the velocity gradients
in the dynamical ejecta. The same approach is used here for both
the wind and the dynamical ejecta as their velocity gradients
are similar for all the models we simulate.
This approach preserves the integral of the wavelength-dependent opacities,
which can generate significantly larger values than those produced via the
expansion opacity formalism.
A practical advantage of this line-smeared approach is that the
wavelength-dependent opacities can be represented with a reasonable number
of photon energy points, making possible the generation of opacity tables
that can be used in an efficient look-up approach in radiation transport
simulations.

Of relevance for the application of the line-smeared opacities in~\SuperNu\ is
the extent of line smearing relative to the group sizes in the wavelength grid.
The group structure used in the radiative transfer step is logarithmic, with
$\Delta\lambda/\lambda\sim 0.05$ for each group.
Consequently, our group structure does not resolve the smeared lines.
The groups, however, should resolve P-Cygni features that may appear in the
spectra (these arise from the separation of line absorption and emission features
in the spectra due to the velocity of the ejecta).
For instance, at day 5 of the expansion for the density profile in equation
\eqref{eq:rho_sphsym}, $\kappa=10$ cm$^2$/g gives a photosphere at about
$v=0.16c$.
For a P-Cygni line feature at this photosphere, the span of wavelength between
the emission feature and the absorption feature is
$\Delta \lambda/\lambda = v_{\rm photo}/c = 0.16$, where $\lambda$ is the
line center.
This value of $\Delta \lambda/\lambda$ is a factor of $\sim3$ larger than
that of the multigroup grid, which in turn is a factor of $\sim5$ larger
than the effective broadening from the line smearing.
These wavelength scales provide some justification for the use of line-smeared
opacities and multigroup in the present simulations.

For some numerical justification, \cite{fontes17a} compare light curves from
\SuperNu\ simulations of a pure-iron W7-type ejecta, using broadened LANL
opacities, or~\SuperNu's default opacity calculation (see~\ref{sec:radiative_transfer}).
For \SuperNu's default opacity calculation, Fe lines were obtained from the Kurucz line
list\footnote{\url{http://kurucz.harvard.edu/atoms.html}}.
The light curves show a discrepancy of $\sim$20\% in the peak luminosities of
broadband and bolomentric light curves.

\subsubsection{Time resolution test}
\label{sec:time_res_test}

\begin{figure}
  \begin{center}
    \begin{tabular}{c}
    \includegraphics[width=0.49\textwidth]{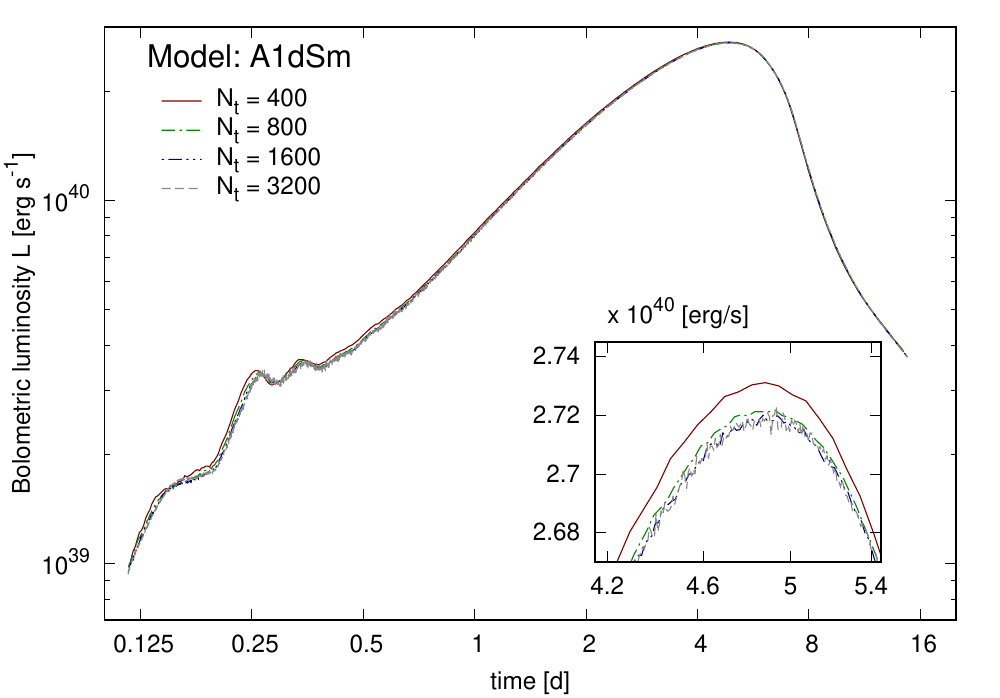}
    \end{tabular}
  \end{center}
  \caption{
     Bolometric light curves for four different time resolutions used in
     spherically-symmetric dynamical ejecta model A with LTE opacities of
     elemental Sm plasma ({\tt A1dSm}, see Table~\ref{tab:dyn_models}).
     Number of timesteps covers 20 days in the comoving frame.}
\label{fig:conv_A1dSm}
\end{figure}

With the opacity and radiative transfer methods in place, we performed
several resolution tests.
For instance, to properly select temporal numerical resolution, we tested
the sensitivity of our results to timestep size on spherically-symmetric model
of dynamical ejecta with LTE opacities of elemental Sm (see Table~\ref{tab:dyn_models}).
Fig.~\ref{fig:conv_A1dSm} shows the bolometric luminosity for four different
resolutions with progressively smaller timesteps, covering time period of 20
days with $N_t=400$, $800$, $1600$ and $3200$ timesteps.
The number of photon wavelength groups ($N_{\lambda}=100$) and the number of
radial cells ($N_r = 128$) were kept constant.
Under these conditions, the light curves show clear first-order convergent
behavior, as expected from the numerical scheme. We have therefore selected
the highest resolution $N_t=3200$ per 20 days (in comoving frame) everywhere
in this study.

\section{Sensitivity studies}
\label{sec:sensitivity}

\begin{table*}
\caption{Parameters of the models used in this study.}
\begin{adjustbox}{width=\textwidth}
\begin{tabular}{rlccccccc}
\hline
\hline
      &        & density profile
                       & $m_{\rm wind}/m_{\rm ej}$
                               & $\bar{v}_{\rm wind}/{\bar{v}_{\rm ej}}^\dagger$
                               & $\varkappa_{\rm wind}/\varkappa_{\rm ej}$
                                                          & nuclear & thermali- & $\gamma$-ray\\
      &Model   & wind + dyn. ejecta & $[M_\odot]$
                               & $[c]$
                               & $[{\rm cm}^2\;g^{-1}]$
                                                          & heating &  zation   & treatment  \\
\hline
               & SAm1  & 1D analytic & 0.001 & 0.125 & 10 &       &                          & -- \\
               & SAm2  & 1D analytic & 0.01  & 0.125 & 10 &       &                          & -- \\
               & SAm3  & 1D analytic & 0.1   & 0.125 & 10 &       &                          & -- \\
Section~\ref{sec:sa_grey}:
               & SAv1  & 1D analytic & 0.014 & 0.05  & 10 &       &                          & -- \\
 spherical     & SAv2  & 1D analytic & 0.014 & 0.10  & 10 & power & $\epsilon_{\rm th}=0.25$ & -- \\
  symmetry,    & SAv3  & 1D analytic & 0.014 & 0.15  & 10 &  law  &                          & -- \\
grey opacity   & SAk0  & 1D analytic & 0.014 & 0.125 & 1  &       &                          & -- \\
               & SAk1  & 1D analytic & 0.014 & 0.125 & 10 &       &                          & -- \\
               & SAk2  & 1D analytic & 0.014 & 0.125 &100 &       &                          & -- \\
               & SAk3  & 1D analytic & 0.014 & 0.125 &1000&       &                          & -- \\
\hline
               & SASe  & 1D analytic & 0.014 & 0.125 & $\varkappa_{\rm Se}$    &       &                          & -- \\
               & SABr  & 1D analytic & 0.014 & 0.125 & $\varkappa_{\rm Br}$    &       &                          & -- \\
               & SATe  & 1D analytic & 0.014 & 0.125 & $\varkappa_{\rm Te}$    &       &                          & -- \\
               & SAPd  & 1D analytic & 0.014 & 0.125 & $\varkappa_{\rm Pd}$    &       &                          & -- \\
Section~\ref{sec:sa_multi}:
               & SAZr  & 1D analytic & 0.014 & 0.125 & $\varkappa_{\rm Zr}$    &       &                          & -- \\
  spherical    & SACr  & 1D analytic & 0.014 & 0.125 & $\varkappa_{\rm Cr}$    &       &                          & -- \\
  symmetry,    & SACe  & 1D analytic & 0.014 & 0.125 & $\varkappa_{\rm Ce}$    & power & $\epsilon_{\rm th}=0.25$ & -- \\
 multrigroup   & SASm  & 1D analytic & 0.014 & 0.125 & $\varkappa_{\rm Sm}$    &  law  &                          & -- \\
 opacity       & SANd  & 1D analytic & 0.014 & 0.125 & $\varkappa_{\rm Nd}$    &       &                          & -- \\
               & SAU   & 1D analytic & 0.014 & 0.125 & $\varkappa_{\rm U }$    &       &                          & -- \\
               & SAw1  & 1D analytic & 0.014 & 0.125 & $\varkappa_{\rm wind1}$ &       &                          & -- \\
               & SAw2  & 1D analytic & 0.014 & 0.125 & $\varkappa_{\rm wind2}$ &       &                          & -- \\
               & SAd   & 1D analytic & 0.014 & 0.125 & $\varkappa_{\rm dyn}$   &       &                          & -- \\
\hline
               & A1dSm & 1D spherically-averaged A  & 0.013 & 0.132 & $\varkappa_{\rm Sm}$ &       &                          & -- \\
               & B1dSm & 1D spherically-averaged B  & 0.014 & 0.125 & $\varkappa_{\rm Sm}$ &       &                          & -- \\
Section~\ref{sec:dynamical}:
               & C1dSm & 1D spherically-averaged C  & 0.033 & 0.132 & $\varkappa_{\rm Sm}$ &       &                          & -- \\
dyn. ejecta,   & D1dSm & 1D spherically-averaged D  & 0.034 & 0.136 & $\varkappa_{\rm Sm}$ & power & $\epsilon_{\rm th}=0.25$ & -- \\
spherical      & A1dmSm& 1D spherically-averaged A  & 0.013 & 0.066 & $\varkappa_{\rm Sm}$ &  law  &                          & -- \\
symmetry       & B1dmSm& 1D spherically-averaged B  & 0.014 & 0.080 & $\varkappa_{\rm Sm}$ &       &                          & -- \\
               & C1dmSm& 1D spherically-averaged C  & 0.033 & 0.055 & $\varkappa_{\rm Sm}$ &       &                          & -- \\
               & D1dmSm& 1D spherically-averaged D  & 0.034 & 0.058 & $\varkappa_{\rm Sm}$ &       &                          & -- \\
\hline
               & A2dSm & 2D axisymmetric A          & 0.013 & 0.095 & $\varkappa_{\rm Sm}$ &       &                            & -- \\
Section~\ref{sec:dynamical}:
               & B2dSm & 2D axisymmetric B          & 0.014 & 0.086 & $\varkappa_{\rm Sm}$ & power & $\epsilon_{\rm th} = 0.25$ & -- \\
axisymmetry    & C2dSm & 2D axisymmetric C          & 0.033 & 0.119 & $\varkappa_{\rm Sm}$ &  law  &                            & -- \\
               & D2dSm & 2D axisymmetric D          & 0.034 & 0.121 & $\varkappa_{\rm Sm}$ &       &                            & -- \\
\hline
               & W2A (W2)  & 1D analytic + 2D axisym. A & $0.005+0.013$ & $0.08+0.095$ & $\varkappa_{\rm Zr}+\varkappa_{\rm Sm}$ &       &                             & -- \\
               & W2B       & 1D analytic + 2D axisym. B & $0.005+0.014$ & $0.08+0.086$ & $\varkappa_{\rm Zr}+\varkappa_{\rm Sm}$ &       &                             & -- \\
               & W2C       & 1D analytic + 2D axisym. C & $0.005+0.033$ & $0.08+0.119$ & $\varkappa_{\rm Zr}+\varkappa_{\rm Sm}$ &       &                             & -- \\
               & W2D       & 1D analytic + 2D axisym. D & $0.005+0.034$ & $0.08+0.121$ & $\varkappa_{\rm Zr}+\varkappa_{\rm Sm}$ &       &                             & -- \\
               & W2Se      & 1D analytic + 2D axisym. A & $0.005+0.013$ & $0.08+0.095$ & $\varkappa_{\rm Se}+\varkappa_{\rm Sm}$ &       &                             & -- \\
Section~\ref{sec:combined}:
               & W2Br      & 1D analytic + 2D axisym. A & $0.005+0.013$ & $0.08+0.095$ & $\varkappa_{\rm Br}+\varkappa_{\rm Sm}$ &       &                             & -- \\
dyn. ejecta    & W2Te      & 1D analytic + 2D axisym. A & $0.005+0.013$ & $0.08+0.095$ & $\varkappa_{\rm Te}+\varkappa_{\rm Sm}$ & power &  $\epsilon_{\rm th} = 0.25$ & -- \\
+ wind         & W2Pd      & 1D analytic + 2D axisym. A & $0.005+0.013$ & $0.08+0.095$ & $\varkappa_{\rm Pd}+\varkappa_{\rm Sm}$ &  law  &                             & -- \\
               & W2Zr (W2) & 1D analytic + 2D axisym. A & $0.005+0.013$ & $0.08+0.095$ & $\varkappa_{\rm Zr}+\varkappa_{\rm Sm}$ &       &                             & -- \\
               & W2Cr      & 1D analytic + 2D axisym. A & $0.005+0.013$ & $0.08+0.095$ & $\varkappa_{\rm Cr}+\varkappa_{\rm Sm}$ &       &                             & -- \\
               & W2light   & 1D analytic + 2D axisym. A & $0.001+0.013$ & $0.08+0.095$ & $\varkappa_{\rm Zr}+\varkappa_{\rm Sm}$ &       &                             & -- \\
               & W2heavy   & 1D analytic + 2D axisym. A & $0.02 +0.013$ & $0.08+0.095$ & $\varkappa_{\rm Zr}+\varkappa_{\rm Sm}$ &       &                             & -- \\
               & W2slow    & 1D analytic + 2D axisym. A & $0.005+0.013$ & $0.04+0.095$ & $\varkappa_{\rm Zr}+\varkappa_{\rm Sm}$ &       &                             & -- \\
               & W2fast    & 1D analytic + 2D axisym. A & $0.005+0.013$ & $0.16+0.095$ & $\varkappa_{\rm Zr}+\varkappa_{\rm Sm}$ &       &                             & -- \\
\hline
                    & X$_1$     & 1D analytic + 2D axisym. A & $0.005+0.013$ & $0.08+0.095$ & $\varkappa_{\rm wind 1}+\varkappa_{\rm dyn}$ & power       & $\epsilon_{\rm th}= 0.25$ & -- \\
Section~\ref{sec:realistic_models}:
                    & X$_2$     & 1D analytic + 2D axisym. A & $0.005+0.013$ & $0.08+0.095$ & $\varkappa_{\rm wind 2}+\varkappa_{\rm dyn}$ &  law        & $\epsilon_{\rm th}= 0.25$ & -- \\
detailed composi-   & DZ$_1$    & 1D analytic + 2D axisym. A & $0.005+0.013$ & $0.08+0.095$ & $\varkappa_{\rm wind 1}+\varkappa_{\rm dyn}$ & $\times 10$ & $\epsilon_{\rm th}= 0.25$ & -- \\
tion and nuclear    & DZ$_2$    & 1D analytic + 2D axisym. A & $0.005+0.013$ & $0.08+0.095$ & $\varkappa_{\rm wind 2}+\varkappa_{\rm dyn}$ & $\times 10$ & $\epsilon_{\rm th}= 0.25$ & -- \\
heating             & Xnh$_1$   & 1D analytic + 2D axisym. A & $0.005+0.013$ & $0.08+0.095$ & $\varkappa_{\rm wind 1}+\varkappa_{\rm dyn}$ &  from       & species-                  & -- \\
                    & Xnh$_2$   & 1D analytic + 2D axisym. A & $0.005+0.013$ & $0.08+0.095$ & $\varkappa_{\rm wind 2}+\varkappa_{\rm dyn}$ & network     & dependent                 & -- \\
\hline
\hline
            & $\gamma A_1$ & 1D analytic + 2D axisym. A & $0.005+0.013$ & $0.08+0.095$ & $\varkappa_{\rm wind 1}+\varkappa_{\rm dyn}$ &         &           & grey \\
            & $\gamma B_1$ & 1D analytic + 2D axisym. B & $0.005+0.014$ & $0.08+0.086$ & $\varkappa_{\rm wind 1}+\varkappa_{\rm dyn}$ &         &           & grey \\
            & $\gamma C_1$ & 1D analytic + 2D axisym. C & $0.005+0.033$ & $0.08+0.119$ & $\varkappa_{\rm wind 1}+\varkappa_{\rm dyn}$ &         &           & grey \\
Section~\ref{sec:realistic_models}:
            & $\gamma D_1$ & 1D analytic + 2D axisym. D & $0.005+0.034$ & $0.08+0.121$ & $\varkappa_{\rm wind 1}+\varkappa_{\rm dyn}$ &  from   & species-  & grey \\
realistic models
            & $\gamma A_2$ & 1D analytic + 2D axisym. A & $0.005+0.013$ & $0.08+0.095$ & $\varkappa_{\rm wind 2}+\varkappa_{\rm dyn}$ & network & dependent & grey \\
            & $\gamma B_2$ & 1D analytic + 2D axisym. B & $0.005+0.014$ & $0.08+0.086$ & $\varkappa_{\rm wind 2}+\varkappa_{\rm dyn}$ &         &           & grey \\
            & $\gamma C_2$ & 1D analytic + 2D axisym. C & $0.005+0.033$ & $0.08+0.119$ & $\varkappa_{\rm wind 2}+\varkappa_{\rm dyn}$ &         &           & grey \\
            & $\gamma D_2$ & 1D analytic + 2D axisym. D & $0.005+0.034$ & $0.08+0.121$ & $\varkappa_{\rm wind 2}+\varkappa_{\rm dyn}$ &         &           & grey \\
\hline
\hline
\end{tabular}
\end{adjustbox}
\begin{flushleft}
{\small ($^\dagger$) For 1D analytic density profiles, $\bar{v}=v_{\rm max}/2$,
 and for the numerical density distributions it is median velocity, namely such
 that half the mass moves faster, while the other half is slower than $\bar{v}$.}
\end{flushleft}
\label{tab:allmodels}
\end{table*}

We covered a range of models with progressively increasing levels of
sophistication, gradually adding ingredients and observing their impact on the
light curves and spectra.
Model parameters are summarized in Table~\ref{tab:allmodels}.

We start with simple models with wavelength-independent "grey" opacity and
spherically-symmetric analytic density profiles (see Sect.~\ref{sec:sa_grey}).
For nuclear heating, the analytic power law fitting
formula~(\ref{eq:analytic_nuclear_heating}) is adopted with thermalization
efficiency $\epsilon_{\rm th}=0.25$.
These models explore a range of masses ({\tt SAm1}-{\tt SAm3}), median expansion
velocities ({\tt SAv1}-{\tt SAv3}) and grey opacities ({\tt SAk0}-{\tt SAk2}).
We then explore the impact of composition (Sect.~\ref{sec:sa_multi}) by
upgrading to multigroup opacity for single-element LTE plasmas, for ten
representative elements (models {\tt SASe} - {\tt SAU}).
In all models with multigroup opacity, the opacity is binned into
$N_{\lambda}=100$ logarithmically spaced bins covering a wavelength range from
$\lambda_{\rm min}=0.1$ to $\lambda_{\rm max}=12.8$ micron.
Then, we use a simple mixing scheme to simulate multi-species composition models
(introduced in Sect.~\ref{sec:composition}) of dynamical ejecta ({\tt SAd}) and
two representative types of wind ({\tt SAw1} and {\tt SAw2}).

Next, Section~\ref{sec:dynamical} describes models of just the dynamical ejecta.
Here we use different averaging for the dynamical ejecta simulations:
spherically-symmetric ({\tt A1dSm}-{\tt D1dSm}), axisymmetric
({\tt A2dSm}-{\tt D2dSm}) and another set of spherically-symmetric models with
a different type of averaging ({\tt A1dmSm}-{\tt D1dmSm}). In all these models,
the same multigroup LTE opacities of Sm are employed to represent the
lanthanides.

Combined models of the dynamical ejecta and wind are detailed in
Sect.~\ref{sec:combined}. {\tt W2A} - {\tt W2D} combine
spherically-symmetric wind with different morphologies of dynamical ejecta to
simulate the impact of a "lanthanide curtain" on potential blue transients from
the wind. {\tt W2Se}-{\tt W2Cr} demonstrate variation in the macronova signature
depending on the composition of the wind, while {\tt W2light}/{\tt W2heavy}
and {\tt W2slow}/{\tt W2fast} explore sensitivity to the wind mass and velocity.
The mixed multi-species composition of wind ("wind 1" and "wind 2", see
Table~\ref{tab:mfs_repr}) and dynamical ejecta are employed in models
{\tt X1}/{\tt X1}.

In models {\tt X1nh} and {\tt X2nh}, macronova signals are calculated with
upgraded detailed nuclear heating output and separate density-dependent
thermalization efficiencies, as described in Sect.~\ref{sec:composition}.
In these models, a simple ray-trace is used with a calibrated grey opacity
of 0.1~$\cmg$ to estimate the thermalization efficiency for $\gamma$-rays
\citep{barnes16a}.
Finally, in the most sophisticated set of models, $\gamma A_1$-$\gamma D_2$, the
$\gamma$-ray thermalization is replaced with energy deposition calculated
from a grey, pure absorbing, Monte Carlo treatment~\citep{swartz95}
\citep[again using the calibrated grey opacity of][]{barnes16a}.

\subsection{Semianalytic models: grey opacity}
\label{sec:sa_grey}

The simplest models that we explore have grey opacity and
spherically-symmetric analytic density distributions (described in
Sect.~\ref{sec:selfsim}).  These models can be characterized by only three
parameters: ejecta mass $m_{\rm ej}$, grey opacity $\kappa$ and expansion
velocity $v$.  We compare these models to the ones studied in
\cite{grossman14}.  Parameters of these models are listed in
Table~\ref{tab:sa_models}. The baseline model ({\tt SAk1}) implements dynamical
ejecta mass and expansion velocity from simulations of a most typical neutron
star binary with masses $1.4\ M_\odot+1.3\ M_\odot$ \citep[model B
from][]{rosswog14a}.

\begin{table}
\caption{Summary of spherically-symmetric analytic ({\tt SA*}) models.
All models use a logarithmic time grid with $N_t=1600$ time steps.
Columns specify: total ejecta mass, half the maximum ejecta velocity
$v_{\rm max}/2\approx\bar{v}$ (see Eq.\ref{eq:mas_vave}), opacity model,
bolometric peak time, peak bolometric luminosity, and the effective
blackbody temperature.}
\begin{tabular}{lcccccc}
\hline
      & $m_{\rm ej}$ & ${v_{\rm max}}/{2}$
      &  $\kappa$    & $t_p$ & $L_p$ & $T_{{\rm eff},p}$ \\
Model & $[M_\odot]$  & $[c]$
      &  $[\frac{{\rm cm}^2}{g}]$
                     & $[d]$ & $[\frac{10^{40}{\rm erg}}{{\rm s}}]$
                                     & $[K]$             \\
\hline
SAm1  & 0.001 & 0.125 & 10 & 0.46 & 1.18  & 4092 \\
SAm2  & 0.01  & 0.125 & 10 & 0.92 & 3.32  & 3724 \\
SAm3  & 0.1   & 0.125 & 10 & 2.00 & 8.41  & 3173 \\
SAv1  & 0.014 & 0.05  & 10 & 1.74 & 1.90  & 3668 \\
SAv2  & 0.014 & 0.10  & 10 & 1.17 & 3.17  & 3622 \\
SAv3  & 0.014 & 0.15  & 10 & 0.89 & 4.50  & 3730 \\
SAk0  & 0.014 & 0.125 & 1  & 0.51 & 13.84 & 7233 \\
SAk1  & 0.014 & 0.125 & 10 & 1.01 &  3.82 & 3728 \\
SAk2  & 0.014 & 0.125 &100 & 2.22 & 0.958 & 1742 \\
SAk3  & 0.014 & 0.125 &1000& 5.68 & 0.228 &  753 \\
\hline
SASe  & 0.014 & 0.125 & $\kappa_{\rm Se}$ & 0.276 & 56.03 & 16000${}^*$ \\
SABr  & 0.014 & 0.125 & $\kappa_{\rm Br}$ & 0.364 & 47.72 & 16000${}^*$ \\
SATe  & 0.014 & 0.125 & $\kappa_{\rm Te}$ & 0.383 & 37.78 & 16000${}^*$ \\
SAPd  & 0.014 & 0.125 & $\kappa_{\rm Pd}$ & 0.393 & 29.91 & 14500${}^*$ \\
SAZr  & 0.014 & 0.125 & $\kappa_{\rm Zr}$ & 0.261 & 19.07 & 11000${}^*$ \\
SACr  & 0.014 & 0.125 & $\kappa_{\rm Cr}$ & 1.093 & 13.11 & 8500${}^*$  \\
SACe  & 0.014 & 0.125 & $\kappa_{\rm Ce}$ & 3.779 &  2.61 & 5500${}^*$  \\
SASm  & 0.014 & 0.125 & $\kappa_{\rm Sm}$ & 4.973 &  2.60 & 1400${}^*$  \\
SANd  & 0.014 & 0.125 & $\kappa_{\rm Nd}$ & 6.388 &  1.23 & 1200${}^*$  \\
SAU   & 0.014 & 0.125 & $\kappa_{\rm U }$ & 3.008 &  4.42 & 1700${}^*$  \\
\hline
SAw1  & 0.014 & 0.125 & $\kappa_{\rm wind1}$ & 0.295 & 21.09 & 16000${}^*$ \\
SAw2  & 0.014 & 0.125 & $\kappa_{\rm wind2}$ & 0.402 & 30.82 & 16000${}^*$ \\
SAd   & 0.014 & 0.125 & $\kappa_{\rm dyn}$   & 3.335 &  3.32 &  1400${}^*$ \\
\hline \\
\end{tabular}
{\small ($^*$) Temperature of a blackbody spectrum with the closest fit
              (as a function of wavelength $\lambda$).}
\label{tab:sa_models}
\end{table}

\begin{figure}
\begin{center}
\begin{tabular}{c}
\includegraphics[width=0.49\textwidth]{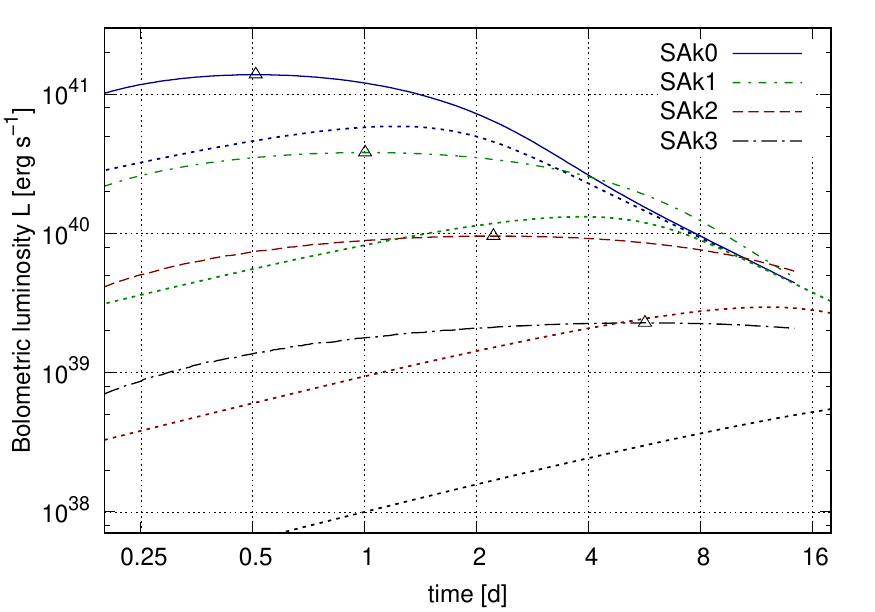}
\end{tabular}
\end{center}
\caption{Bolometric luminosity for the analytic spherically-symmetric density
distribution models {\tt SAk0} -- {\tt SAk3}, computed with \SuperNu\ (thick
solid and dashed lines), compared to luminosity estimates based on
grey-opacity Grossman models (thin dotted lines).
Black triangles indicate luminosity maxima.}
\label{fig:lum_SAkX}
\end{figure}

Figure~\ref{fig:lum_SAkX} displays time evolution of bolometric luminosity for
the four models with the range of grey opacities, {\tt SAk0} -- {\tt SAk3}
(thick lines), along with the light curves produced with a simple semianalytic
model from \cite{grossman14} (thin dotted lines on the plot).
Triangle marks show locations of luminosity maxima for each of the models.
As can be seen from the plot and more clearly in Fig.~\ref{fig:powerfits},
the peak epochs $t_p$ and peak luminosities $L_p$ clearly follow a power law
$L_p\propto t_p^{-1.7}$ with power index $\approx1.7$
that is close to the analytic result $\alpha=1.3$ for the Grossman
models.
All calculations with full radiative transfer show an extended
plateau with very small
variation in luminosity, while the Grossman models instead exhibit a steeper rise
and later peak times. Grossman models also underestimate bolometric
luminosity, especially for high values of grey opacity, where the discrepancy
exceeds one order of magnitude.
This underestimate is likely related, in part, to the fact that the thermal
contribution of the ejecta is completely neglected in the Grossman models.

\begin{figure}
  \begin{center}
  \begin{tabular}{c}
  \includegraphics[width=0.49\textwidth]{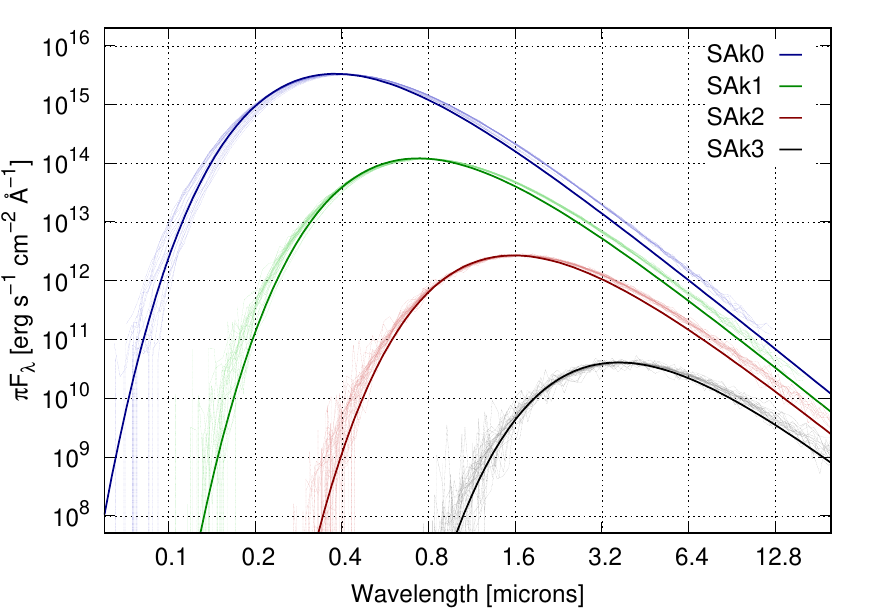}\\
  \includegraphics[width=0.49\textwidth]{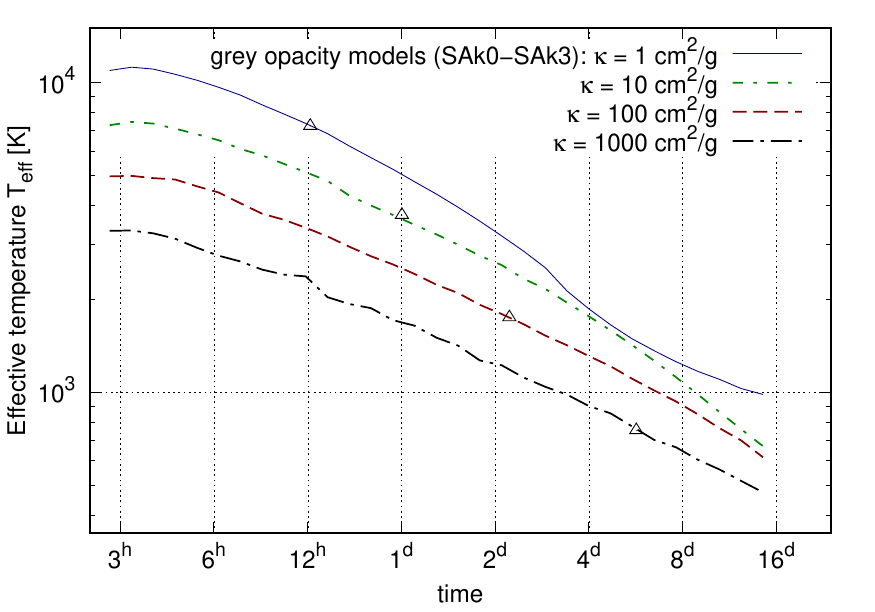}
  \end{tabular}
  \end{center}
  \caption{ Top: spectra of grey-opacity spherically-symmetric analytic models
     {\tt SAk0} - {\tt SAk3} overplotted with best-fit blackbody spectra.
     Bottom: temperature evolution for the same four models,
     calculated from fits to the blackbody spectra. Triangles mark temperatures
     at the epochs of the peaks.
  } 
\label{fig:spectra_SAkX}
\end{figure}

\begin{figure}
  \begin{center}
  \begin{tabular}{c}
  \includegraphics[width=0.49\textwidth]{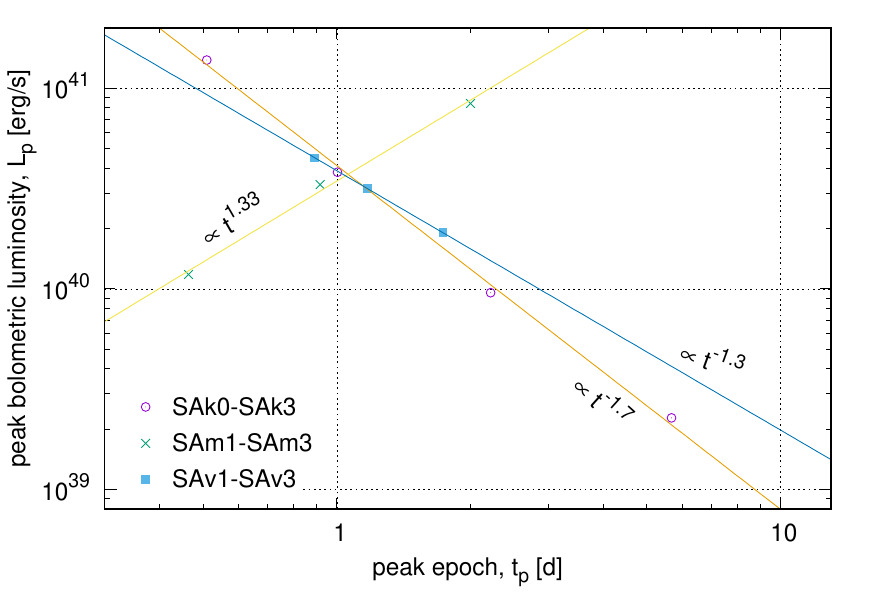}\\
  \end{tabular}
  \end{center}
  \caption{ Peak luminosities $L_p$ plotted against peak epochs $t_p$ for models
     with varying grey opacities {\tt SAk0}-{\tt SAk3}, masses
     {\tt SAm1}-{\tt  SAm3} and velocities {\tt SAv1}-{\tt SAv3}, and
     corresponding power law fits.
  } 
\label{fig:powerfits}
\end{figure}

The spectrum of grey opacity models turns out to be very close to Planck, as
shown in Fig.~\ref{fig:spectra_SAkX} (top), where for each model {\tt SAk0} --
{\tt SAk3} we plotted stacked spectra at different times (thin dashed
lines), shifted and rescaled to match the Planck spectrum with the temperature
at the peak epoch $t_p$.
The temperature can be determined from the spectral peak location
$\lambda_{\rm max}$ through the Wien law:
$T_p\cdot\lambda_{\rm max} = 0.28977\ {\rm cm}\cdot{\rm K}$.
The spectrum is wider than Planck by $\Delta\lambda/\lambda\approx0.1$, which
corresponds to a Doppler shift with the characteristic expansion velocity.
Thus, the spectral evolution in the grey opacity models can be well described
with a single evolution variable, such as effective temperature
$T_{\rm eff}(t)$, as shown in Fig.~\ref{fig:spectra_SAkX} (bottom).

Figure~\ref{fig:powerfits} illustrates behaviour of peak epochs and
luminosities for models with varying ejecta masses, expansion velocities and
opacities in $t_p$ -- $L_p$ plane.
Combining these individual fits, we obtain the following expressions for the
peak epoch, peak bolometric luminosity and effective temperature at the peak:
\begin{align}
  &t_{\rm p}= 1.0\ {\rm d}\;\kappa_{10}^{0.35}\ M_2^{0.318}\ v_1^{-0.60},
  \label{eq:tpeak_scaling}
  \\
  &L_{\rm p}= 2.8\times10^{40}\ {\rm erg}\ {\rm s}^{-1}\;
                             \kappa_{10}^{-0.60}\ M_2^{0.426}\ v_1^{0.776},
  \label{eq:Lpeak_scaling}
  \\
  &T_{\rm p}= 3720\ {\rm K}\;\kappa_{10}^{-0.33}\ M_2^{-0.055}\ v_1^{0.011}.
  \label{eq:Tpeak_scaling}
\end{align}
and ${M_2:=M_{\rm ej}/10^{-2}M_{\odot}}$,
    ${v_1:=\frac{v_{\rm max}/2}{0.1\ c}}$,
    ${\kappa_{10}:= \kappa/{10\ {\rm cm}^2{\rm g}^{-1}}}$.

These fits are qualitatively similar to semianalytic scaling laws derived in \cite{grossman14}:
\begin{align}
  &t_{\rm p}= 4.9\ {\rm d}\;\kappa_{10}^{0.5}\ M_2^{0.5}\ v_1^{-0.5},
  \\
  &L_{\rm p}= 2.5\times10^{40}\ {\rm erg}\ {\rm s}^{-1}\;
                             \kappa_{10}^{-0.65}\ M_2^{0.35}\ v_1^{0.65},
  \\
  &T_{\rm p}= 2200\ {\rm K}\;\kappa_{10}^{-0.41}\ M_2^{-0.16}\ v_1^{-0.08}.
\end{align}

If peak magnitude is computed for a model with given ejecta mass $m_0$, expansion
velocity $v_0$ and opacity $\kappa_0$, then similar empirical fits can be used
to find peak parameters for a model with a different mass $m_{\rm ej}$,
velocity $v$ and opacity $\kappa$. In particular, we find the following trends
for wavelength bands:
\begin{align}
 m_g &= m_{g,0} - 1.13\ log_{10}\ m_{\rm ej}/m_0 - 1.28\ log_{10}\ v/v_0 + 2.65\ log_{10}\ \kappa/\kappa_0\nonumber \\
 m_r &= m_{r,0} - 1.01\ log_{10}\ m_{\rm ej}/m_0 - 1.60\ log_{10}\ v/v_0 + 2.27\ log_{10}\ \kappa/\kappa_0\nonumber \\
 m_i &= m_{i,0} - 0.94\ log_{10}\ m_{\rm ej}/m_0 - 1.52\ log_{10}\ v/v_0 + 2.02\ log_{10}\ \kappa/\kappa_0\nonumber \\
 m_z &= m_{z,0} - 0.94\ log_{10}\ m_{\rm ej}/m_0 - 1.56\ log_{10}\ v/v_0 + 1.87\ log_{10}\ \kappa/\kappa_0
 \label{eq:bb_scaling}
 \\
 m_y &= m_{y,0} - 0.93\ log_{10}\ m_{\rm ej}/m_0 - 1.61\ log_{10}\ v/v_0 + 1.76\ log_{10}\ \kappa/\kappa_0\nonumber \\
 m_J &= m_{J,0} - 0.93\ log_{10}\ m_{\rm ej}/m_0 - 1.61\ log_{10}\ v/v_0 + 1.56\ log_{10}\ \kappa/\kappa_0\nonumber \\
 m_H &= m_{H,0} - 0.95\ log_{10}\ m_{\rm ej}/m_0 - 1.55\ log_{10}\ v/v_0 + 1.33\ log_{10}\ \kappa/\kappa_0\nonumber \\
 m_K &= m_{K,0} - 0.99\ log_{10}\ m_{\rm ej}/m_0 - 1.53\ log_{10}\ v/v_0 + 1.13\ log_{10}\ \kappa/\kappa_0\nonumber
\end{align}

Magnitudes and peak times for all our models are listed in tables given in
Appendix~\ref{app:tables}.

\subsection{Semianalytic models: multigroup opacity}
\label{sec:sa_multi}

As a next step, we replace grey opacities with multigroup LTE opacities,
calculated for the representative r-process elements listed in
Table~\ref{tab:mfs_repr}.
Fig.~\ref{fig:lum_SACrPdSm} shows bolometric and broadband light curves for
all non-grey models from Table~\ref{tab:sa_models}, in the LSST $grizy$ and
2MASS $JHK$ bands.
The apparent variety of broadband light curves can be classified
into three types:
(i) bright early blue transients, peaking in optical bands on a
timescale of a few hours (Fig.~\ref{fig:lum_SACrPdSm}c);
(ii) intermediate red transients, featuring double peaks in the $izy$ bands
on a timescale of a day or two (Fig.~\ref{fig:lum_SACrPdSm}e,b), and
(iii) late near-infrared (nIR) transients, showing very little emission in
optical and peaking in $HK$ bands on a timescale of a week
(Fig.~\ref{fig:lum_SACrPdSm}d,f).
The difference in behavior originates from electronic configurations of the
outer shells of corresponding elements, generating opacities that differ by
orders of magnitude.
These types correspond to the elements with open $p$-shell, $d$-shell and
$f$-shell.
Indeed, the first type includes models with elements Se, Te and Br, which only
have electrons in the outer $p$-shell in both neutral and the few first ionization
stages that we consider.
The second type includes Cr with outer shell configuration ${3d^54s^1}$, Zr
with ${4d^25s^2}$ and Pd with closed outer $d$-shell ${4d^{10}}$ in a neutral
state, but open $d$-shell in ionized states.
The third type includes lanthanides Sm, Ce and Nd and actinide U, all with
open $f$-shell.
A higher orbital quantum number increases the amount of bound-bound transitions,
and, consequently, opacity, by an order of magnitude, causing the computed
qualitative differences in the light curves.
Numerical values of peak times and peak magnitudes for each band and each
spherically-symmetric model can be found in Table~\ref{tab:peak_mags_1d},

Bolometric light curves shown in Fig.~\ref{fig:lum_SACrPdSm}, panel (a), also
exhibit distinctive features that allow them to be classified into one of the three
types. The three brightest models (Se, Br, Te) have a single peak at
$t_p\sim6^h$. Models with open $d$-shell elements (Cr, Pd, Zr) are a factor of
a few dimmer, last longer ($\sim1^d)$ and show an extended plateau or a second
peak in bolometric luminosity. Finally, models with open $f$-shell are more
than one order of magnitude dimmer with a distinct rising phase and a peak at
$t_p\sim4^d-8^d$.

The model with Ce stands out among open $f$-shell models with its early bright
peak around $t_p\sim0.15^d$. As can be seen on Fig.~\ref{fig:lum_SACrPdSm},
panel (d), optical light curves for this model have a peak which is much
brighter than for other open $f$-shell models shown on panel (f). Going back
to the opacity plot (Fig.~\ref{fig:opac_sm_loglog} in
Sect.~\ref{sec:opacities}, bottom panel) we can see how this can be explained
by element Ce having almost two orders of magnitude lower smeared opacity in
blue optical wavelengths than other open $f$-shell elements.
At late time (around one week) the light curve for Ce has a second
peak in nIR $HK$ bands just like other lanthanide/actinide-based models.

\setlength{\unitlength}{1cm}
\begin{figure*}
\begin{center}
\begin{adjustbox}{width=0.88\textwidth}
\begin{tabular}{cc}
\begin{picture}(8.0,6.0)
  \put(0,0){\includegraphics[width=0.48\textwidth]{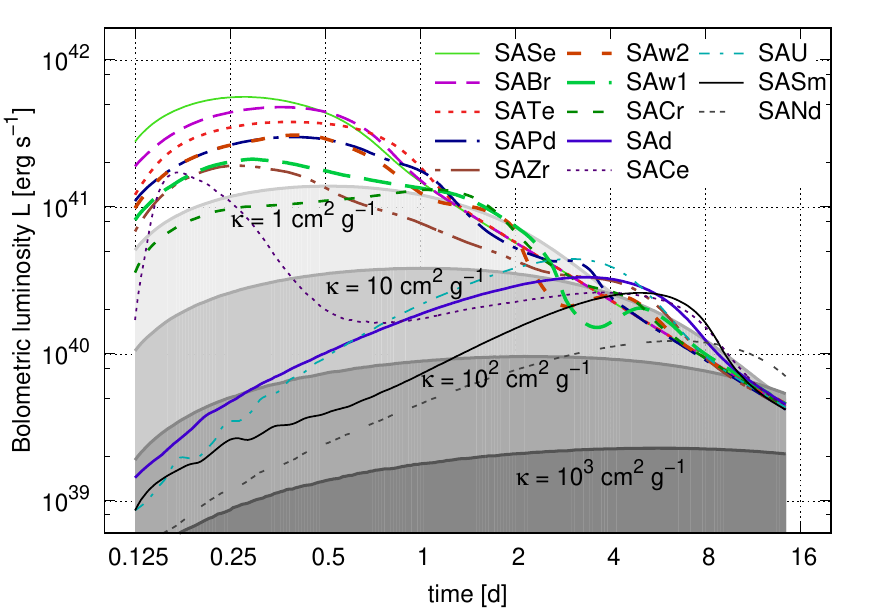}}
  \put(1.3,0.9){{\color{white}{\large (a)}}}
\end{picture} &
\begin{picture}(8.0,6.0)
  \put(0,0){\includegraphics[width=0.48\textwidth]{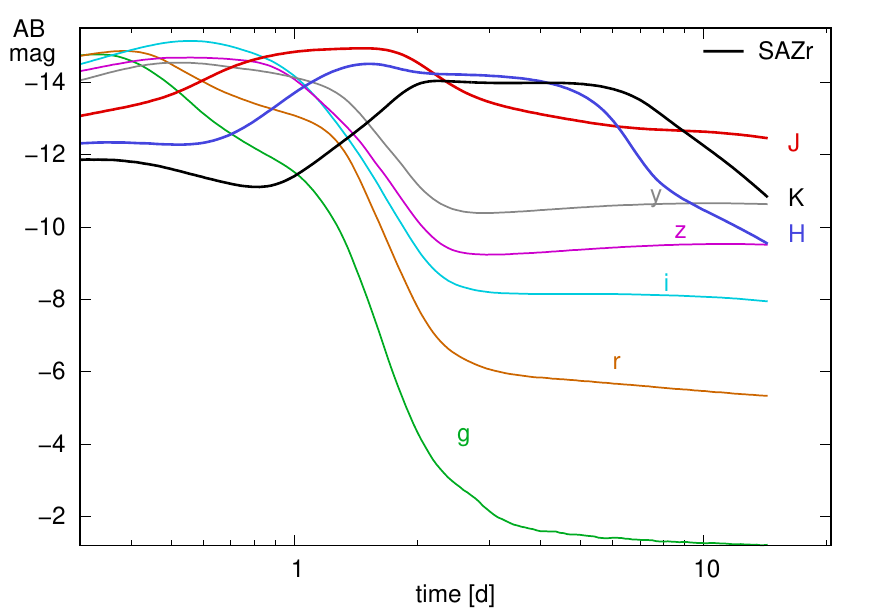}}
  \put(1.1,0.8){{\large (b)}}
\end{picture}
\\
\begin{picture}(7.5,5.5)
  \put(0,0){\includegraphics[width=0.47\textwidth]{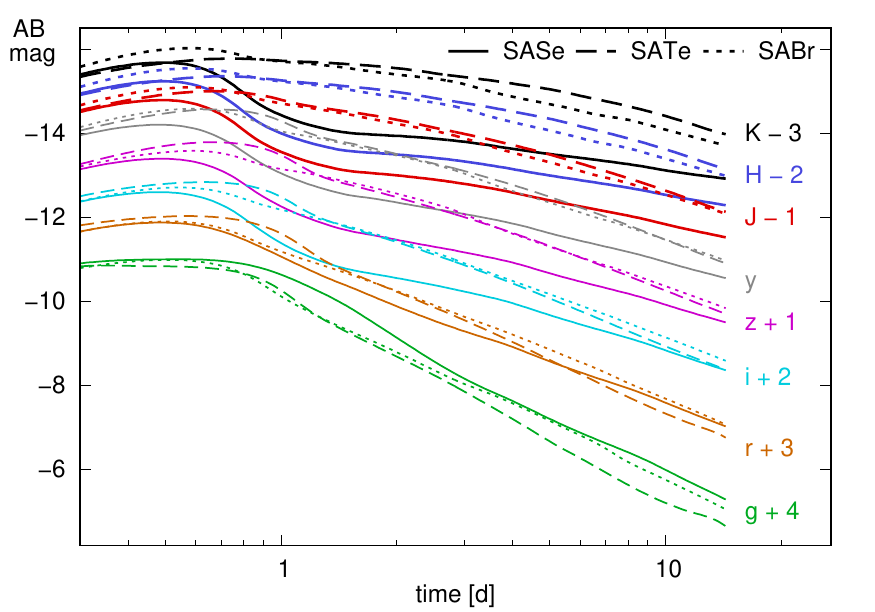}}
  \put(1.1,0.8){{\large (c)}}
\end{picture} &
\begin{picture}(7.5,5.5)
  \put(0,0){\includegraphics[width=0.47\textwidth]{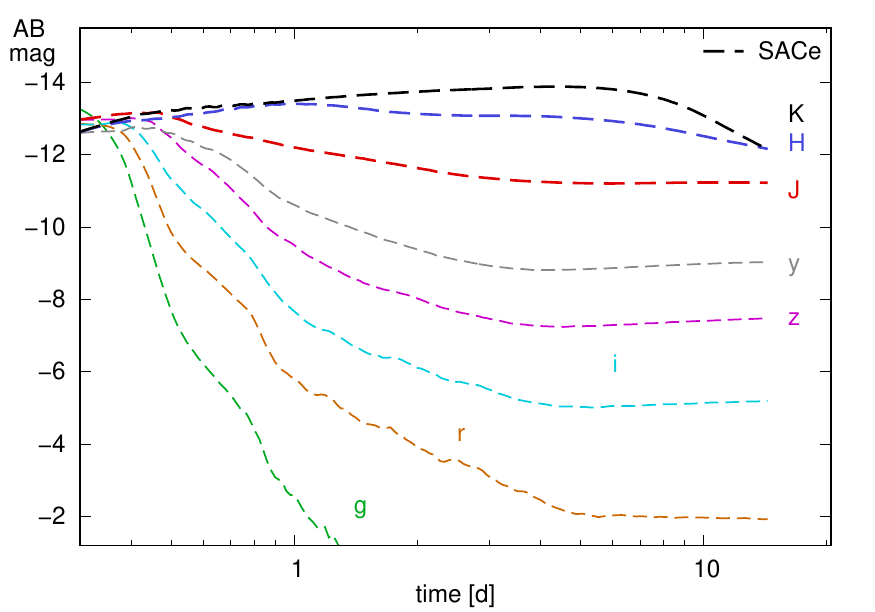}}
  \put(1.1,0.8){{\large (d)}}
\end{picture}
\\
\begin{picture}(7.5,5.5)
  \put(0,0){\includegraphics[width=0.47\textwidth]{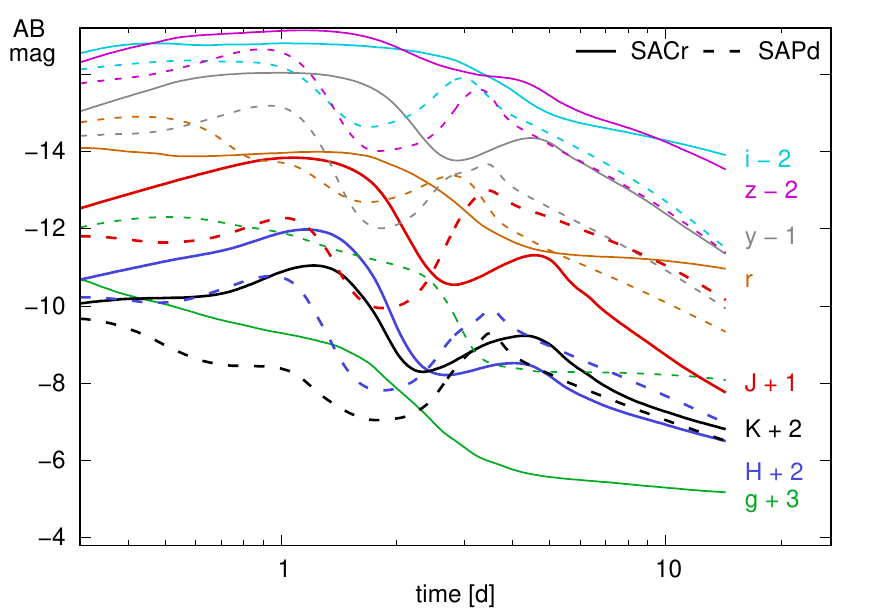}}
  \put(1.1,0.8){{\large (e)}}
\end{picture} &
\begin{picture}(7.5,5.5)
  \put(0,0){\includegraphics[width=0.47\textwidth]{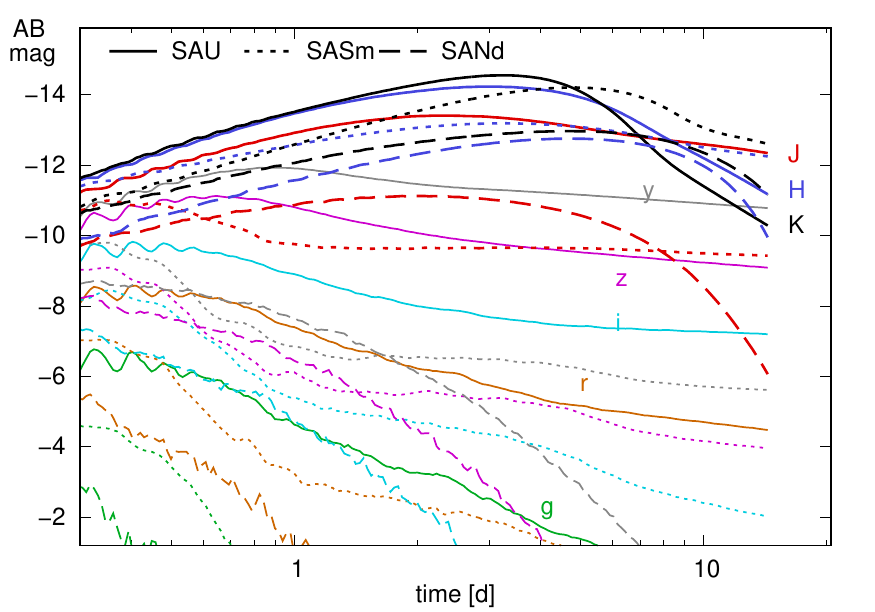}}
  \put(1.1,0.8){{\large (f)}}
\end{picture}
\\
\begin{picture}(7.5,5.5)
  \put(0,0){\includegraphics[width=0.47\textwidth]{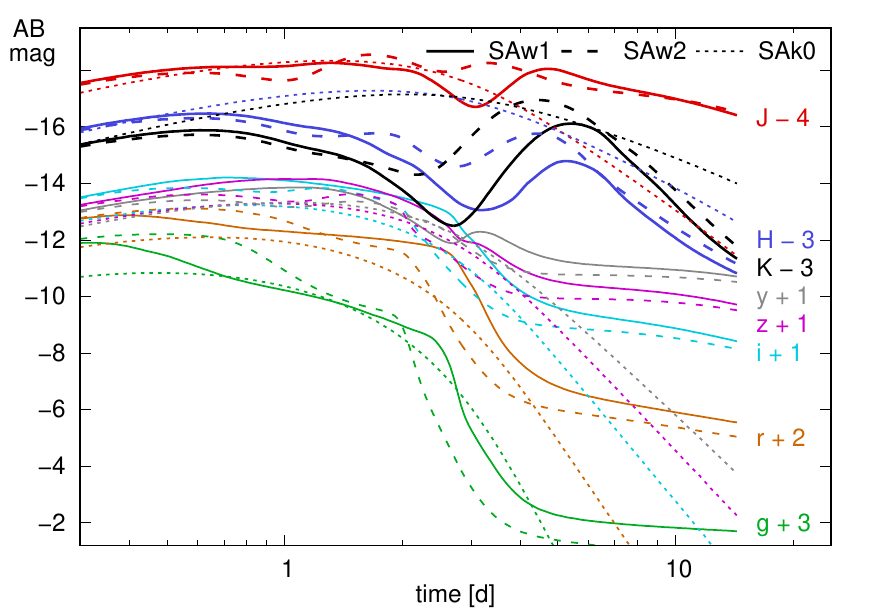}}
  \put(1.1,0.8){{\large (g)}}
\end{picture} &
\begin{picture}(7.5,5.5)
  \put(0,0){\includegraphics[width=0.47\textwidth]{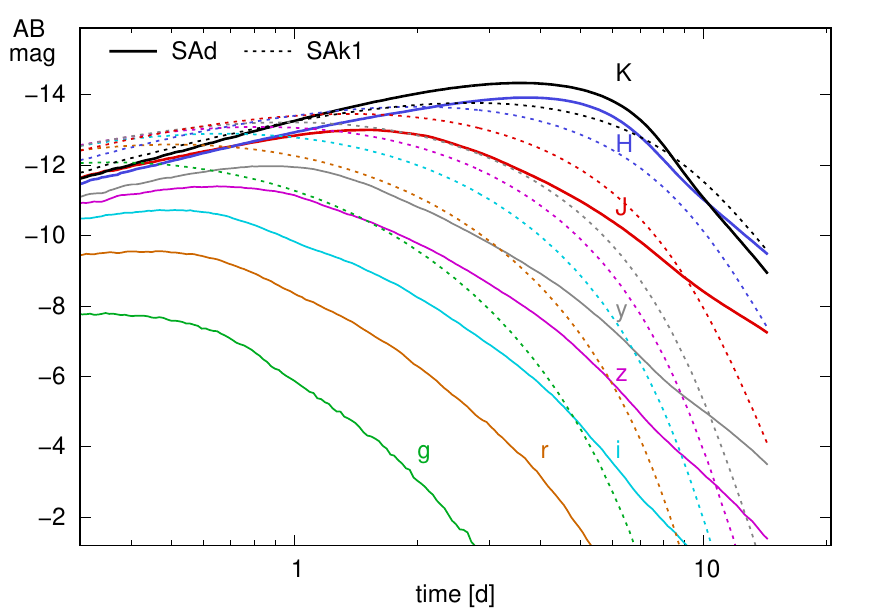}}
  \put(1.1,0.8){{\large (h)}}
\end{picture}
\end{tabular}
\end{adjustbox}
\end{center}
\caption{Bolometric (a) and broadband light curves (b-h) for models with
spherically-symmetric analytic ({\tt SA*}) density distributions
and multigroup LTE opacities for single-element plasma (indicated by the
remaining letters in abbreviations: Se, Br, Te, Pd, Zr, Cr, Ce, U, Sm or Nd),
and mixed composition for dynamical ejecta ({\tt SAd}) and two types of wind
({\tt SAw1} and {\tt SAw2}). See elemental composition for the mixed models
in Table~\ref{tab:mfs_repr}.
The shades of grey in panel (a) indicate bolometric light curves
of the grey opacity models {\tt SAk0} -- {\tt SAk3}.
Different bands are denoted with different colors.
The light curves in the right column in panels (c), (e) and (g) are offset up
or down to make the figures easier to read. Each band is offset by the same
integer, shown on the right, e.g. "$K-3$" indicates that the light curve in
the $K$ band is shifted upward by 3~mag.}
\label{fig:lum_SACrPdSm}
\end{figure*}
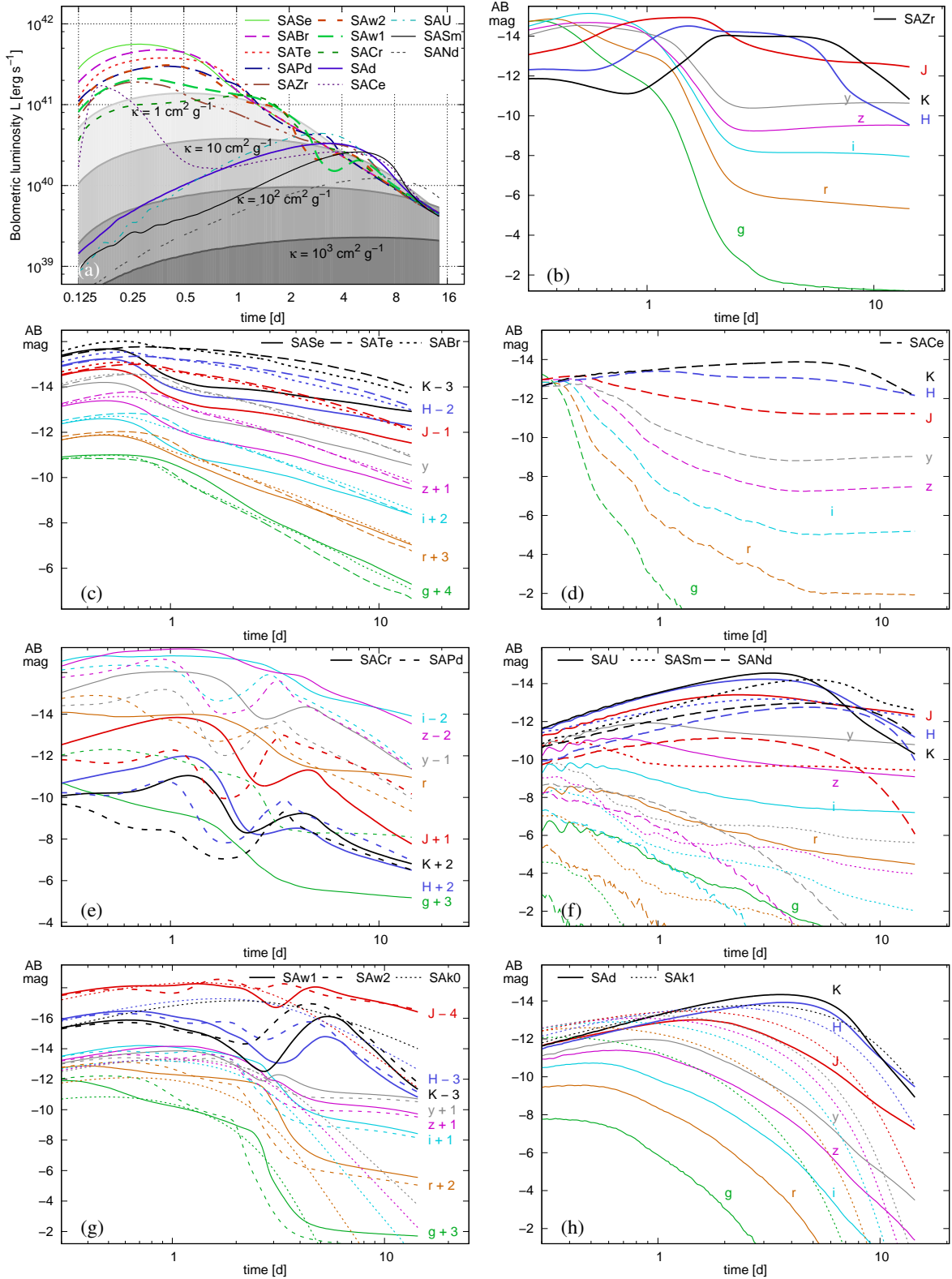

The two bottom panels of Fig.~\ref{fig:lum_SACrPdSm} show the light curves for
mixed compositions: two types of wind ({\tt SAw1}, {\tt SAw2}) and dynamical
ejecta {\tt SAd} (see Table~\ref{tab:mfs_repr} for their composition details).
These are shown in comparison with the two models with grey opacity: {\tt
SAk0} for winds, and {\tt SAk1} for the ejecta.
In all cases, the light curves of mixed models resemble those of the elements
with the highest opacity: the light curves of wind models are closest to Zr
and Cr, while the light curves for dynamical ejecta are closest to the ones of
U and Sm. Note that our prescription for the opacity of mixed compositions by
simple mass-weighted approach, detailed in Sect.~\ref{sec:opacities}, is
probably an underestimate. In reality, highly opaque elements dominate the
opacities even if present in very small amounts
\citep[see discussion on this topic in Sect.~6 in][]{kasen13}.

It is instructive to compare light curves of multigroup opacity models with
those of grey opacity models, and try to infer an "effective opacity" that can
be used as a simple approximation for the wind and dynamical ejecta models.
Previous works \citep[e.g.,][]{rosswog17a,grossman14} used $\kappa=10\ \cmg\ $
for dynamical ejecta and $\kappa=1\ \cmg\ $ for wind outflows.
Bolometric luminosities of wind and dynamical ejecta models ({\tt SAw1}, {\tt
SAw2} and {\tt SAd} in Fig.~\ref{fig:lum_SACrPdSm}a), can be compared with
grey opacity models (displayed as shaded areas on the same plot).
As can be seen from the plot, the wind models agree with the
$\kappa=1\ \cmg\ $ model very roughly, only up to a factor of a few.
Luminosity in dynamical ejecta model {\tt SAd} is suppressed compared to
$\kappa=10\ \cmg\ $ grey model at early times, but matches with the grey model
at the peak, which is the time most relevant for detectability.

On the other hand, as pointed out in \cite{dessart16}, using effective
grey opacity produces inaccurate results, which can differ from detailed
multigroup calculations by as many as 50~\%, both in luminosity, and in peak
time. In general, uniform effective opacity underestimates the opacity in the
remnant core, which makes it transparent earlier and thus leads to earlier peaks
\citep[see][]{dessart16}.

As can be seen from the two bottom panels of Fig.~\ref{fig:lum_SACrPdSm}, this
effective opacity analogy can only be reasonably extended for the wind models
{\tt SAw1} and {\tt SAw2}, but not for the dynamical ejecta model.
More specifically, the light curve of the dynamical ejecta model
{\tt SAd} only agrees with the grey model {\tt SAk1} in the infrared $H$ and
$K$ bands. The analogy fails in optical bands and in the $J$ band: for instance,
the flux for the {\tt SAd} model in the $y$ band (Fig.~\ref{fig:lum_SACrPdSm}h)
around $t\sim4^d$ is almost four magnitudes dimmer than what is predicted by
the effective opacity. This behavior is attributed to a peculiar "spectral
cliff" in the $f$-shell elements spectra, leading to abrupt suppression of the
optical bands. This is explained in more detail in the analysis of the spectra
below.

\begin{figure*}
\begin{center}
\begin{tabular}{cc}
\includegraphics[width=0.45\textwidth]{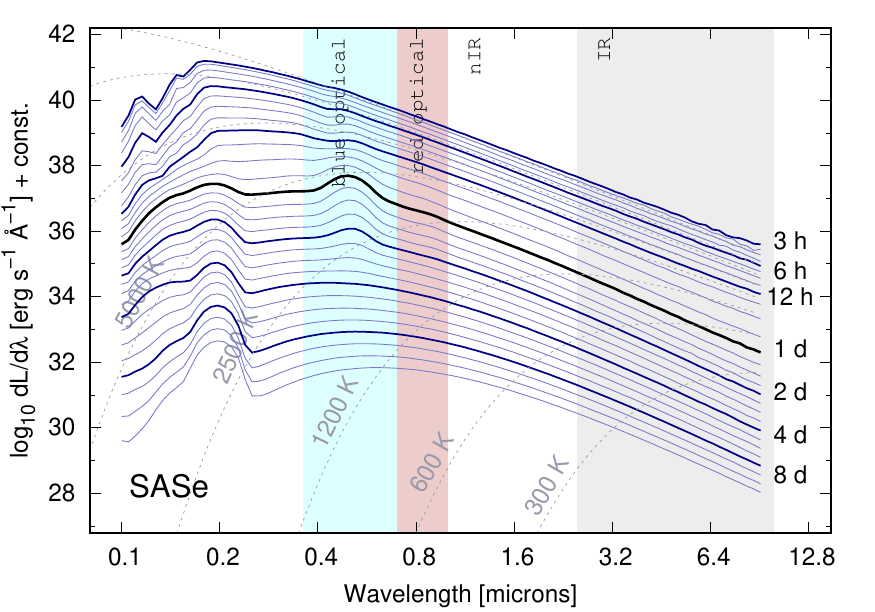} &
\includegraphics[width=0.45\textwidth]{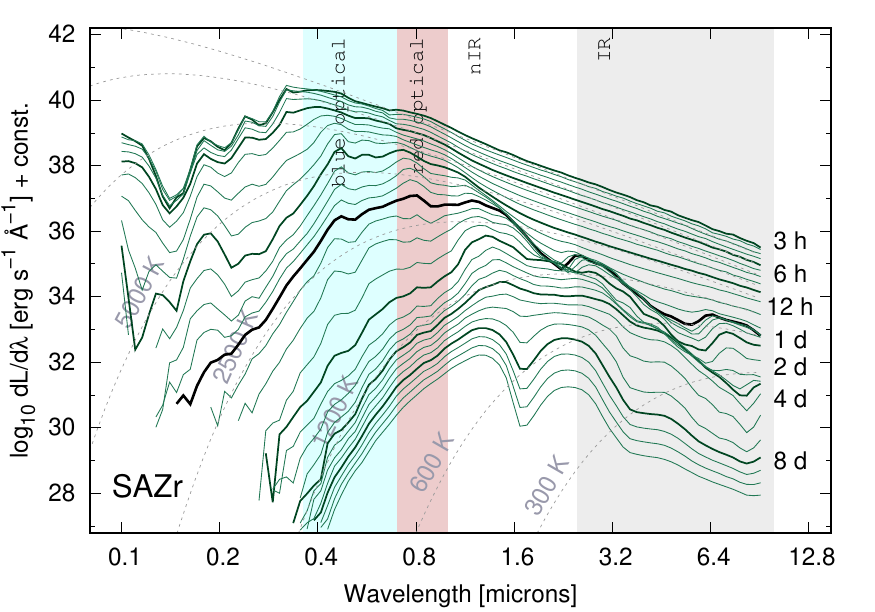}
\\
\includegraphics[width=0.45\textwidth]{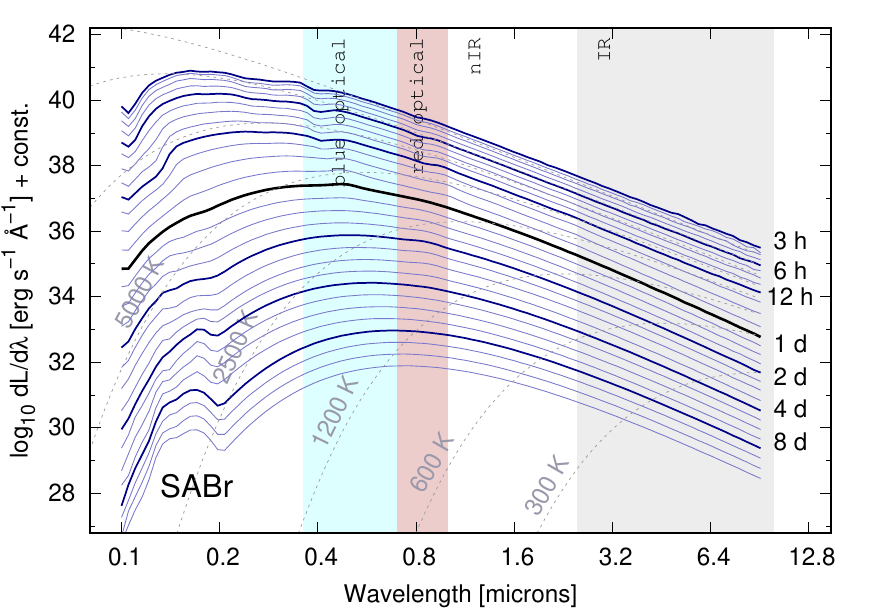} &
\includegraphics[width=0.45\textwidth]{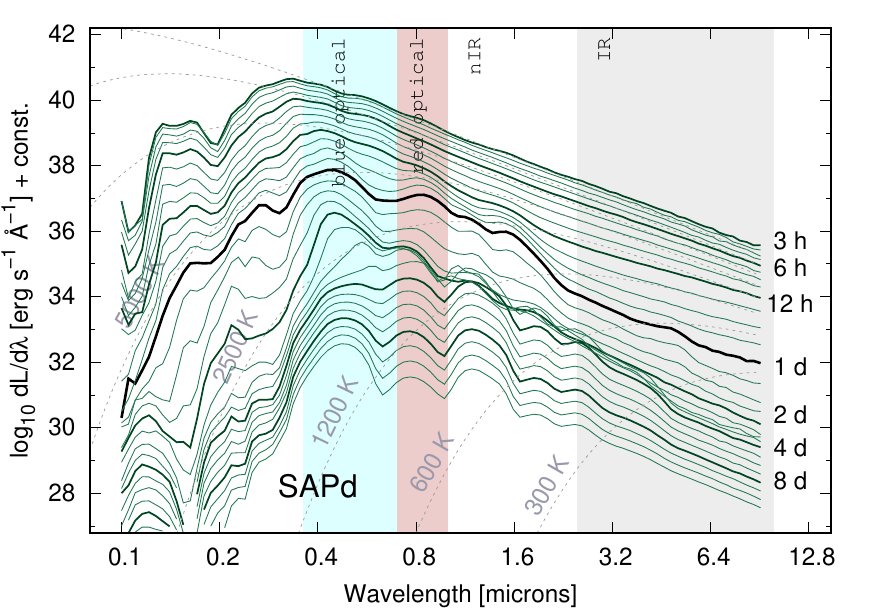}
\\
\includegraphics[width=0.45\textwidth]{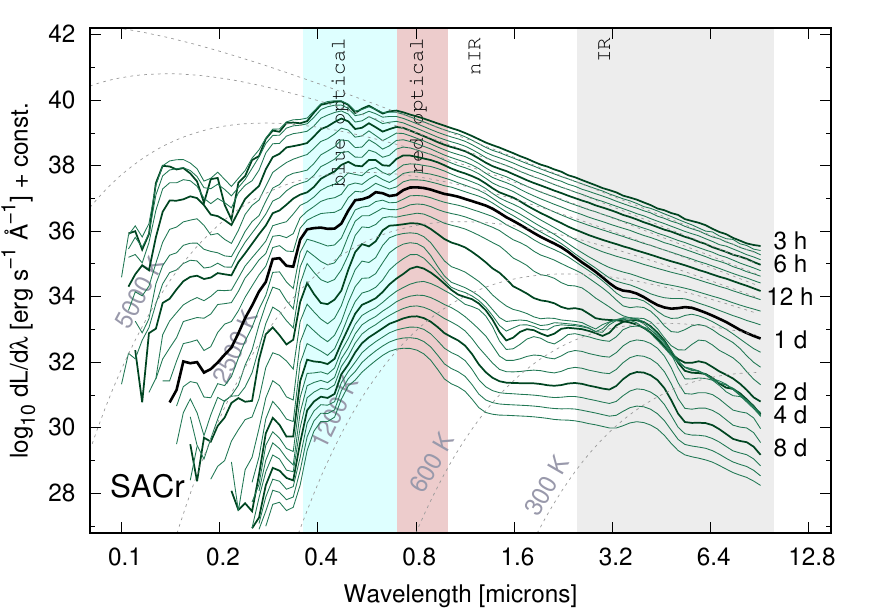} &
\includegraphics[width=0.45\textwidth]{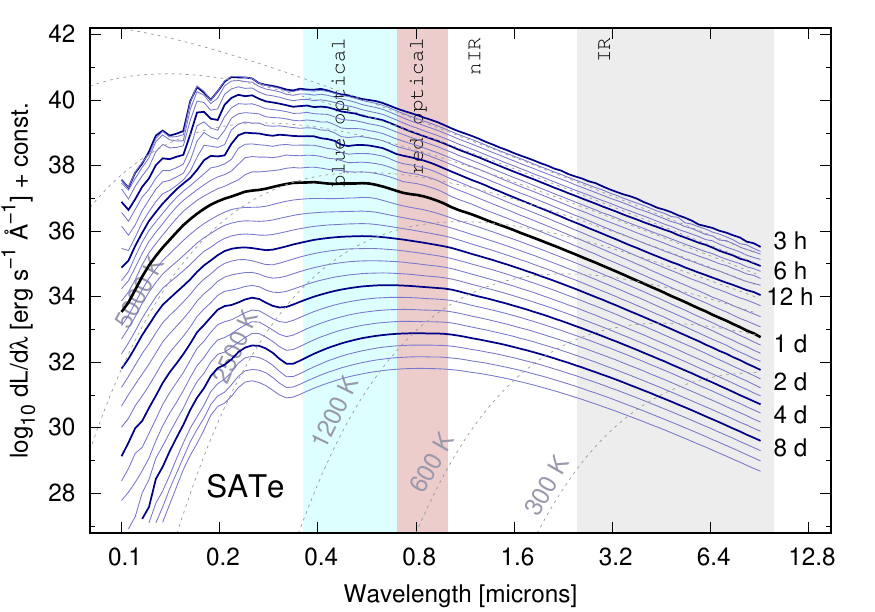}
\\
\includegraphics[width=0.45\textwidth]{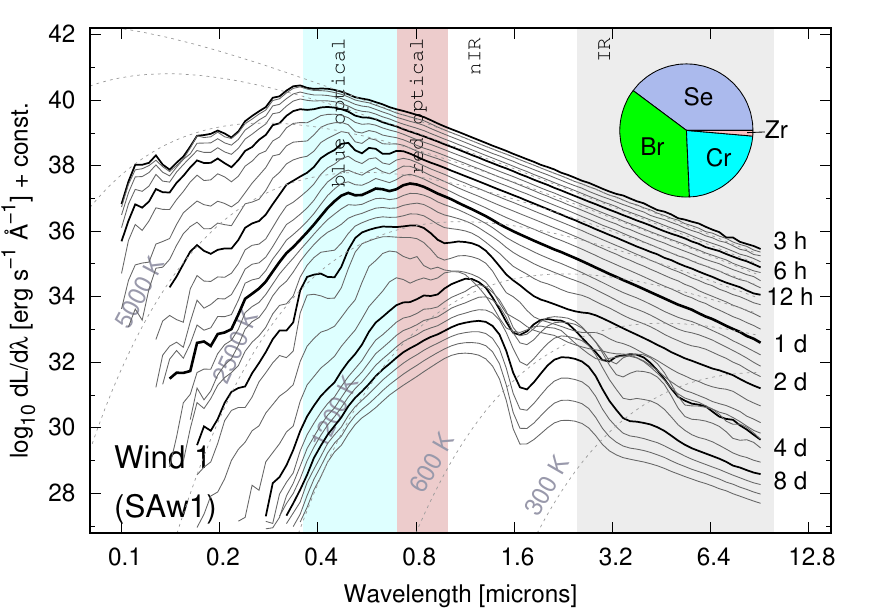} &
\includegraphics[width=0.45\textwidth]{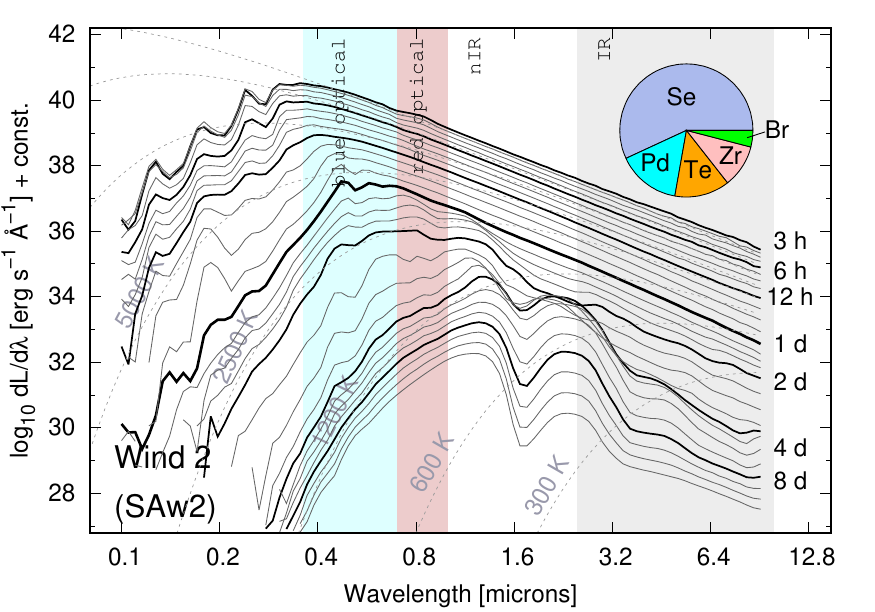}
\end{tabular}
\end{center}
\caption{Evolution of macronova spectra for spherically-symmetric models with
LTE opacities formed by lighter non-lanthanide elements Se, Te, Br, Pd, Zr, Cr
and two types of wind, representing their mixture: {\tt SAw1} and {\tt SAw2}
(see Table~\ref{tab:sa_models}). For clarity, spectral curves for different
times are offset by ${{\rm const.}(t)=-\log_2{t[{\rm d}]}}$ (i.e. no offset for
{$t=1$ day}).
Thin dashed lines show blackbody spectra for a range of temperatures.
}
\label{fig:spec_SACrPdSm}
\end{figure*}

\begin{figure*}
\begin{center}
\begin{tabular}{cc}
\includegraphics[width=0.45\textwidth]{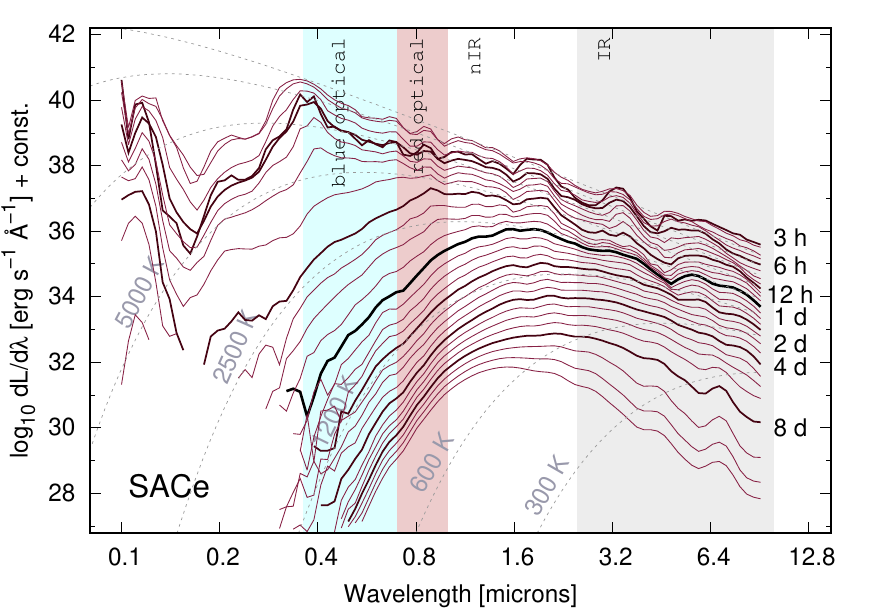} &
\includegraphics[width=0.45\textwidth]{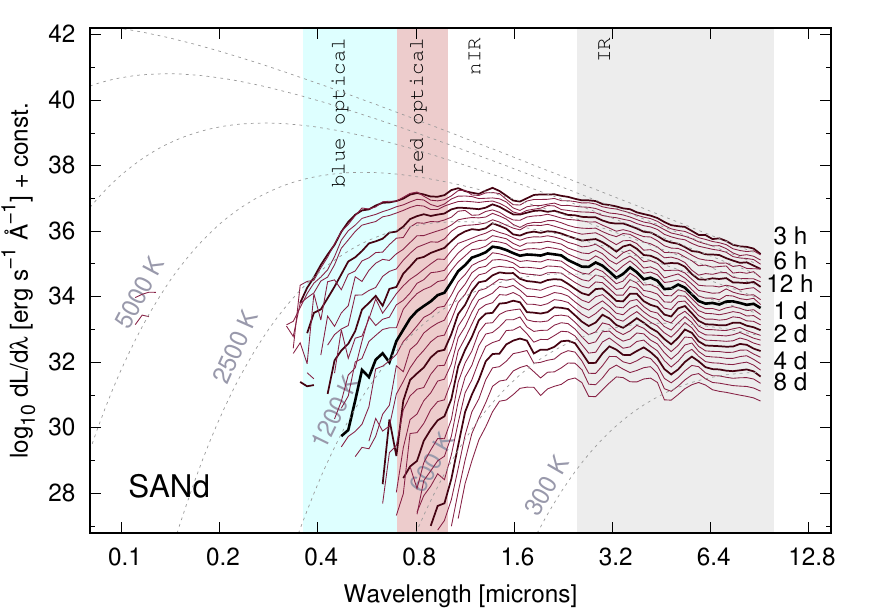}
\\
\includegraphics[width=0.45\textwidth]{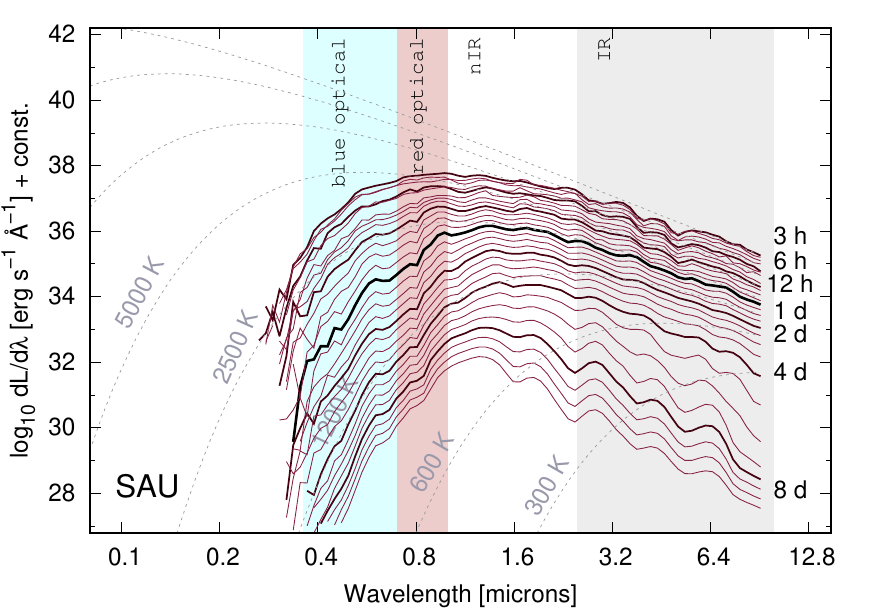} &
\includegraphics[width=0.45\textwidth]{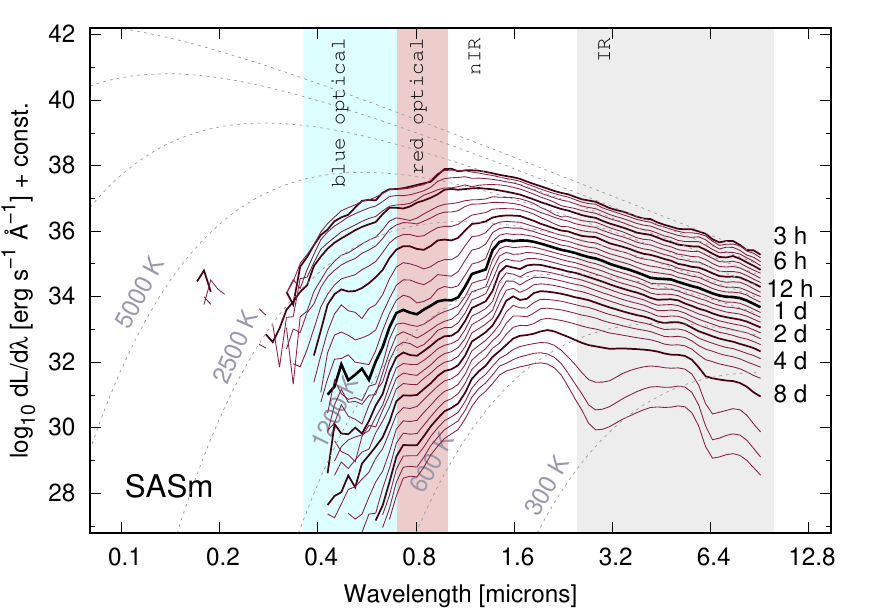}
\end{tabular}
\includegraphics[width=0.45\textwidth]{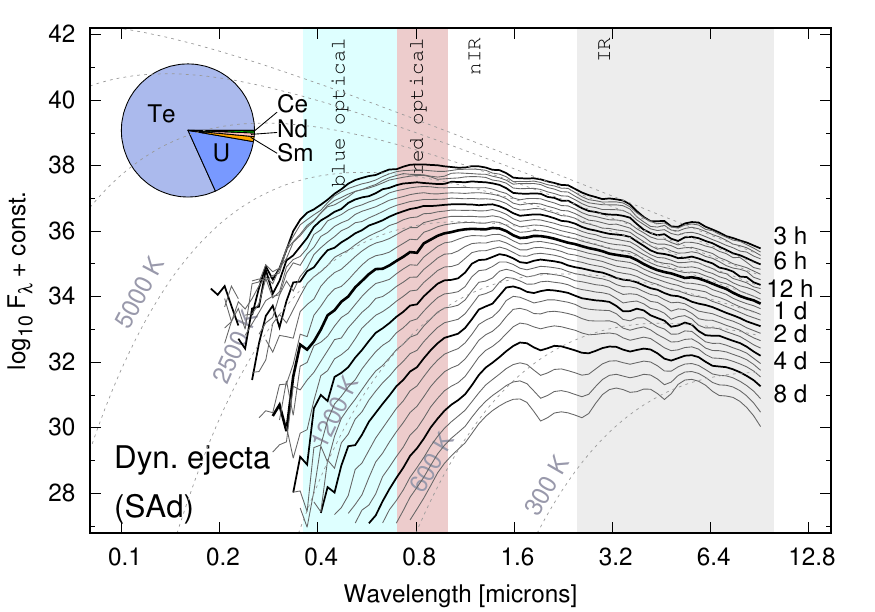}
\end{center}
\caption{Evolution of macronova spectra for models with LTE opacities of
lanthanides Ce, Nd, Sm, actinide U, and dynamical ejecta mixture.
(see Table~\ref{tab:sa_models}). Spectra for different times $t$ are shifted
by a time-dependent constant ${{\rm const.}(t)=-\log_2{t[{\rm d}]}}$.
Thin dashed lines show blackbody spectra for a range of temperatures.}
\label{fig:spec_SACeNdU}
\end{figure*}

Figures~\ref{fig:spec_SACrPdSm}-\ref{fig:spec_SACeNdU} present corresponding
spectral evolutions for each model in terms of luminosity per wavelength,
${dL_{\lambda}/d\lambda}$
(in units of ${{\rm erg}\ {\rm s}^{-1}\ \mathring{A}^{-1}}$).
Note that the spectra on these plots, for a time $t$, are shown shifted by a
time-dependent constant ${const(t)=-\log_2{t[d]}}$, positive for $t<1^d$,
negative for $t>1^d$ and vanishing at one day. This gives a clearer picture of
the evolution of spectral shapes as time progresses from top to bottom of the
plot, forming a "spectral landscape" in the $\{\lambda,t\}$-space.
Thin grey lines on the plots show Planck spectra for a range of temperatures.
Finally, to assist with comparing the spectra, we added charts showing the
composition for mixed model plots.

Spectra for all models are similar to the blackbody spectrum in the sense that
there is just one maximum with a steep rise at short wavelengths and a
gradual ${\propto\lambda^{-4}}$ power law decay (Rayleigh-Jeans law) at long
wavelengths. This shape is indented with broad spectral features, but no
distinct lines or multiple peaks are present.
One important feature in the spectra of $f$-shell elements (and to some extent
$d$-shell elements) is the presence of a peculiar "spectral cliff" at late
times, where the blue part of the spectrum is very strongly suppressed past a
certain wavelength (for instance, around $1.6$ microns for Sm). This is what
leads to a drastic difference of the light curves in optical bands and even
in the $J$ band in comparison with the grey opacity models.

Models with $p$-shell elements Se, Br and Te do not have this feature at all,
but rather exhibit very smooth spectra, which is close to Planck, with only a
few spectral features (see Fig.~\ref{fig:spec_SACrPdSm}, all plots in blue
color).
Models with elements with open $d$-shells (plots in green color in
Fig.~\ref{fig:spec_SACrPdSm}) have much more distorted spectra that are
suppressed in UV. One particular opacity feature seen in nIR at wavelengths around
$2-3\ {\mu}m$ at the times about $\sim1^d$ is responsible for the second peaks
seen in the nIR light curves in Fig.~\ref{fig:lum_SACrPdSm}b,e.

Figure~\ref{fig:spec_SACeNdU} shows spectral evolution for lanthanides and
actinide U (all plots in red color). Spectra of models with U, Nd and Sm are
completely suppressed in UV and strongly suppressed in optical bands. This is
due to much higher opacity in optical, which also explains why these three
models do not have a plateau in bolometric luminosity compared to grey
opacity models.  (see Fig.~\ref{fig:lum_SACrPdSm}a). Instead, the models
exhibit a gradual increase in bolometric luminosity as the remnant cools and its
thermal radiation moves to longer wavelengths where the opacity is lower.

In the nIR and IR, spectra of lanthanide and actinide models deviate from the
Rayleigh-Jeans power law, while also displaying persistent wavelike patterns,
unique for each element. An interesting observation is that for three out of
four $f$-shell elements, the peak in the spectrum, which is generally expected to
evolve towards longer wavelengths, does the opposite at times $t>4^d$. These
models essentially become slightly bluer at late epochs, which is a
distinctive feature that can potentially be exploited to identify macronovae.

The model with Ce, possessing the simplest electronic structure among the
lanthanides considered, initially has a spectrum which extends through the optical
range all the way to UV (see Fig.~\ref{fig:spec_SACeNdU}, top left).  However,
after $t>1^d$ the spectrum starts behaving similarly to the other lanthanide
spectra.

Figure~\ref{fig:spec_SACrPdSm} (bottom row) shows spectral evolution of mixed
wind models. Both wind models look very similar and closely resemble the spectral
evolution of the Zr model (top left panel in Fig.~\ref{fig:spec_SACrPdSm}). This
is remarkable considering that in model {\tt SAw1}, for Wind 1, only
$\sim1.3\ \%$ of Zr is present.
The mixed dynamical ejecta model {\tt SAd} is presented in the bottom plot of
Fig.~\ref{fig:spec_SACeNdU}. Just like with the wind models, the elements with
the highest opacity dominate the spectrum: the dip around ${\lambda=2}$~microns
specific to Sm and the wave-like pattern of Nd both can be seen at the late epochs
of {\tt SAd} spectra, while no spectral feature of Te can be found, despite
the fact that it constitutes $>80\ \%$ by mass.

\subsection{Dynamical ejecta: spherical symmetry and axisymmetry}
\label{sec:dynamical}

In this section we turn our attention away from composition and focus on the
impact of morphology. To simplify the comparison between the models while
keeping a certain level of realism, we use detailed multigroup opacities of Sm
in all of our models, which is the lanthanide that was explored in previous
works \citep{fontes15a,fontes17a}. Consideration of just the dynamical ejecta models is also
motivated by the fact that, under certain conditions in Nature, secondary wind
outflows can be completely subdominant: either being obscured by dynamical ejecta,
or having too low mass, or too slow expansion velocity.  Thus, results of this
section can be used for constraining theoretical models in which the wind
outflow is not present or can be neglected. By focusing on dynamical ejecta
only, we can explore the effects of the spatial distribution of the ejecta and their
orientation with respect to the observer.

Parameters of our models are summarized in Table~\ref{tab:dyn_models}, and
averaging is described in Sect.~\ref{sec:dyn_ejecta}.
Table~\ref{tab:dyn_models} also lists peak parameters of the bolometric light
curves and the $H$ band magnitudes.
The different types of averaging of the 3D SPH distribution are indicated by
suffixes: "{\tt 2d}" (axisymmetric models), "{\tt 1d}" (axisymmetric models
integrated vertically) and "{\tt 1dm}" (axisymmetric models averaged
with respect to the local density maximum in the equatorial plane).
As can be seen from Table~\ref{tab:dyn_models}, different types of averaging result in different
median expansion velocities.
Models {\tt A1dSm}--{\tt D1dSm} possess the highest expansion
velocity, models averaged around local density maximum, {\tt A1dmSm}--{\tt
D1dSm}, have the lowest expansion, and axisymmetric models have expansion
velocities somewhere in between. As expected from scaling formulae
(\ref{eq:tpeak_scaling}, \ref{eq:Lpeak_scaling}), slower expansion produces
dimmer light curves that peak later in time.

\begin{figure}
\begin{center}
\begin{tabular}{c}
\includegraphics[width=0.49\textwidth]{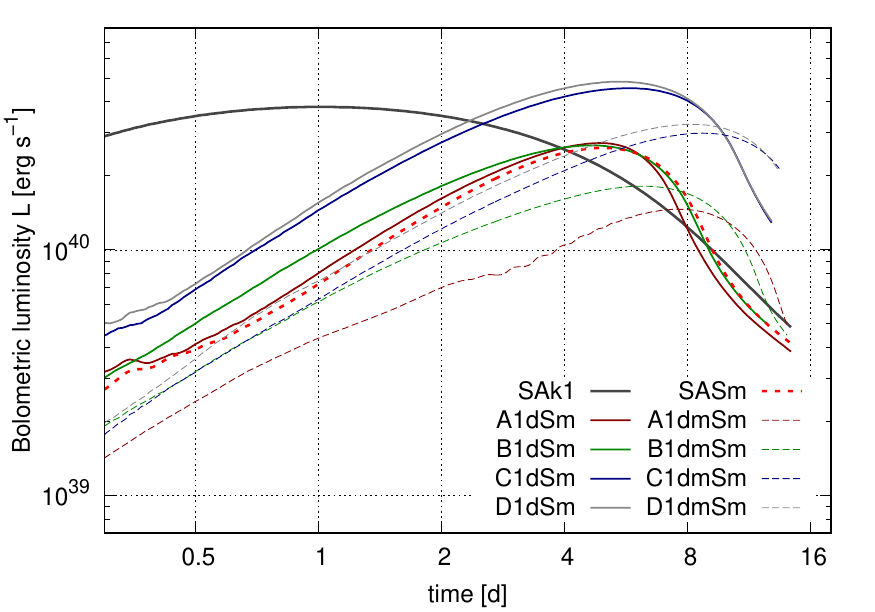} \\
\includegraphics[width=0.49\textwidth]{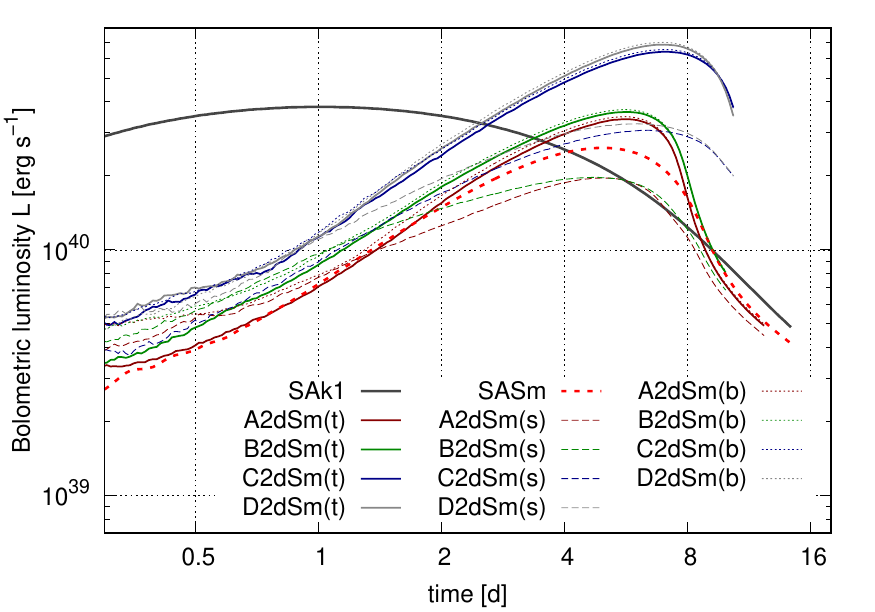}
\end{tabular}
\end{center}
\caption{Top: time evolution of bolometric luminosity for spherically-symmetric
dynamical ejecta models {\tt A1dSm}-{\tt D1dSm} and {\tt A1dmSm}-{\tt D1dmSm}
(top) which use detailed opacity of Sm (see Table~\ref{tab:dyn_models}).
Bottom: same for axisymmetric models {\tt A2dSm}-{\tt D2dSm} as observed
on-axis ("t" for "top view", "b" for bottom view) and "from the side" ("s").
Models {\tt SASm} and {\tt SAk1} are shown for comparison.}
\label{fig:lum_SASmABCD}
\end{figure}

\begin{table}
\caption{Models of dynamical ejecta (based on merger simulations).
Columns specify: total ejecta mass, median velocity, bolometric peak time,
peak bolometric luminosity, time of peak in the $H$ band and maximum magnitude
in the $H$ band.
All models in this table use detailed opacities of Sm.
In the last block of models (2d-models), a letter in brackets indicates
viewing angle: "t" for top view, and "s" for side view
(bottom view is almost identical to the top view for these models).}
\begin{tabular}{lcccccc}
\hline
      & $M_{\rm ej}$ & $\bar{v}_{1/2}$ & $t_p$ & $L_p$ & $t_H$ & $m_H$ \\
Model & $[M_\odot]$  & $[c]$  & $[d]$ & $[10^{40}\frac{\rm erg}{\rm s}]$
      & $[d]$ & ${\rm [mag]}$ \\
\hline
A1dSm    & 0.013 & 0.132 & 4.877 & 2.72 & 3.34 & -13.3 \\
B1dSm    & 0.014 & 0.125 & 4.669 & 2.66 & 2.77 & -13.3 \\
C1dSm    & 0.033 & 0.132 & 5.758 & 4.55 & 2.88 & -13.7 \\
D1dSm    & 0.034 & 0.136 & 5.457 & 4.83 & 2.87 & -14.0 \\
\hline
A1dmSm   & 0.013 & 0.066 & 7.630 & 1.46 & 5.17 & -12.4 \\
B1dmSm   & 0.014 & 0.080 & 6.230 & 1.81 & 3.72 & -12.7 \\
C1dmSm   & 0.033 & 0.055 & 8.723 & 2.98 & 5.49 & -13.1 \\
D1dmSm   & 0.034 & 0.058 & 8.202 & 3.24 & 4.67 & -13.2 \\
\hline
A2dSm(t) & 0.013 & 0.095 & 5.650 & 3.40 & 4.47 & -13.3 \\
A2dSm(s) &       &       & 5.007 & 1.99 & 3.33 & -13.1 \\
B2dSm(t) & 0.014 & 0.086 & 5.671 & 3.65 & 4.37 & -13.4 \\
B2dSm(s) &       &       & 4.783 & 1.98 & 2.57 & -13.2 \\
C2dSm(t) & 0.033 & 0.119 & 7.083 & 6.40 & 4.98 & -13.9 \\
C2dSm(s) &       &       & 6.407 & 3.10 & 3.36 & -13.6 \\
D2dSm(t) & 0.034 & 0.121 & 6.982 & 6.85 & 5.20 & -14.0 \\
D2dSm(s) &       &       & 5.999 & 3.30 & 3.30 & -13.8 \\
\hline
\end{tabular}
\label{tab:dyn_models}
\end{table}

Figure~\ref{fig:lum_SASmABCD} shows bolometric light curves for these models,
along with two models studied in previous sections: {\tt SASm} with analytic
radial density profile, and {\tt SAk1} with the grey opacity
$\kappa=10\ \cmg$.
Several common trends, which reveal the impact of morphology, can be identified.
For a fixed mass and average density, spherical configurations should give the
dimmest possible transients in terms of bolometric power radiated in all
directions. This is due to the lowest possible geometric area of the
photosphere, which keeps the maximum amount of generated heat from escaping.
For the same reason, equivalent ejecta masses with denser distributions are
expected to produce later peaks.
Vice versa, more flattened and irregular matter distributions would produce
brighter and earlier signals (from orientations or ``views'' with a sufficiently
large projected photosphere area).
In accord with previous works \citep{grossman14}, on-axis
orientations produce a transient which is brighter than for a "side"
orientation by a factor of 2-3. Both types of 1D-models fall within the
range between the brighter "top/bottom" and dimmer "side" orientations of the
corresponding 2D-models.
This orientation effect simply reflects the difference in the area of photosphere
projection on the view plane, which is higher for the on-axis case.
The light curve of the {\tt SASm} model, which has a slightly different density
profile but the same mass and expansion velocity as {\tt A1dSm}, agrees with
the light curve of {\tt A1dSm} very well, demonstrating that the exact shape
of the density profile does not have a significant effect.

Our dynamical ejecta configurations are not symmetric with respect
to reflection in the equatorial plane; therefore there is a small difference
between the "top" and "bottom" views. Although this difference is negligible
for pure dynamical ejecta models, it becomes substantial once the wind
component is added (see Sect.~\ref{sec:combined} below).

The light curve of the grey opacity model {\tt SAk1} roughly agrees in its
bolometric luminosity with A/B-models in the vicinity of the peak, but shows
much brighter values for $t<2^d$. This is because our dynamical ejecta models,
calculated with the detailed opacity of Sm, are strongly suppressed in the optical
part of the spectrum. Even though at times $t<2^d$ the temperature in the
radiative layer produces blackbody spectra peaking in the optical, this radiation
is strongly suppressed on the way out compared to the grey opacity models.
As the remnant cools down with time due to expansion, the peak of
thermal emission shifts into infrared where the opacity is much lower. Later, at
${t\sim4^d-8^d}$ the light curves of A/B-models catch up and even exceed the
bolometric luminosity of the grey opacity model with $\kappa=10\ \cmg$.

\begin{figure*}
\begin{adjustbox}{width=0.86\textwidth}
\begin{tabular}{c}
\includegraphics[height=0.23\textheight]{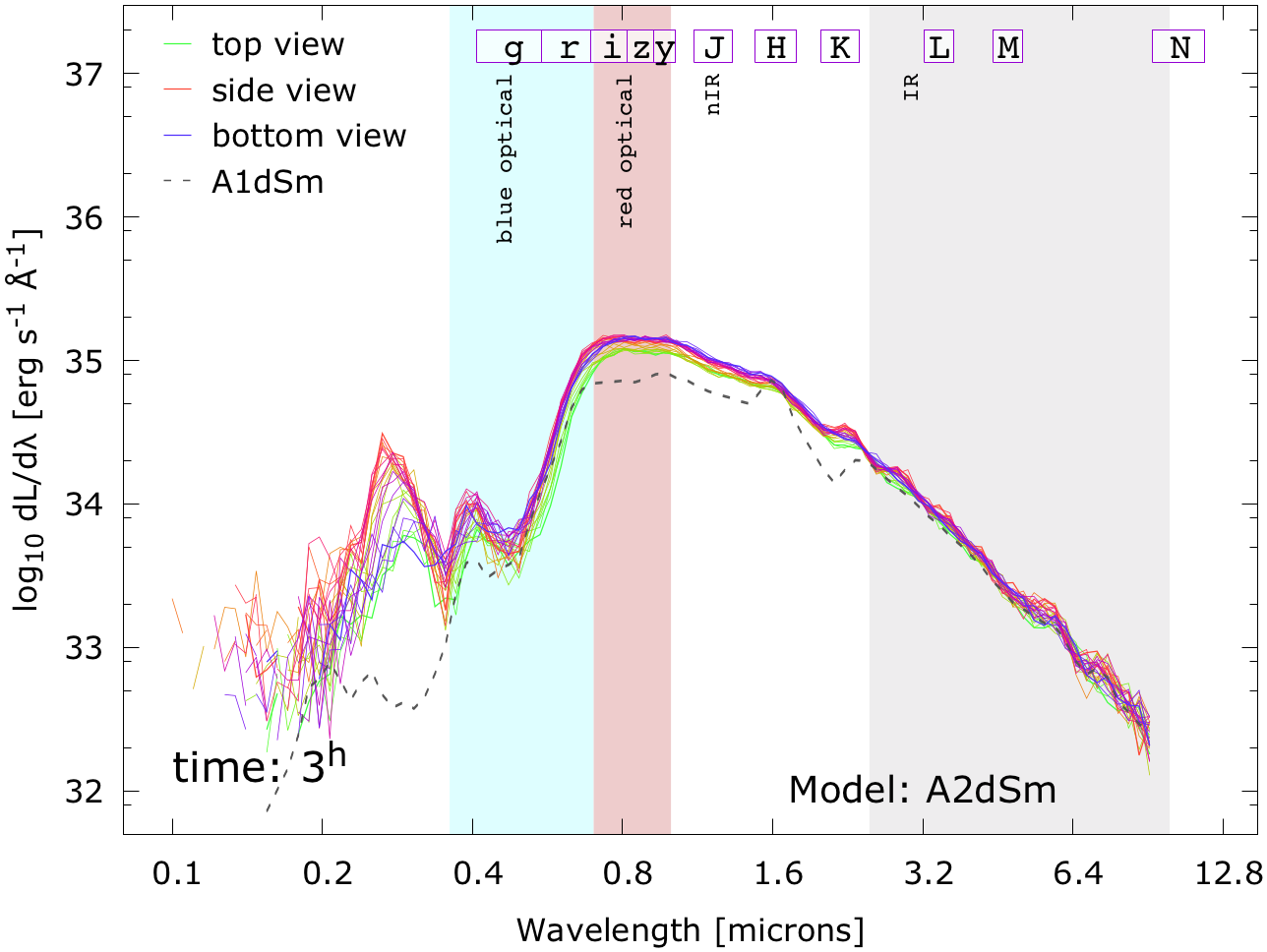}
\includegraphics[height=0.23\textheight]{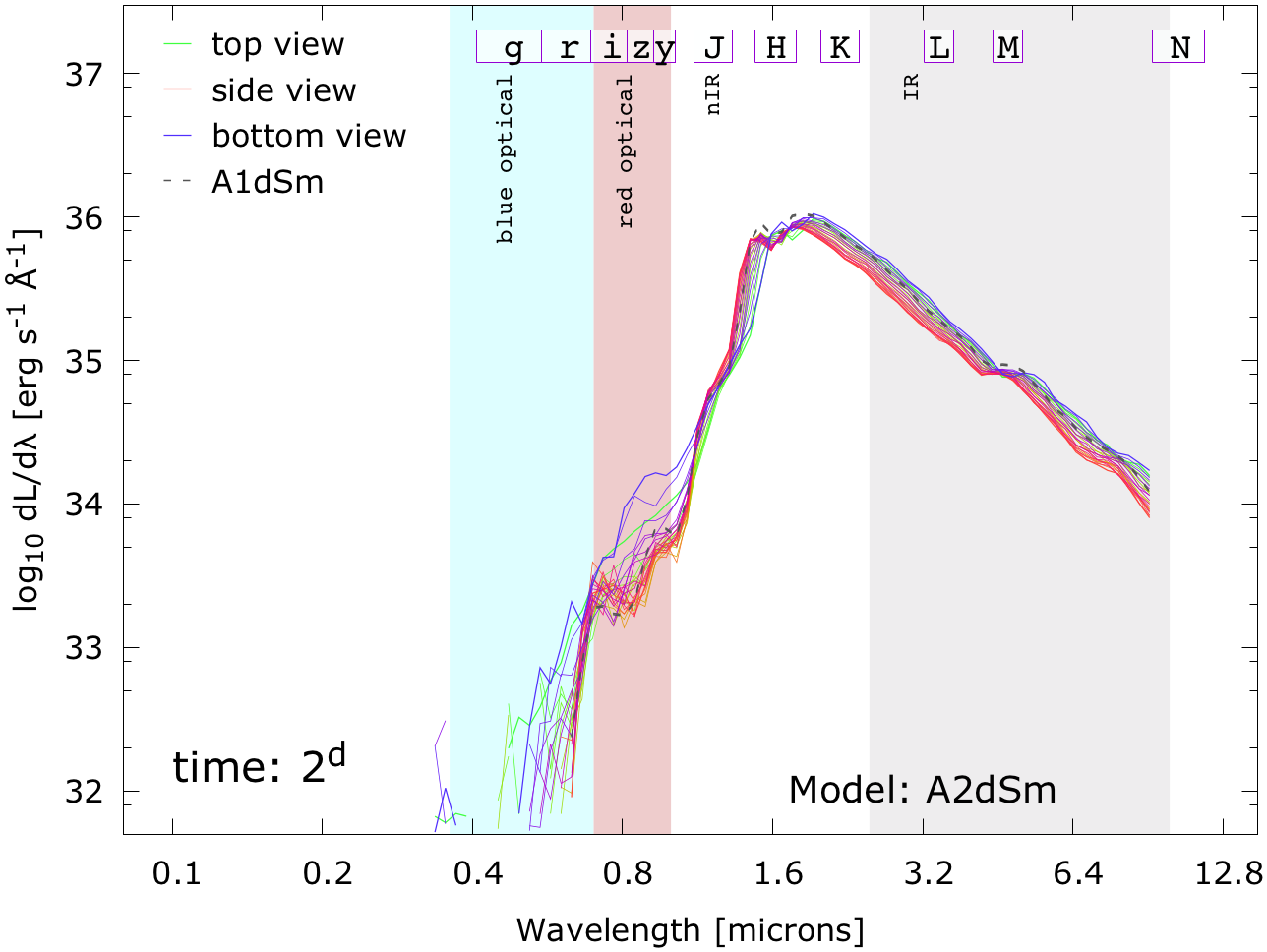}
\\
\includegraphics[height=0.23\textheight]{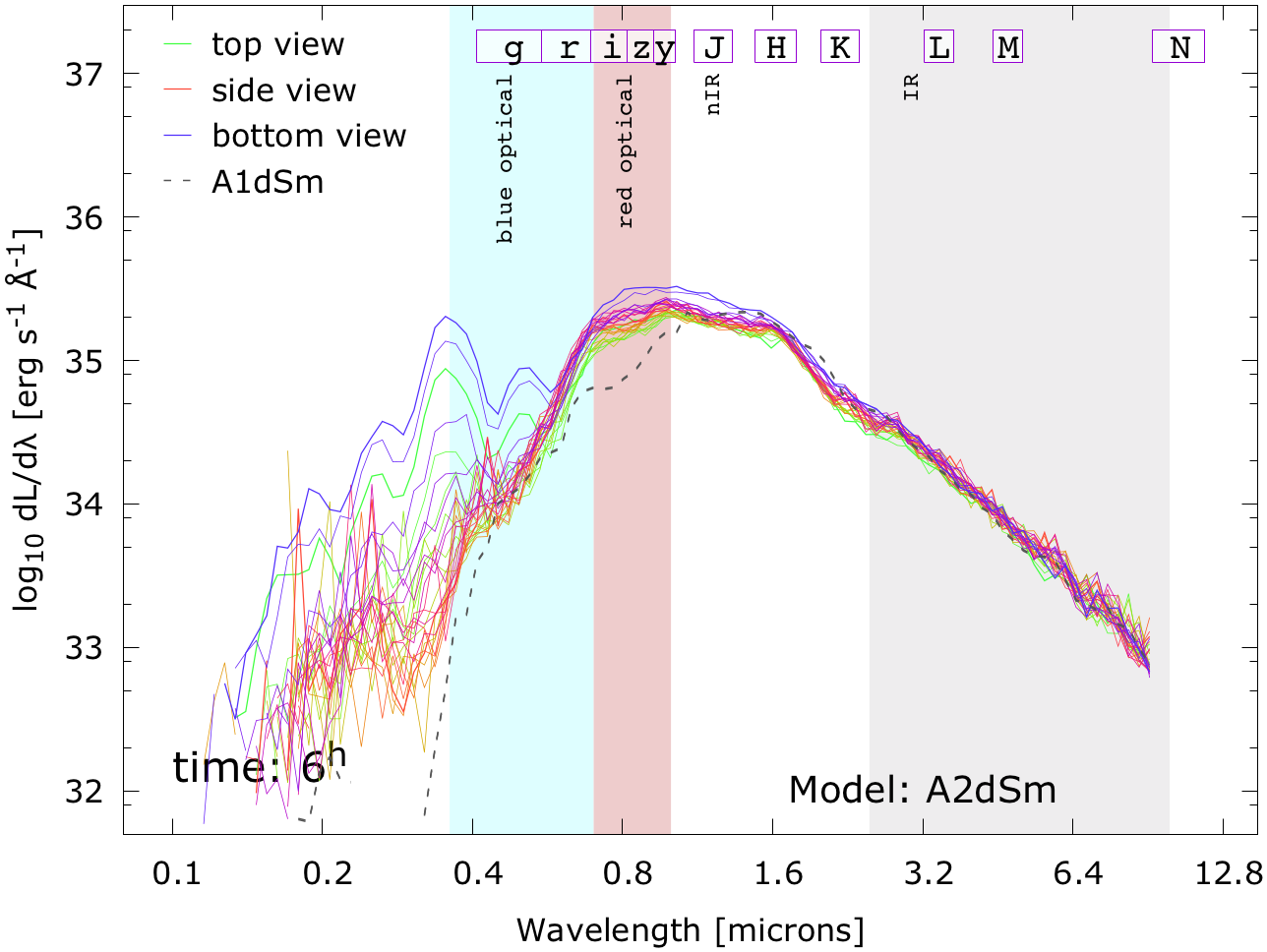}
\includegraphics[height=0.23\textheight]{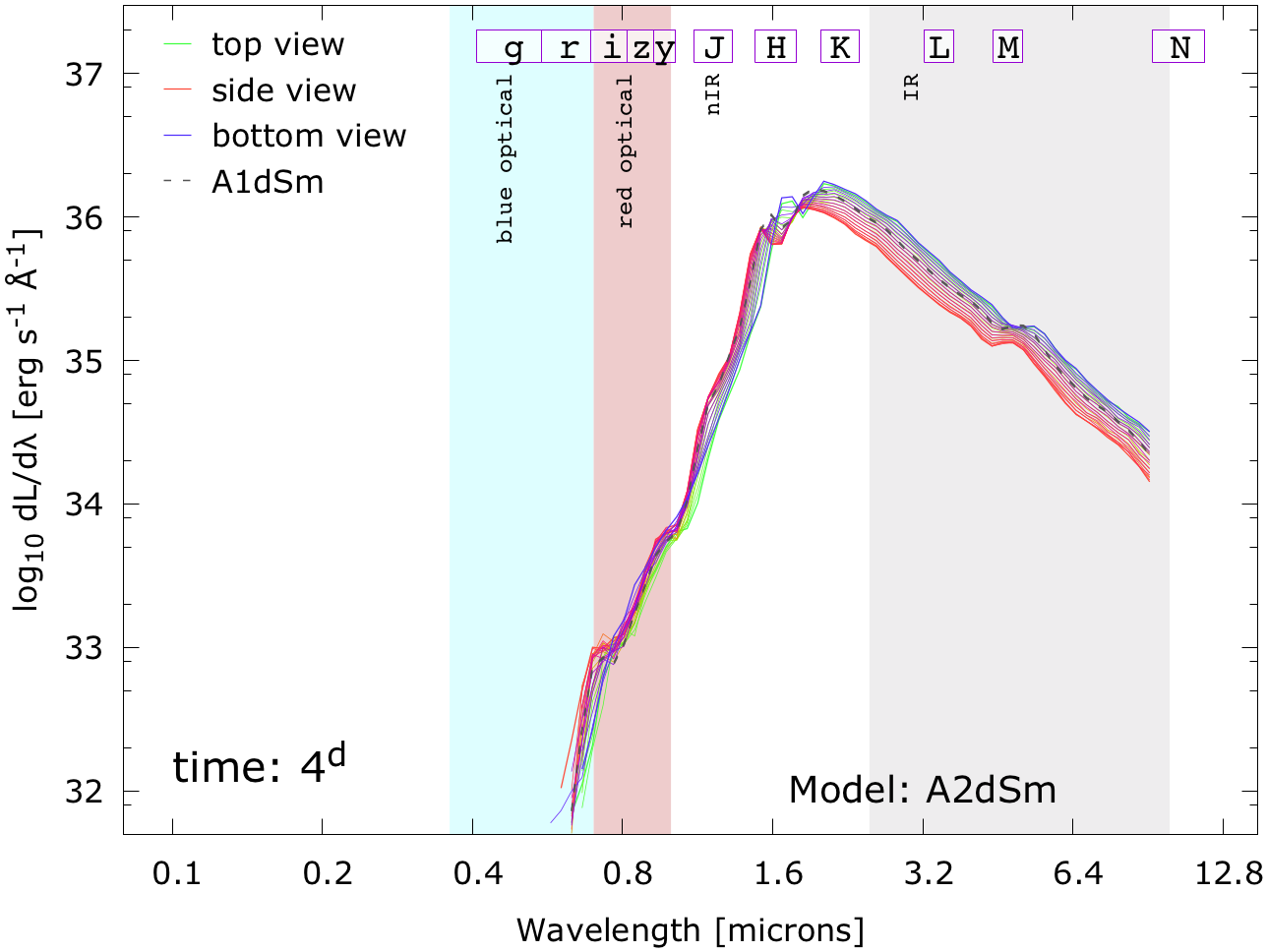}
\\
\includegraphics[height=0.23\textheight]{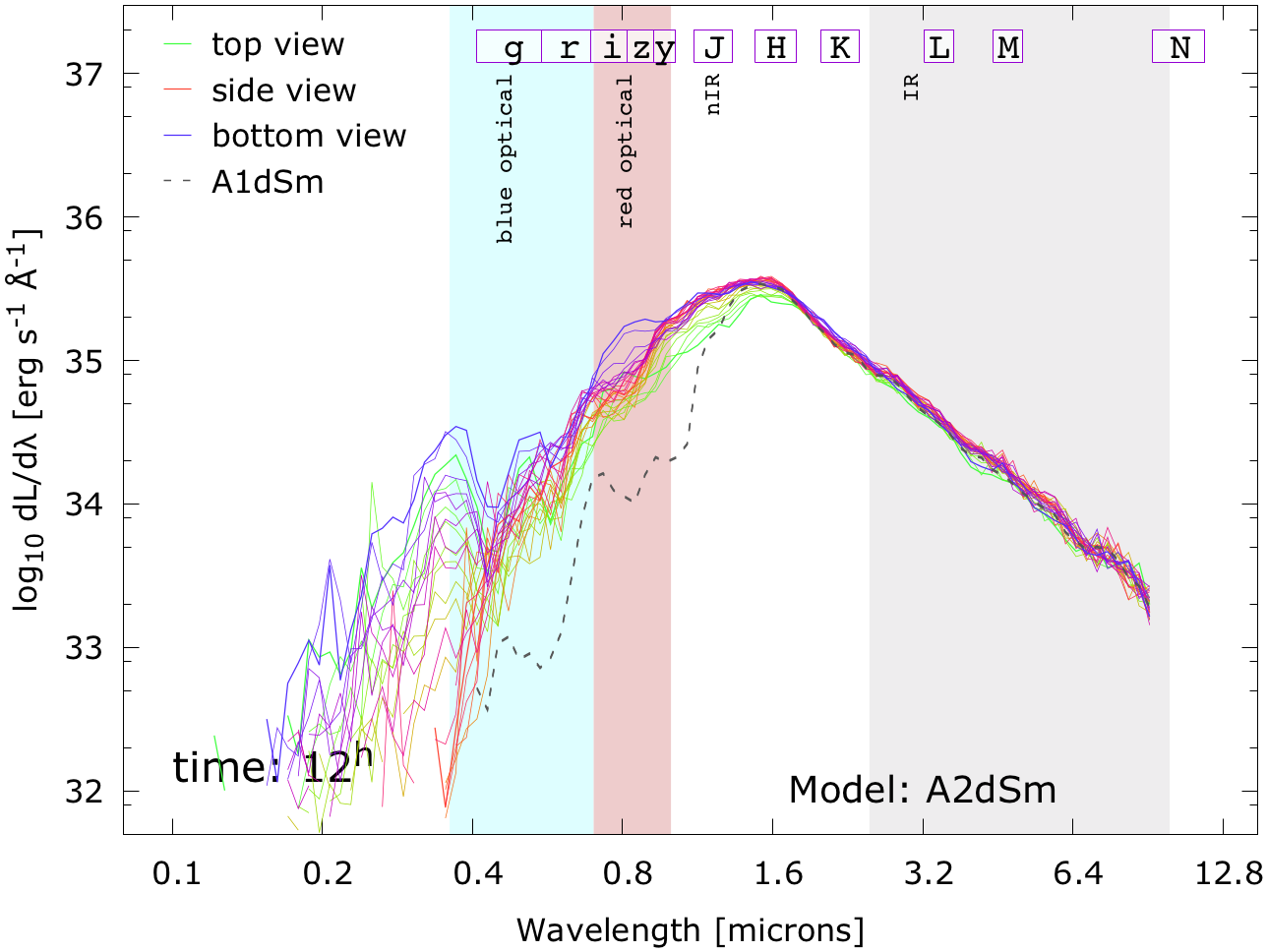}
\includegraphics[height=0.23\textheight]{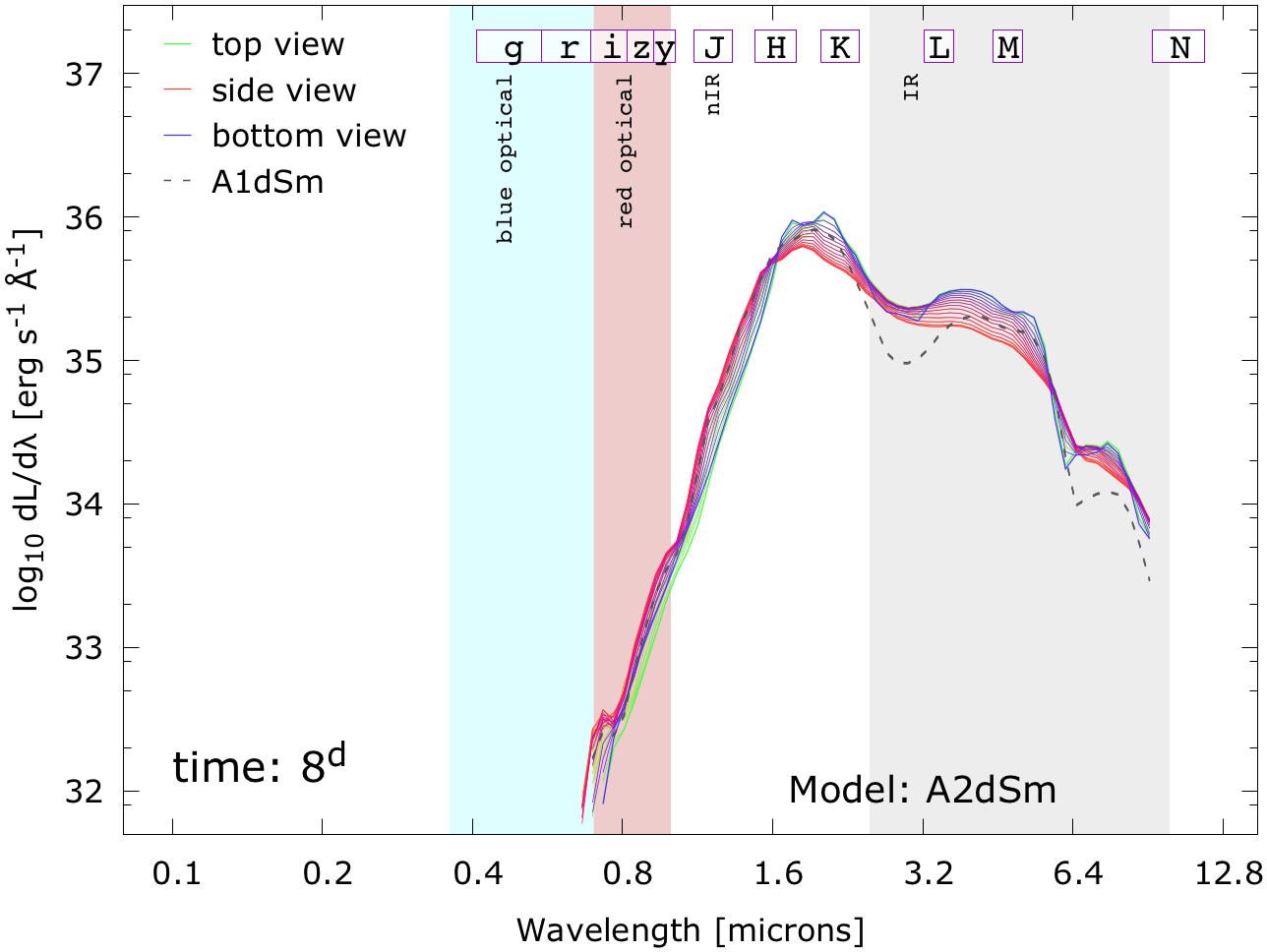}
\\
\includegraphics[height=0.23\textheight]{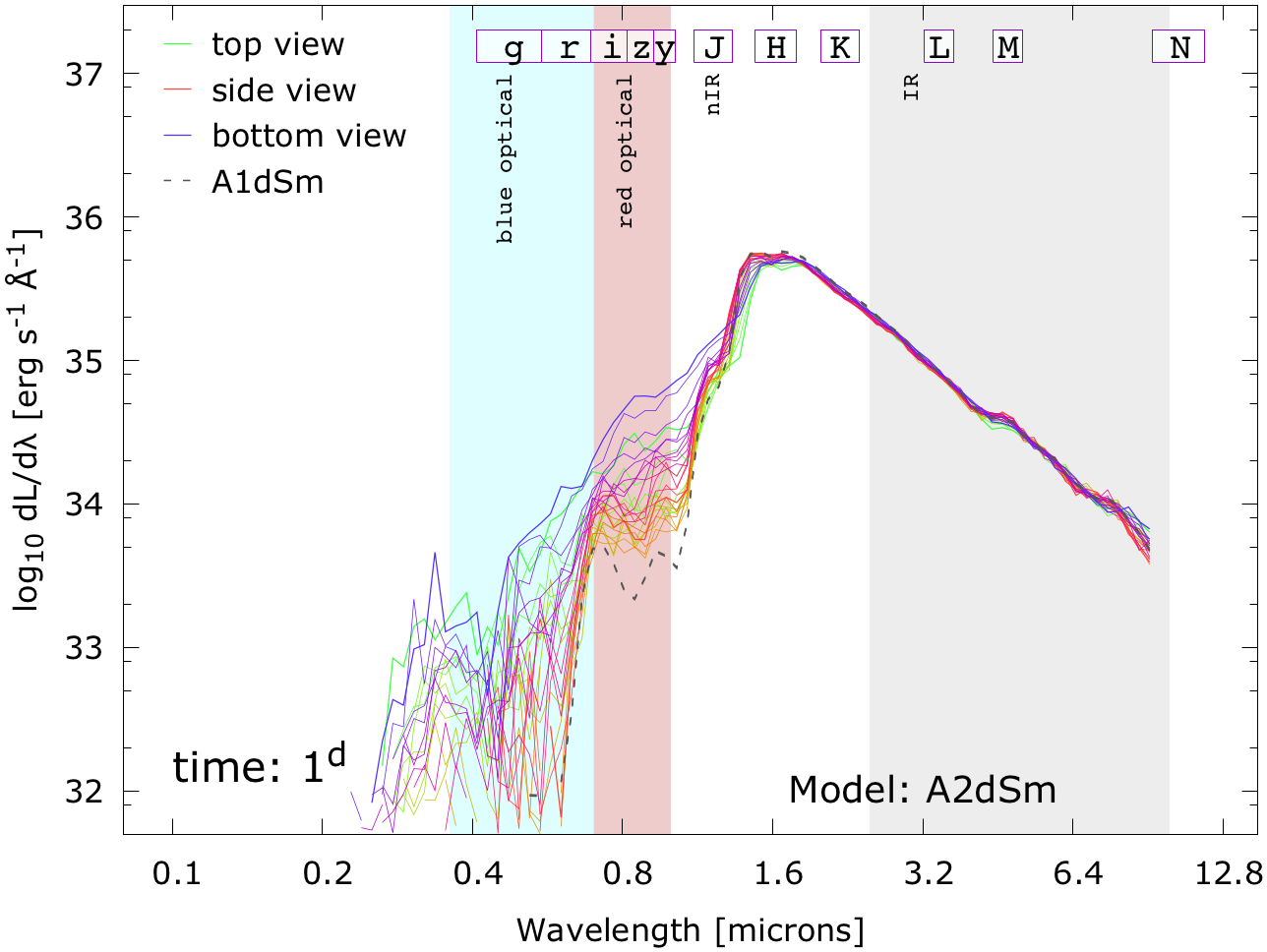}
\includegraphics[height=0.23\textheight]{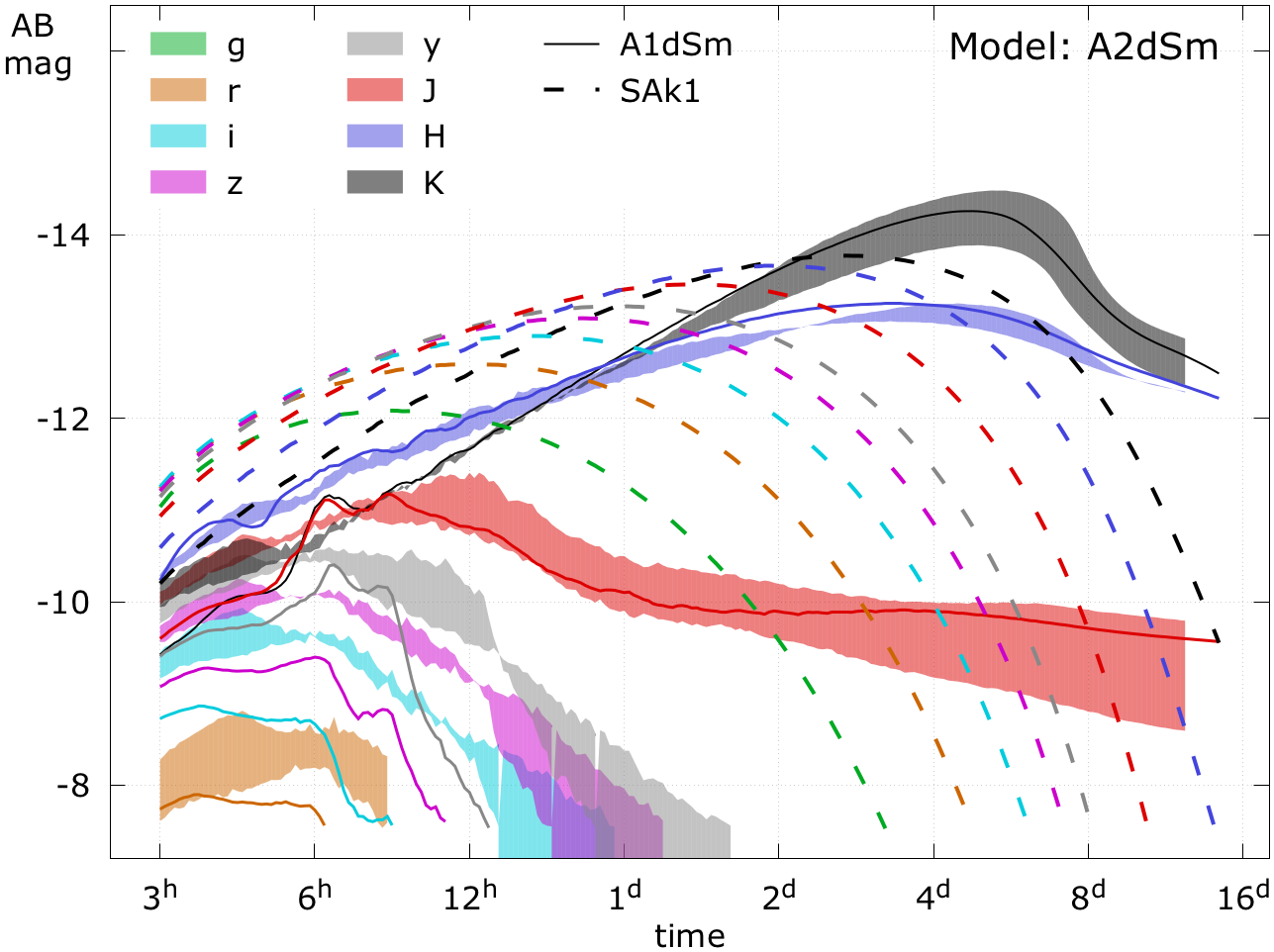}
\\
\end{tabular}
\end{adjustbox}
\caption{Time evolution of synthetic spectra for models {\tt A2dSm} and {\tt
A1dSm}. Bottom right: broadband light curves for model {\tt A2dSm} for two
different orientations (colored ranges) with respect to the observer, compared
to the broadband light curves for the grey opacity model {\tt SAk1} with
{$\kappa=10\ \cmg$} (dashed lines) and spherically-symmetric averaged model
{\tt A1dSm} (solid lines). Note that the light curve in the $g$ band for
the grey opacity model reaches -12 mag, for models {\tt A1dSm} and {\tt A2dSm}
it is far too dim and thus is not shown.}
\label{fig:spectra_A2dSm}
\end{figure*}

Even this rough agreement between the grey opacity light curve and the
Sm-opacity models in the bolometric case breaks down if we consider broadband
light curves, similar to the discrepancy pointed out in Sect.~\ref{sec:sa_multi}.
Due to the presence of the "spectral cliff" in open $f$-shell element models,
all $grizy$- and $J$-band light curves are strongly suppressed in comparison to
the grey opacity model.
This is shown in Fig.~\ref{fig:spectra_A2dSm}, bottom right panel, which
displays broadband light curves for models {\tt A2dSm}, {\tt A1dSm} and the
grey opacity model {\tt SAk1}.
All the light curves except for the longest-wavelength $H$ and $K$ bands are
suppressed by more than 3 mags. Other panels in Fig.~\ref{fig:spectra_A2dSm}
demonstrate snapshots of spectra for model {\tt A2dSm} at different times for
different viewing angles, and the "spectral cliff" at around 1.6 microns
(in the middle of the $H$ band) can be clearly identified. The rest of the
dynamical ejecta models from Table~\ref{tab:dyn_models} show very similar
spectral evolution and light curves.

\begin{figure}
\begin{center}
 \includegraphics[width=0.49\textwidth]{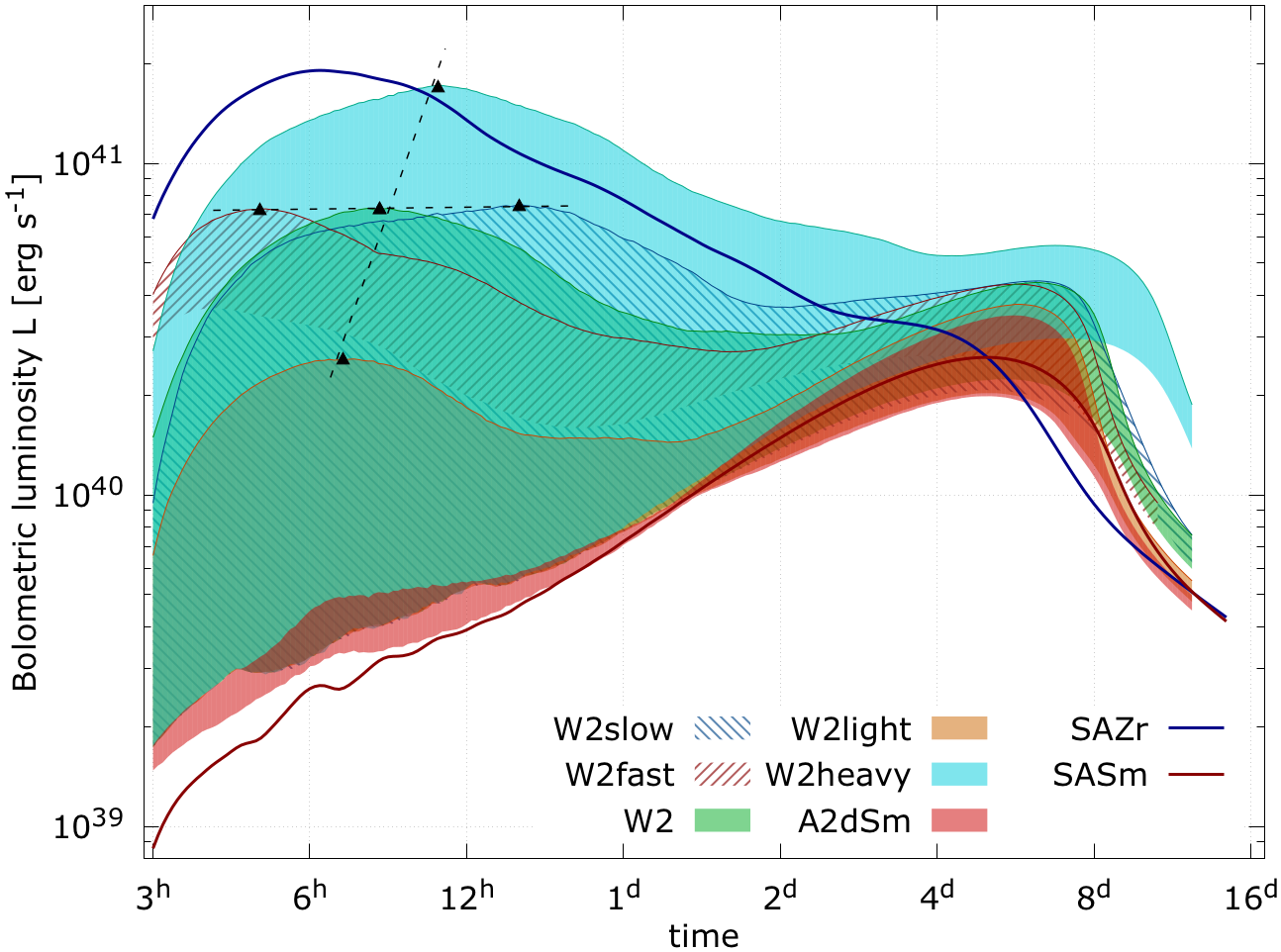}
\end{center}
\caption{Range of bolometric luminosities for combined models with varying
  wind parameters: {\tt W2}, {\tt W2light}-{\tt W2heavy} and
  {\tt W2slow}-{\tt W2fast} (see Table~\ref{tab:models_2d}) spanned by
  different orientations with respect to the observer. Spherically-symmetric
  multigroup opacity models {\tt SAZr} and {\tt SASm}, corresponding to the
  composition of the wind and dynamical ejecta respectively, are shown for
  comparison. The axisymmetric dynamical ejecta model {\tt A2dSm}, which
  is {\tt W2} without wind, is also shown. Upward triangles indicate locations
  of the blue transient peak, and dashed lines show peak trends with varying
  wind mass (semi-vertical) and velocity (horizontal line).}
\label{fig:lums_W2}
\end{figure}

An interesting morphological effect reveals itself in Fig.~\ref{fig:spectra_A2dSm}
when comparing the spectra of 2D model {\tt A2dSm} with corresponding
spherically-symmetric 1D case {\tt A1dSm}.
At early epochs $t<1^d$, the 2D model has non-negligible contributions in the optical
and even UV parts of the spectrum, absent in the 1D spectra. Moreover, this
feature is more pronounced for the on-axis orientations relative to the
"side" view. In other words, the merger remnant appears "bluer" in color if
shaped in toroidal form as opposed to the same mass arranged in a spherical
configuration. This is manifest in the $grizy$ bands, as shown in the bottom
right panel of Fig.~\ref{fig:spectra_A2dSm}: light curves of 2D models are
initially brighter than 1D models by almost 1 mag.

This effect can be explained if we recall that the remnant spectrum is
shaped by thermal emission originating from the depths of the remnant
and viewed through a layer of semi-opaque material, which dents the original
Planck spectrum with its opacity profile. This layer has smaller optical
depth for axisymmetric models, and as a consequence, the blue thermal emission
is less suppressed on the way out. In an extreme case, if ejecta had a
thin shell-like configuration, the spectrum would have been completely
unsuppressed in the optical bands despite lanthanide contamination.

\begin{table*}
\caption{Parameters of the combined axisymmetric models of dynamical ejecta
and wind.
First three columns list wind mass $M_{\rm wind}$,
half the maximum wind velocity $v_{\rm wind,max}$ in the wind density profile
(\ref{eq:rho_sphsym}), and opacity of the wind +
dynamical ejecta. Masses and median velocities of the dynamical ejecta
components A, B, C, D are listed in Table~\ref{tab:dyn_models}.
For dynamical ejecta component in the {\tt W2}-models we use detailed opacities of
Sm ($\kappa_{\lambda,{\rm Sm}}$). For dynamical ejecta in the rest of the models, we
use a mix of detailed ejecta opacities, as described in Sect.~\ref{sec:opacities}.
}
\begin{tabular}{l|ccc|cccc|cccc}
\hline\hline
      & & & \multicolumn{4}{c}{top view}
          & \multicolumn{4}{c}{bottom view}
\\
         & $M_{\rm wind}$ & $v_{\rm wind}/2$ & opacity
                          & $t_p^{(1)}$ & $L_p^{(1)}$ & $t_p^{(2)}$ & $L_p^{(2)}$
                          & $t_p^{(1)}$ & $L_p^{(1)}$ & $t_p^{(2)}$ & $L_p^{(2)}$
\\
Model    & $[M_\odot]$    & $[c]$ & 
                          & $[d]$ & $[10^{40}\frac{\rm erg}{\rm s}]$ & $[d]$ & $[10^{40}\frac{\rm erg}{\rm s}]$
                          & $[d]$ & $[10^{40}\frac{\rm erg}{\rm s}]$ & $[d]$ & $[10^{40}\frac{\rm erg}{\rm s}]$
\\
\hline
W2A (W2) & 0.005 & 0.08 & $\kappa_{\rm Zr}+\kappa_{\rm Sm}$ & 0.48 & 2.429 & 6.24 & 4.266 &  0.34 & 7.309 & 6.20 & 4.366 \\
W2B      & 0.005 & 0.08 & $\kappa_{\rm Zr}+\kappa_{\rm Sm}$ & -    & -     & 5.93 & 4.102 &  -    & -     & 5.93 & 4.162 \\
W2C      & 0.005 & 0.08 & $\kappa_{\rm Zr}+\kappa_{\rm Sm}$ & 0.38 & 1.990 & 7.24 & 6.788 &  -    & -     & 7.20 & 6.898 \\
W2D      & 0.005 & 0.08 & $\kappa_{\rm Zr}+\kappa_{\rm Sm}$ & -    & -     & 7.08 & 7.208 &  -    & -     & 7.02 & 7.324 \\
\hline
W2Se     & 0.005 & 0.08 & $\kappa_{\rm Se}+\kappa_{\rm Sm}$ & 0.49 & 4.873 & 5.93 & 4.389 &  0.35 & 13.92 & 5.78 & 4.546 \\
W2Br     & 0.005 & 0.08 & $\kappa_{\rm Br}+\kappa_{\rm Sm}$ & 0.65 & 4.260 & 5.94 & 4.387 &  0.44 & 11.67 & 5.80 & 4.531 \\
W2Te     & 0.005 & 0.08 & $\kappa_{\rm Te}+\kappa_{\rm Sm}$ & 0.85 & 3.702 & 5.94 & 4.386 &  0.51 & 9.931 & 5.80 & 4.525 \\
W2Pd     & 0.005 & 0.08 & $\kappa_{\rm Pd}+\kappa_{\rm Sm}$ & 0.65 & 3.292 & 5.93 & 4.388 &  0.48 & 9.227 & 5.80 & 4.537 \\
W2Zr (W2)& 0.005 & 0.08 & $\kappa_{\rm Zr}+\kappa_{\rm Sm}$ & 0.48 & 2.429 & 6.24 & 4.266 &  0.34 & 7.309 & 6.20 & 4.366 \\
W2Cr     & 0.005 & 0.08 & $\kappa_{\rm Cr}+\kappa_{\rm Sm}$ & -    & -     & 5.91 & 4.408 &  1.62 & 4.437 & 5.76 & 4.557 \\
\hline
W2light  & 0.001 & 0.08 & $\kappa_{\rm Zr}+\kappa_{\rm Sm}$ & 0.33 & 0.980 & 5.78 & 3.673 &  0.29 & 2.583 & 5.75 & 3.755 \\
W2heavy  & 0.02  & 0.08 & $\kappa_{\rm Zr}+\kappa_{\rm Sm}$ & 0.94 & 5.949 & 6.91 & 5.518 &  0.44 & 17.09 & 6.79 & 5.653 \\
W2slow   & 0.005 & 0.04 & $\kappa_{\rm Zr}+\kappa_{\rm Sm}$ & 0.78 & 3.948 & 6.25 & 4.371 &  0.63 & 7.453 & 6.22 & 4.426 \\
W2fast   & 0.005 & 0.16 & $\kappa_{\rm Zr}+\kappa_{\rm Sm}$ & 0.20 & 7.287 & 5.73 & 4.225 &  0.21 & 5.786 & 5.69 & 4.327 \\
\hline
X$_1$    & 0.005 & 0.08 & $\kappa_{\rm wind 1}+\kappa_{\rm dyn}$ & -    & -     & 2.76 & 6.408 &  0.42 & 5.952 & 2.77 & 6.593 \\
X$_2$    & 0.005 & 0.08 & $\kappa_{\rm wind 2}+\kappa_{\rm dyn}$ & -    & -     & 2.54 & 6.247 &  0.56 & 7.576 & 3.73 & 6.184 \\
DZ$_1$   & 0.005 & 0.08 & $\kappa_{\rm wind 1}+\kappa_{\rm dyn}$ & -    & -     & 5.17 & 28.79 &  0.5* & 11.0* & 5.28 & 29.89 \\
DZ$_2$   & 0.005 & 0.08 & $\kappa_{\rm wind 2}+\kappa_{\rm dyn}$ & -    & -     & 5.24 & 27.92 &  0.5* & 13.0* & 5.28 & 28.97 \\
Xnh$_1$  & 0.005 & 0.08 & $\kappa_{\rm wind 1}+\kappa_{\rm dyn}$ & 0.69 & 4.041 & 4.29 & 6.893 &  0.40 & 11.08 & 4.23 & 7.094 \\
Xnh$_2$  & 0.005 & 0.08 & $\kappa_{\rm wind 2}+\kappa_{\rm dyn}$ & 0.62 & 9.499 & 2.77 & 10.14 &  0.42 & 25.87 & 2.8${}^*$ & 9.8${}^*$ \\
\hline
$\gamma A_1$&0.005&0.08 & $\kappa_{\rm wind 1}+\kappa_{\rm dyn}$ & 0.7* & 4.4*  & 4.12 & 7.408 &  0.40 & 11.75 & 4.07 & 7.679 \\
$\gamma B_1$&0.005&0.08 & $\kappa_{\rm wind 1}+\kappa_{\rm dyn}$ & -    & -     & 4.25 & 7.615 &  -    & -     & 4.25 & 7.705 \\
$\gamma C_1$&0.005&0.08 & $\kappa_{\rm wind 1}+\kappa_{\rm dyn}$ & 0.6* & 3.9*  & 5.29 & 13.60 &  -    & -     & 5.29 & 13.79 \\
$\gamma D_1$&0.005&0.08 & $\kappa_{\rm wind 1}+\kappa_{\rm dyn}$ & -    & -     & 5.30 & 14.12 &  -    & -     & 5.29 & 14.31 \\
$\gamma A_2$&0.005&0.08 & $\kappa_{\rm wind 2}+\kappa_{\rm dyn}$ & 0.61 & 10.06 & 2.87 & 11.05 &  0.42 & 27.57 & 2.7${}^*$ & 10.8${}^*$  \\
$\gamma B_2$&0.005&0.08 & $\kappa_{\rm wind 2}+\kappa_{\rm dyn}$ & -    & -     & 4.28 & 9.422 &  -    & -     & 4.37 & 9.363 \\
$\gamma C_2$&0.005&0.08 & $\kappa_{\rm wind 2}+\kappa_{\rm dyn}$ & 0.7* & 6.2*  & 5.35 & 14.98 &  -    & -     & 5.40 & 15.08 \\
$\gamma D_2$&0.005&0.08 & $\kappa_{\rm wind 2}+\kappa_{\rm dyn}$ & -    & -     & 5.33 & 15.47 &  -    & -     & 5.40 & 15.57 \\
\hline
\end{tabular}
\\
\begin{flushleft}
{\small ($^*$) Distinct peak is missing in these models.}
\end{flushleft}
\label{tab:models_2d}
\end{table*}

\subsection{Combined models of wind and dynamical ejecta}
\label{sec:combined}

Combined models (as introduced in Sect.~\ref{sec:dyn_wind}) superimpose
axisymmetric configurations of the dynamical ejecta with various
parameterized spherically-symmetric profiles for the wind
(see Fig.~\ref{fig:superimposed}, bottom panel).
These models were designed to assess visibility of a potential bright blue
transient originating from the wind.

The first 14 entries in Table~\ref{tab:models_2d} list combined models
(denoted with a prefix {\tt W2}, "W" for "wind", and "2" for "2D-models").
The rest of the models in the Table improve on these models by adding more
physics, and will be considered in the Sect.~\ref{sec:realistic_models} below.
For the purpose of comparison with previous sections, all {\tt W2}-models are
calculated with the multigroup opacity of Sm (same as for the model {\tt A2dSm}).
All models use dynamical ejecta configuration A, except for {\tt W2B},
{\tt W2C} and {\tt W2D}, which use use configurations B, C and D respectively.
Opacity for the wind in models {\tt W2A}-{\tt W2D} and models
{\tt W2heavy}/{\tt W2light}, {\tt W2fast}/{\tt W2slow} is the multigroup opacity
of Zr, which was selected because it is the most opaque element shaping the
spectra and light curves in mixtures (see Sect.~\ref{sec:sa_multi}).

Figure~\ref{fig:lums_W2} displays bolometric luminosities for the baseline
model {\tt W2}, models departing from the baseline in wind mass
{\tt W2light}/{\tt W2heavy}, and models departing from the baseline in wind
expansion velocity {\tt W2slow}/{\tt W2fast}.
The plot also shows comparative luminosities of {\tt A2dSm} and single-element
spherically-symmetric models {\tt SAZr} and {\tt SASm} with uniform
composition, corresponding to that of the wind and ejecta only.
To reflect the luminosity range due to different orientations, each 2D model
is displayed as a stripe, with the upper stripe bound corresponding to the on-axis
view, and the lower bound showing the "side" view. At times $t<1^d$, the
on-axis luminosities approach those of {\tt SAZr}, while the "side" view
luminosities always stay close to  {\tt SASm}.

\begin{figure*}
\begin{adjustbox}{width=0.90\textwidth}
\begin{tabular}{cc}
\includegraphics[height=0.24\textheight]{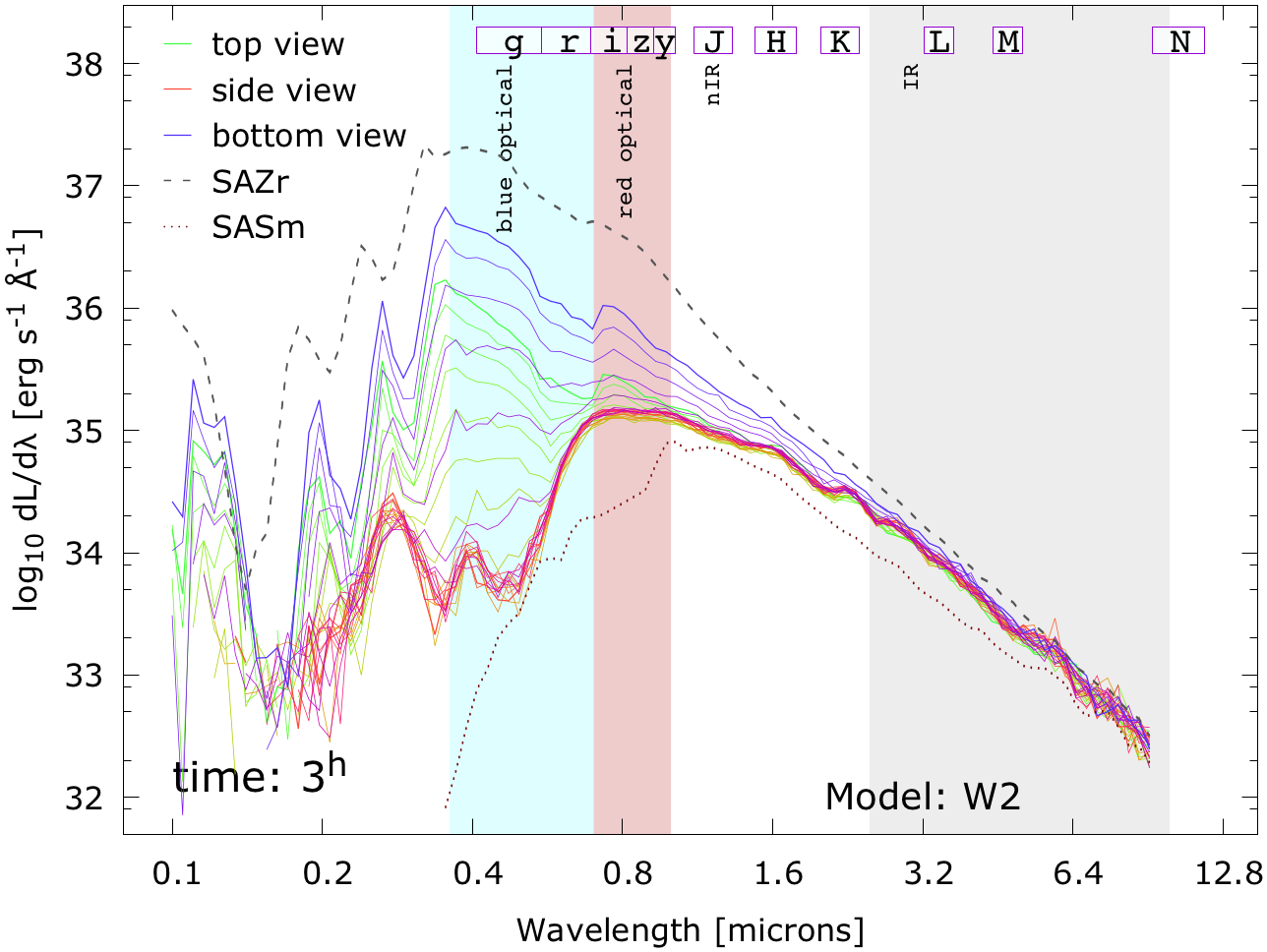} &
\includegraphics[height=0.24\textheight]{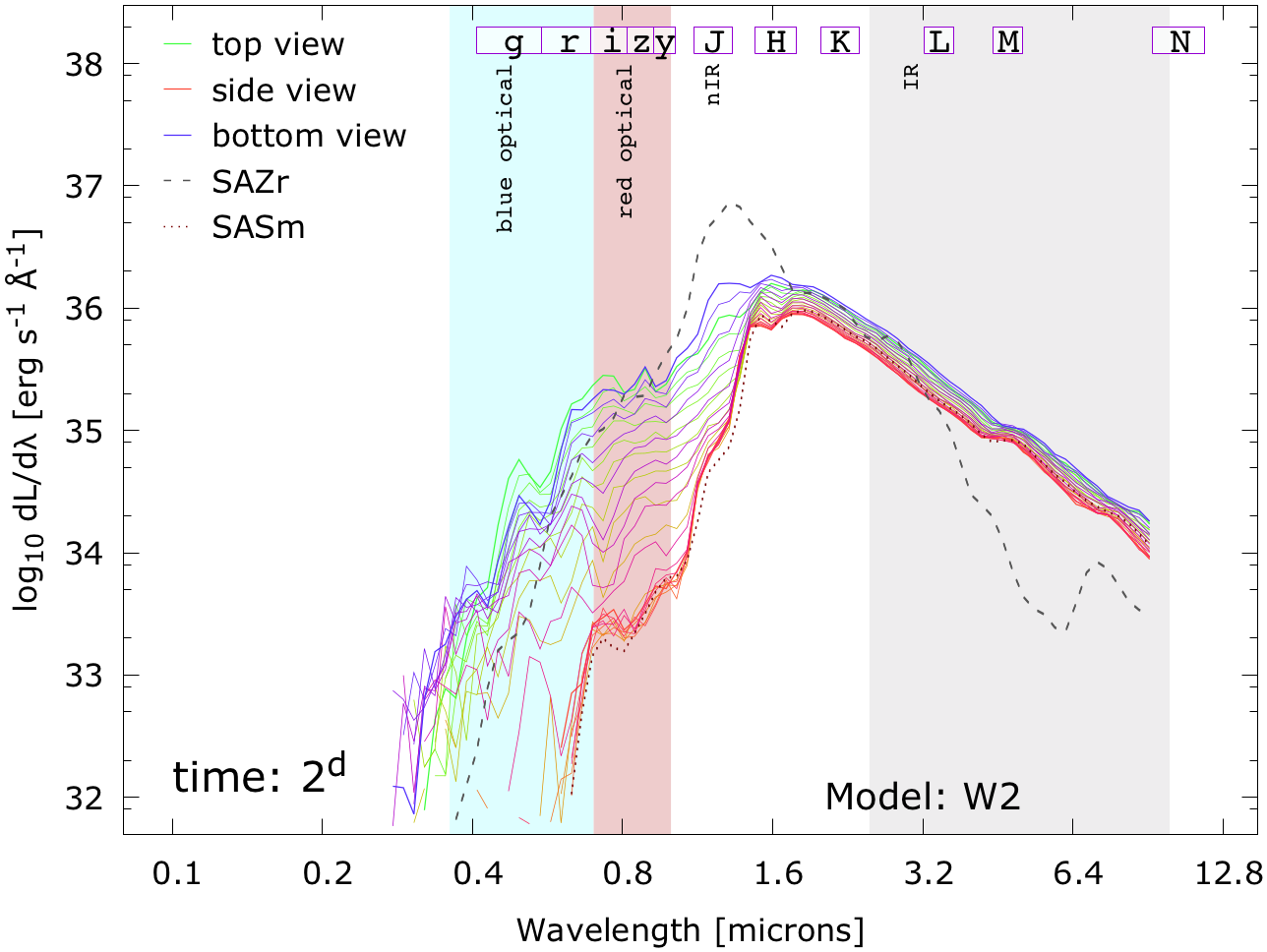}
\\
\includegraphics[height=0.24\textheight]{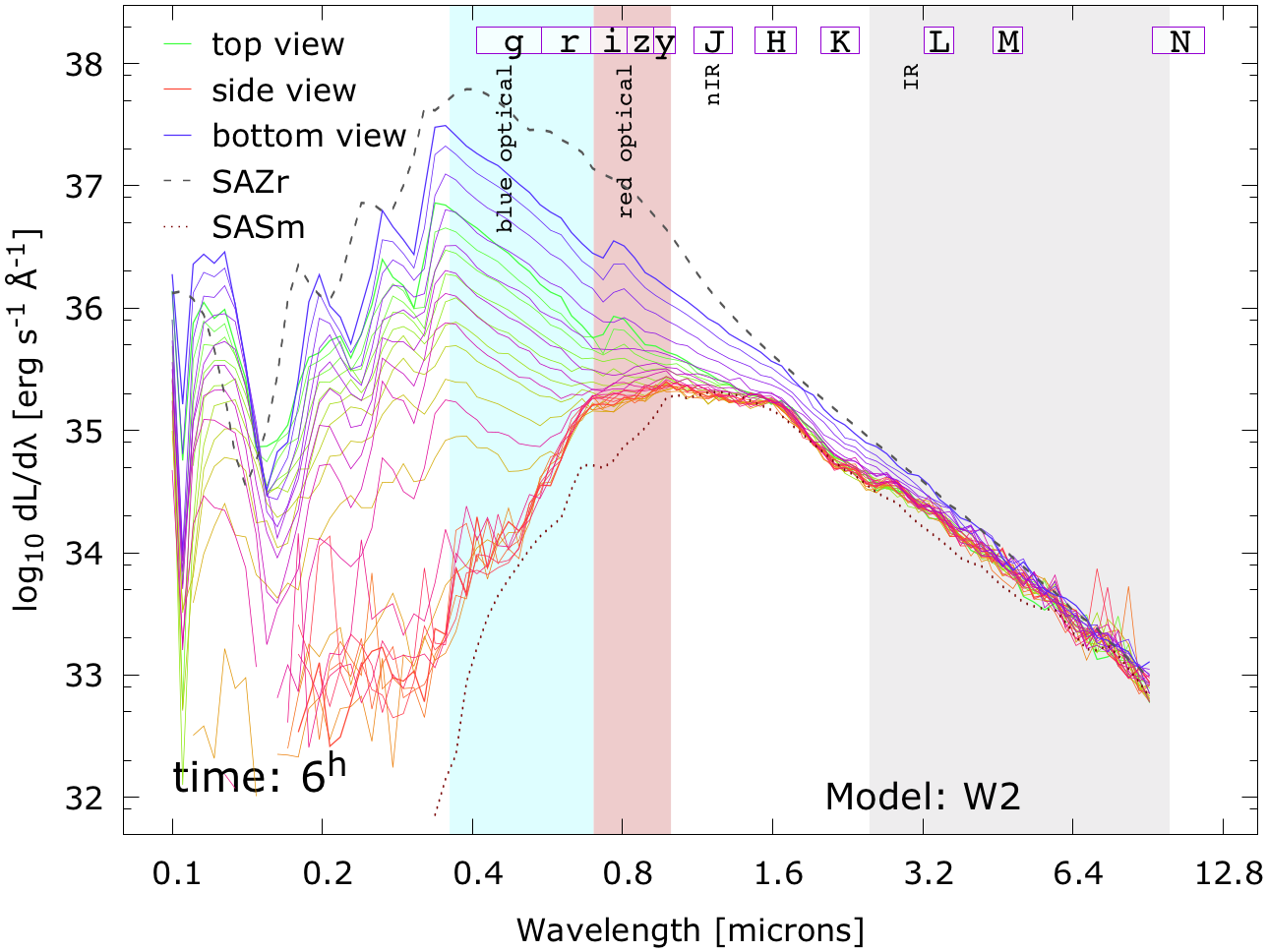} &
\includegraphics[height=0.24\textheight]{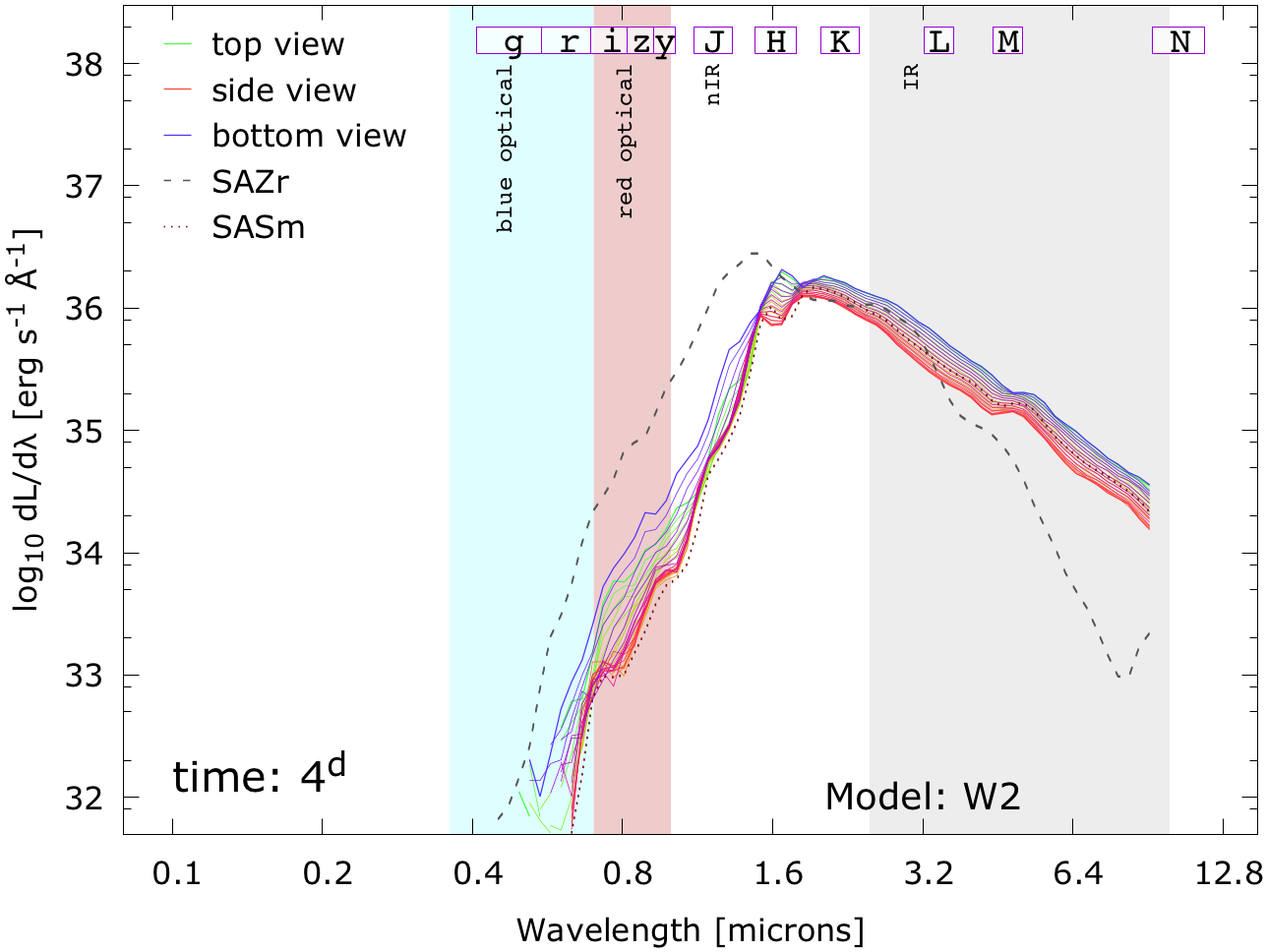}
\\
\includegraphics[height=0.24\textheight]{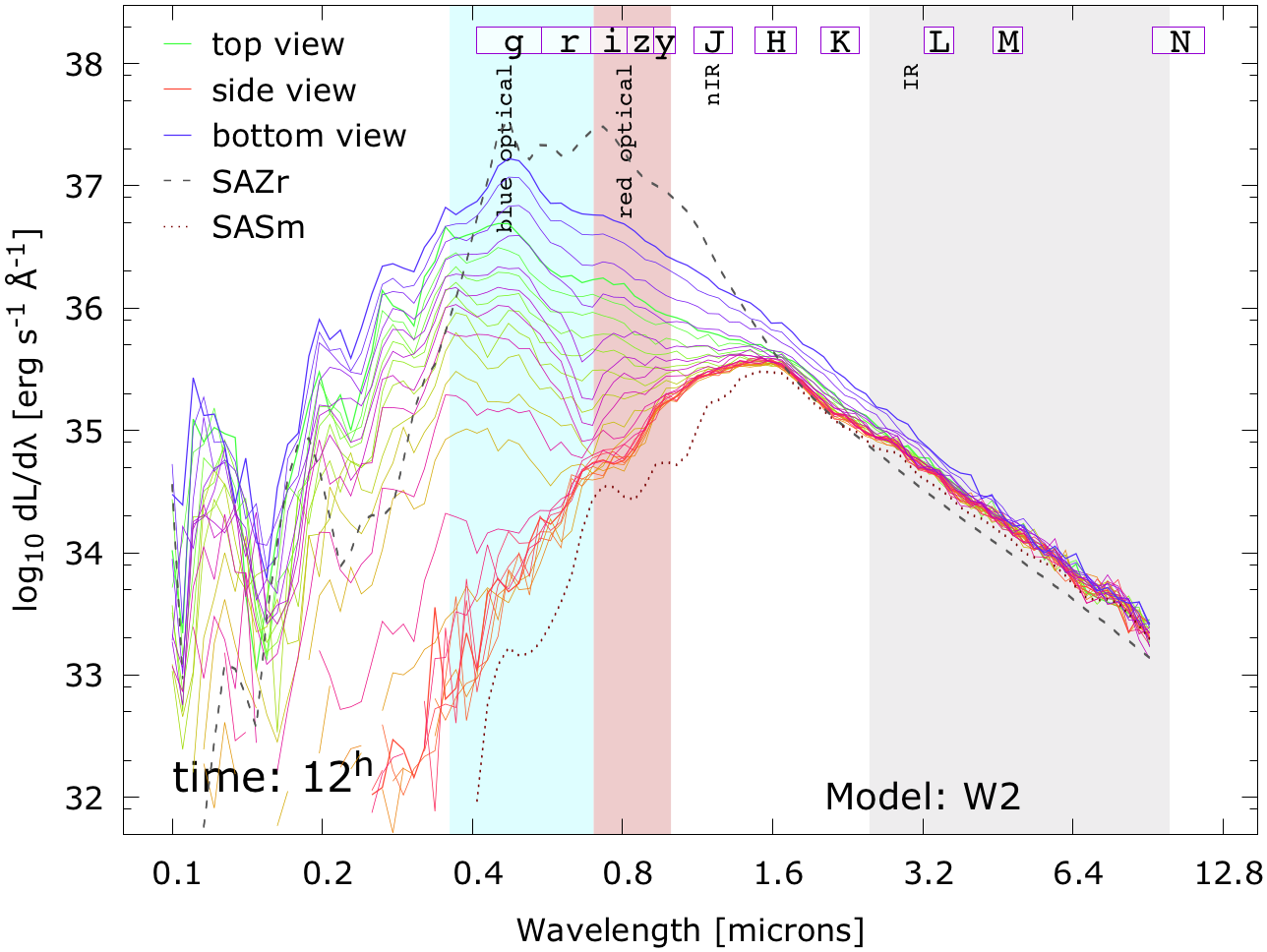} &
\includegraphics[height=0.24\textheight]{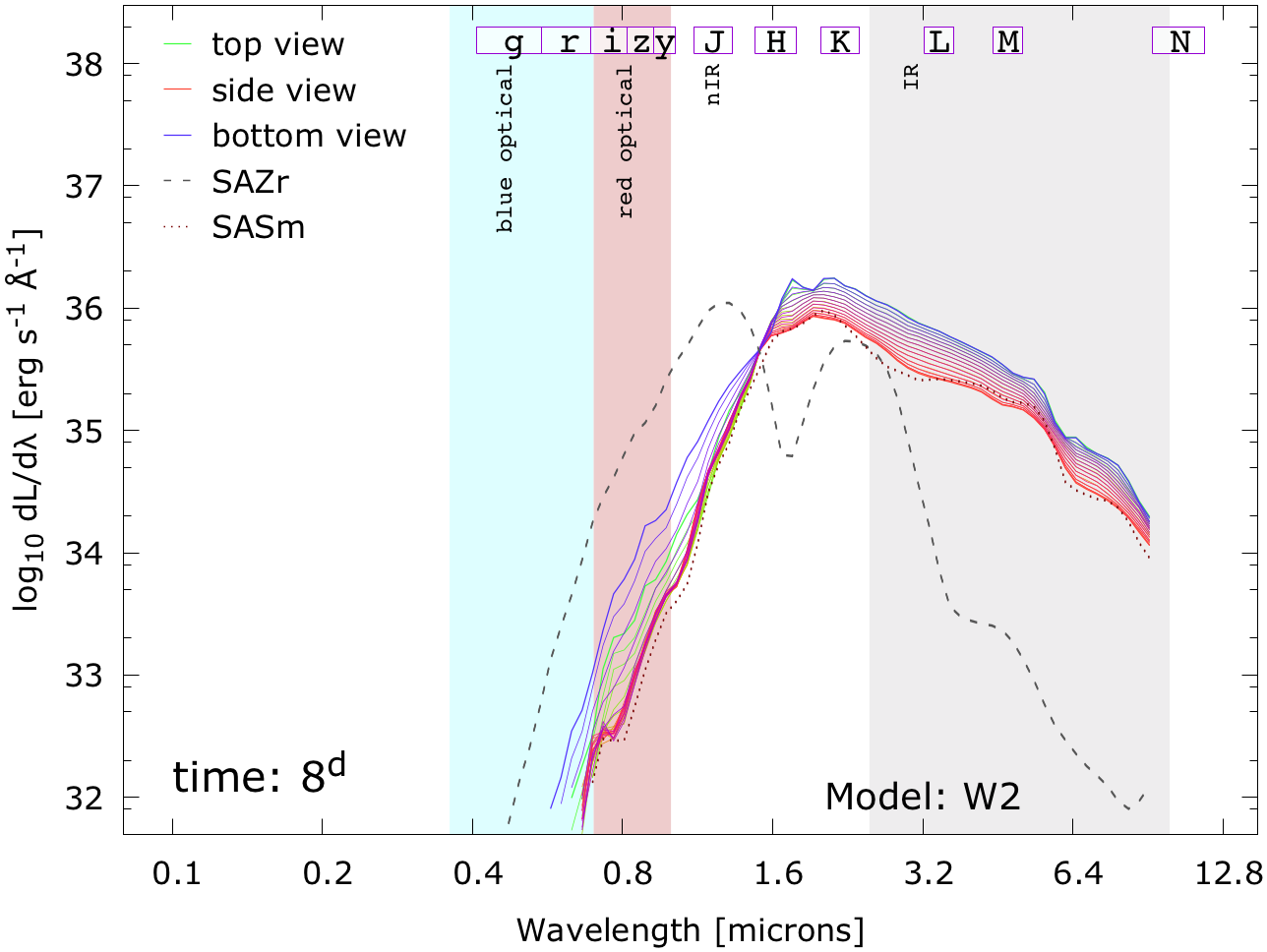}
\\
\includegraphics[height=0.24\textheight]{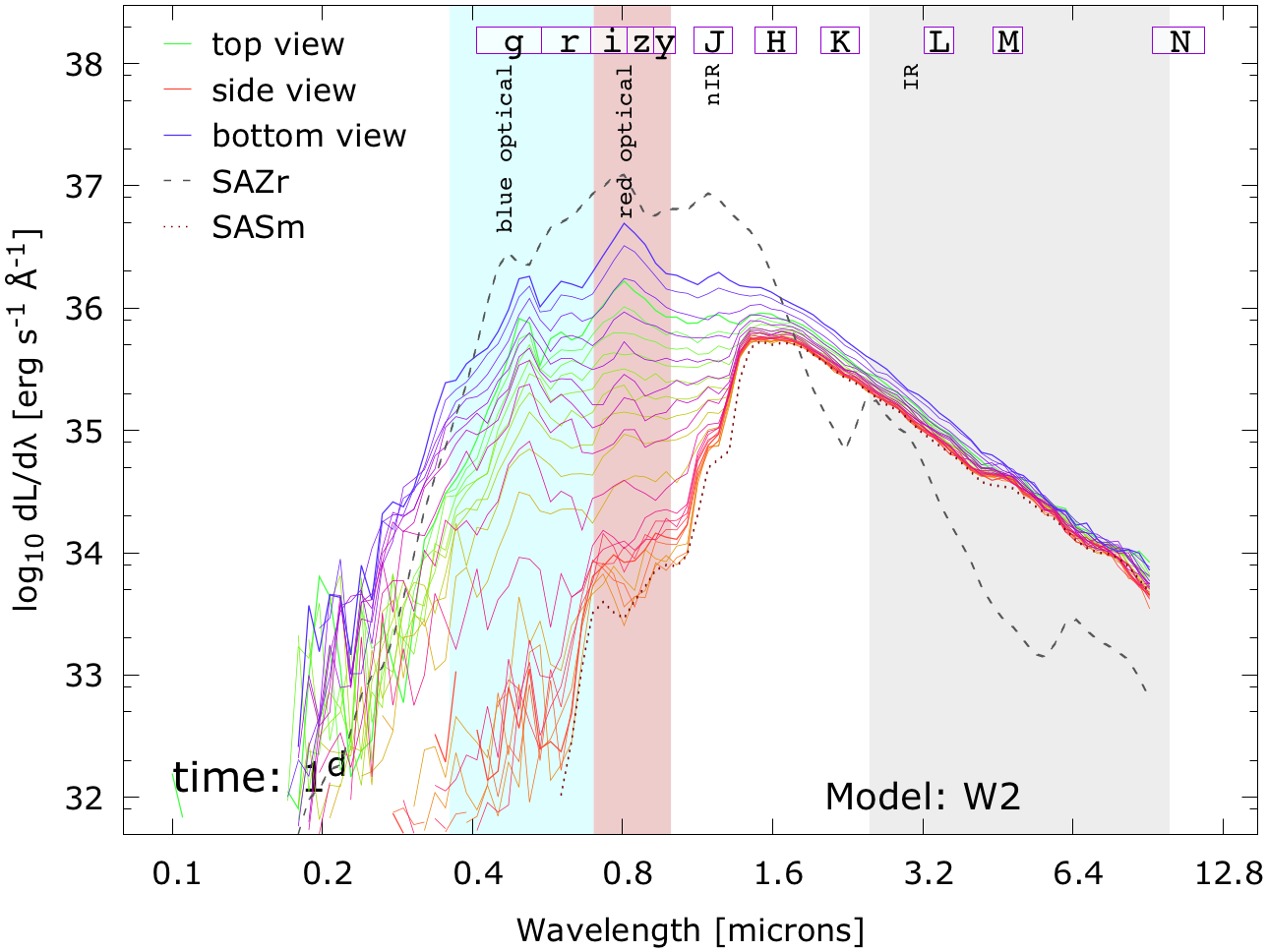} &
\includegraphics[height=0.24\textheight]{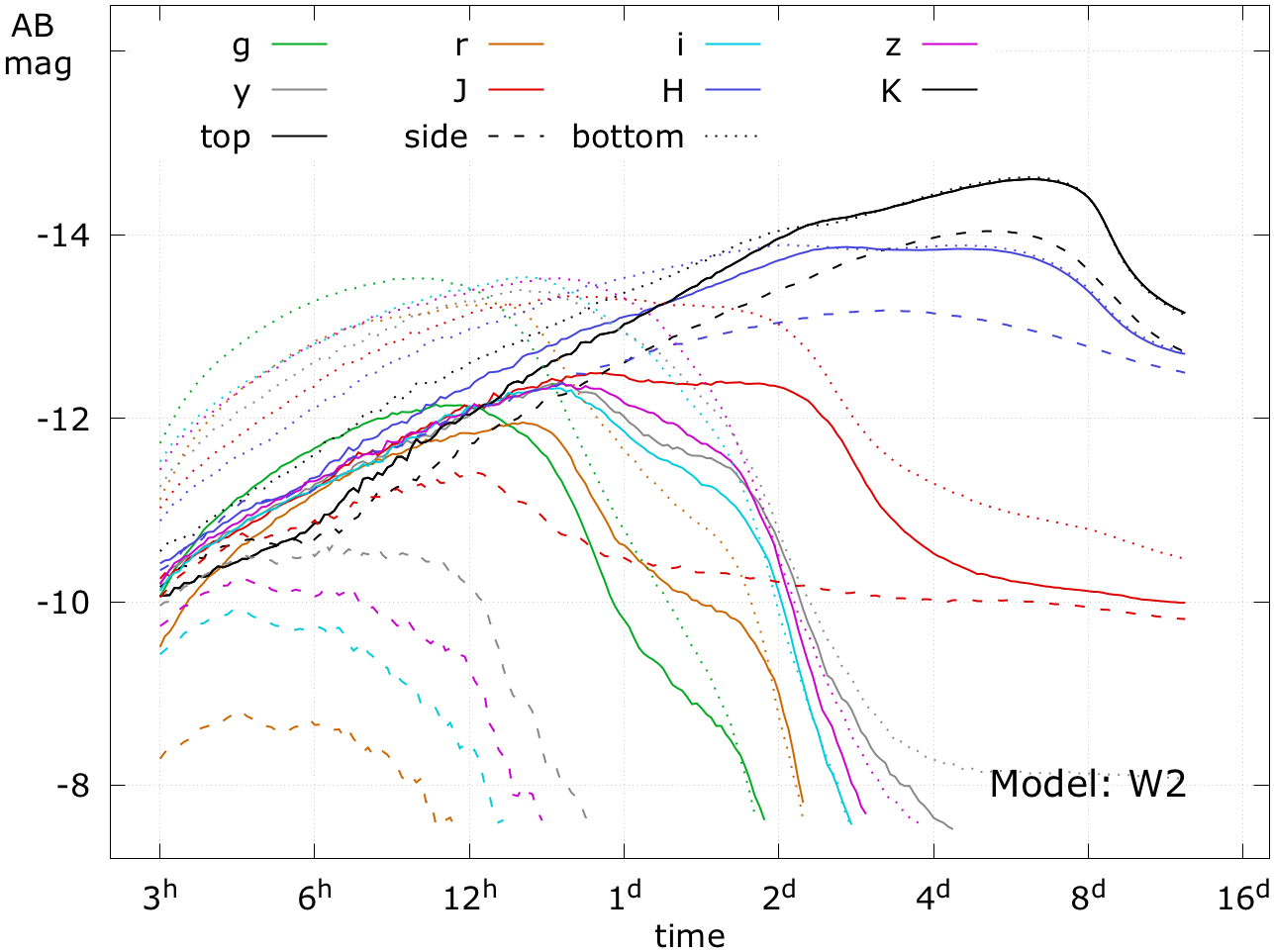}
\\
\end{tabular}
\end{adjustbox}
\caption{Time evolution of angle-dependent spectra for the baseline combined
model {\tt W2} (also denoted as {\tt W2A} or {\tt W2Zr}). Bottom right:
bolometric light curves of the same. Color gradient indicates polar angle
$\theta$, and spans 27 angular bins from "top" ($\theta=0$, blue) to "bottom"
($\theta=\pi$, green), spaced equally in $\cos\theta$.
The spike at $\lambda\sim0.16$ at early times is unphysical and caused by
artificial windows in our opacity profile at the same wavelengths.
}
\label{fig:spectra_W2Zr}
\end{figure*}

\begin{figure*}
\begin{tabular}{cc}
\includegraphics[height=0.24\textheight]{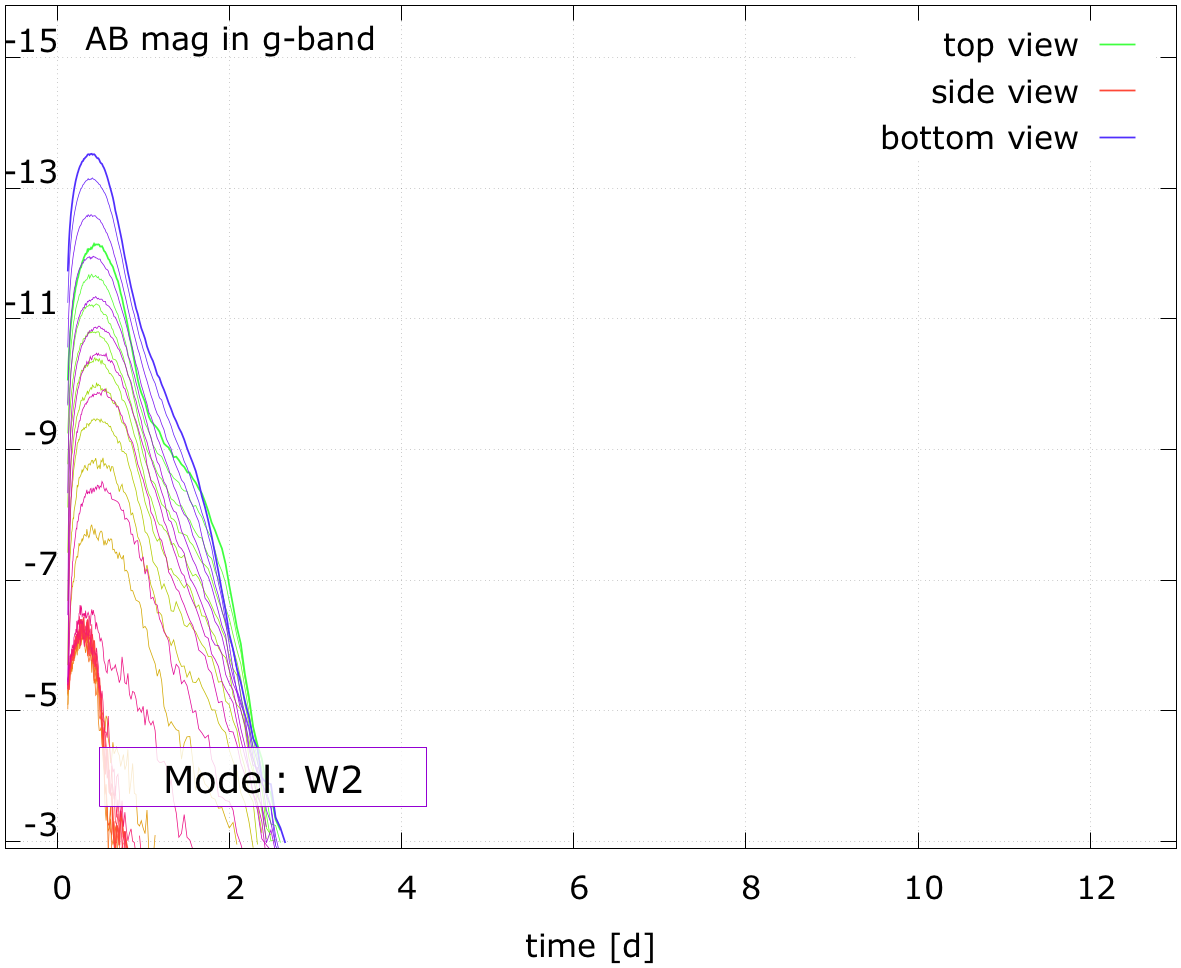} &
\includegraphics[height=0.24\textheight]{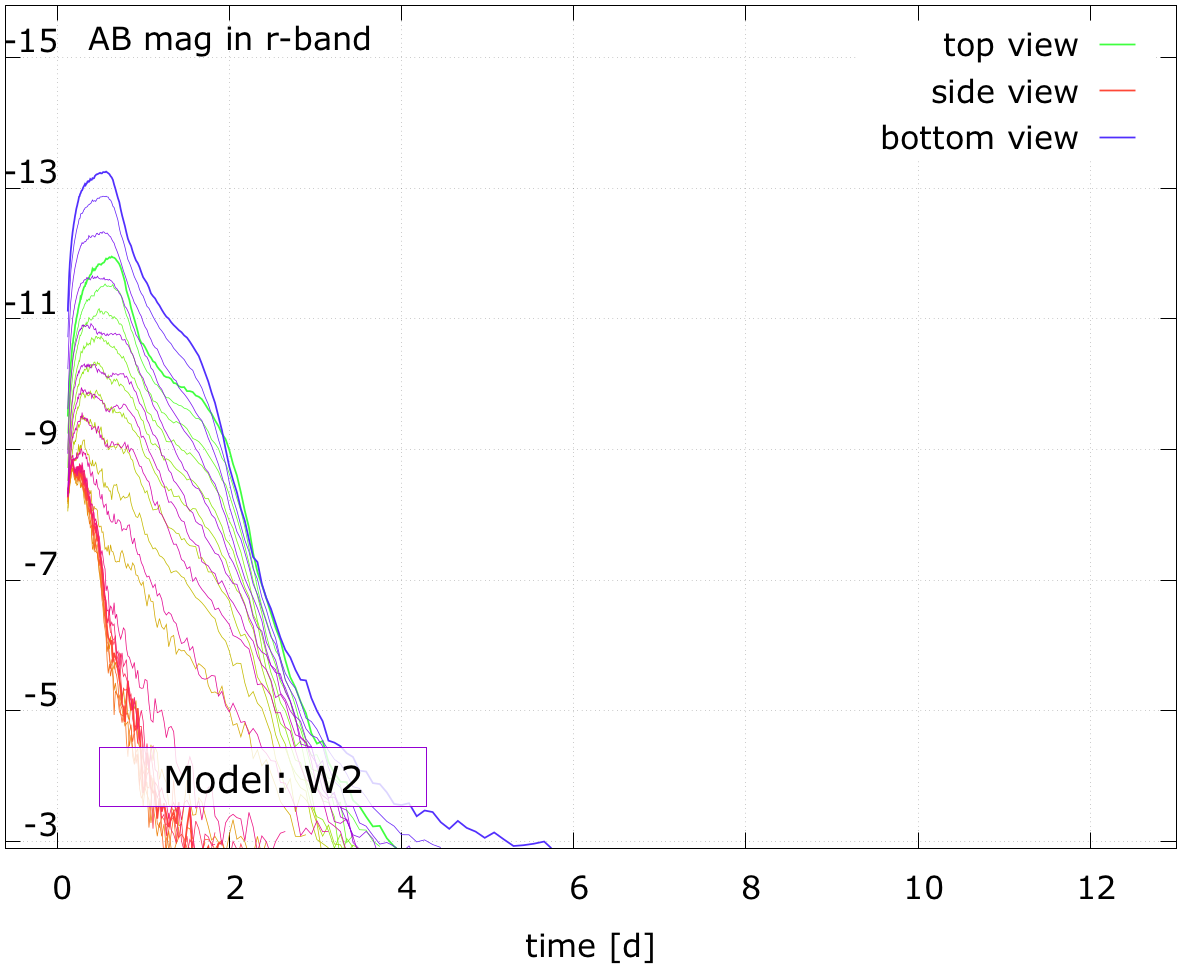}
\\
\includegraphics[height=0.24\textheight]{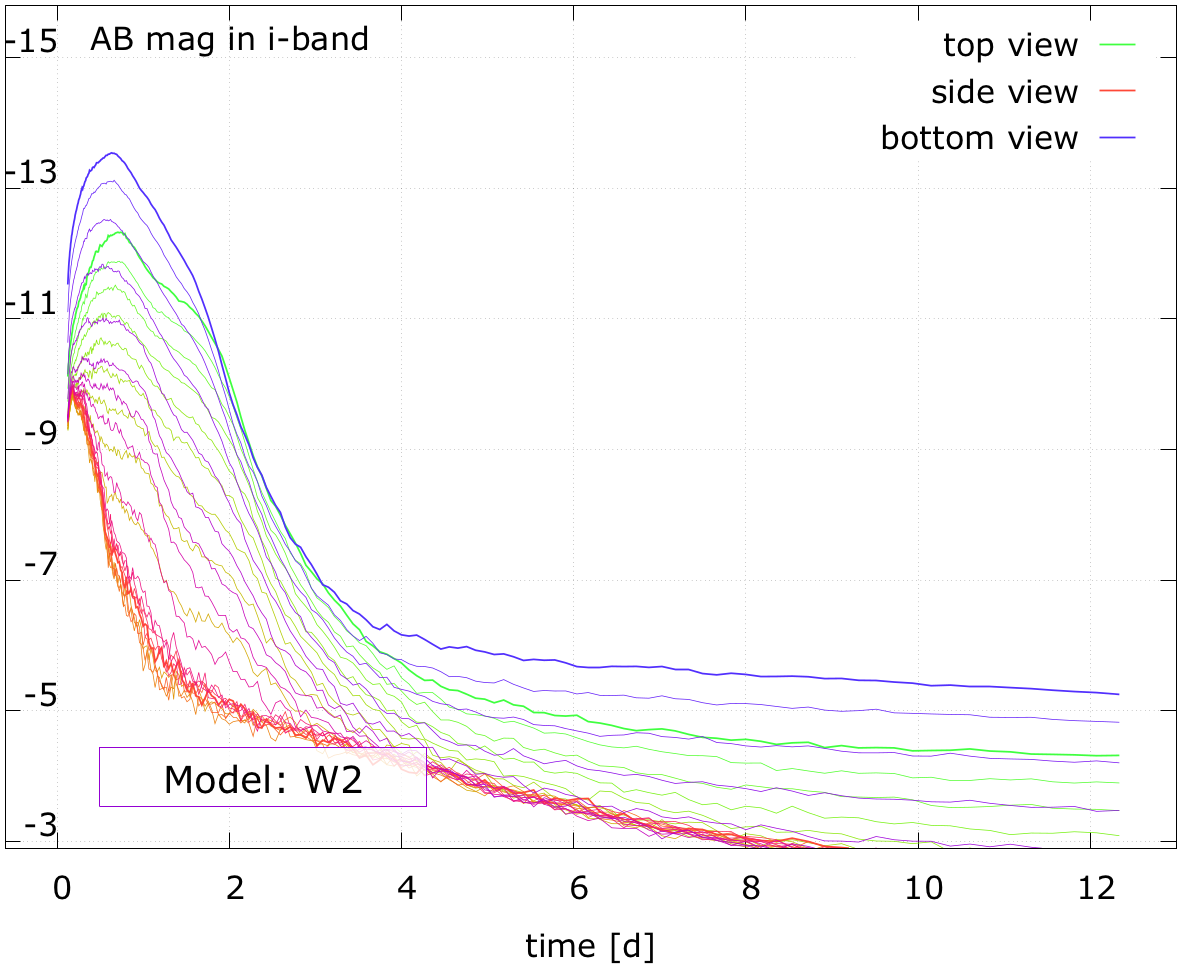} &
\includegraphics[height=0.24\textheight]{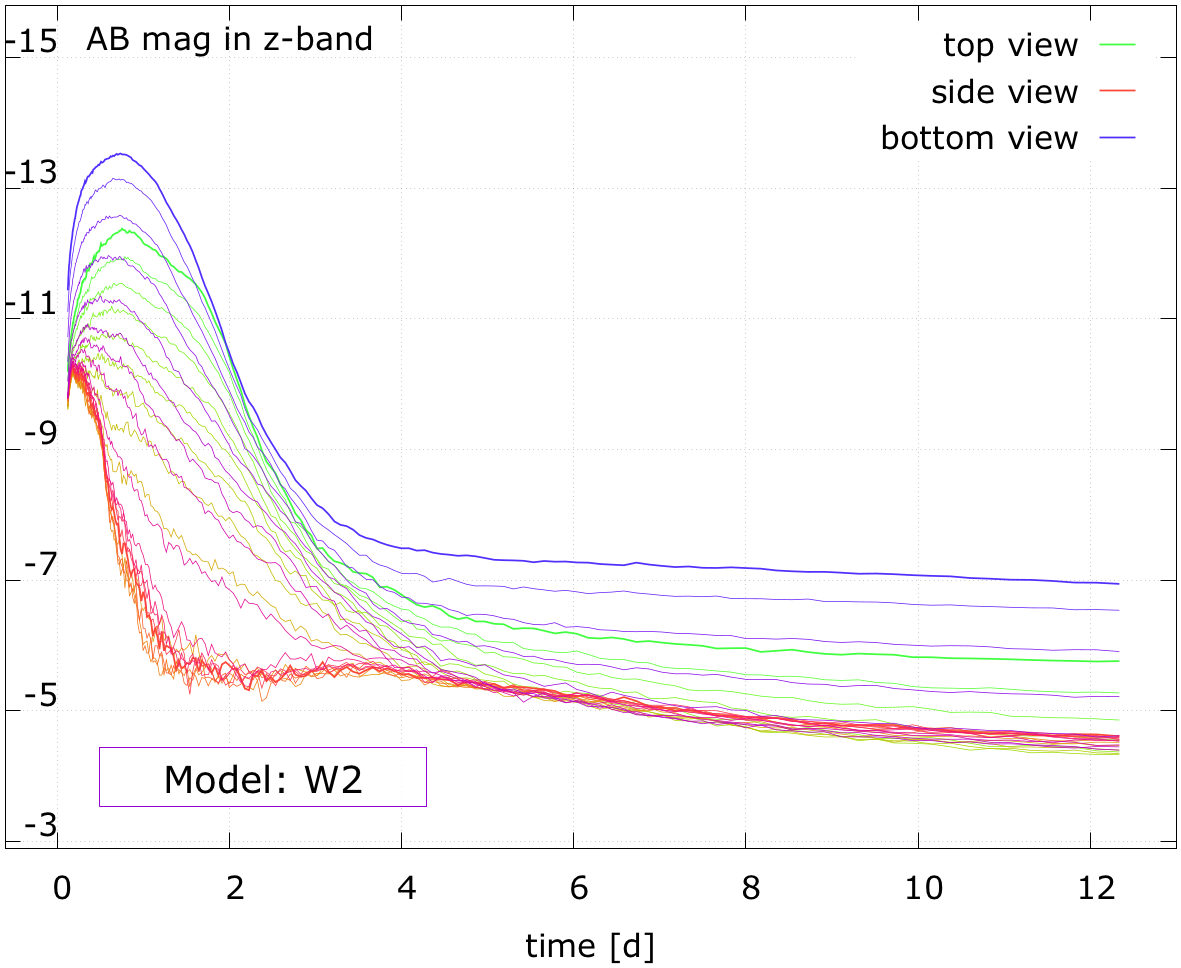}
\\
\includegraphics[height=0.24\textheight]{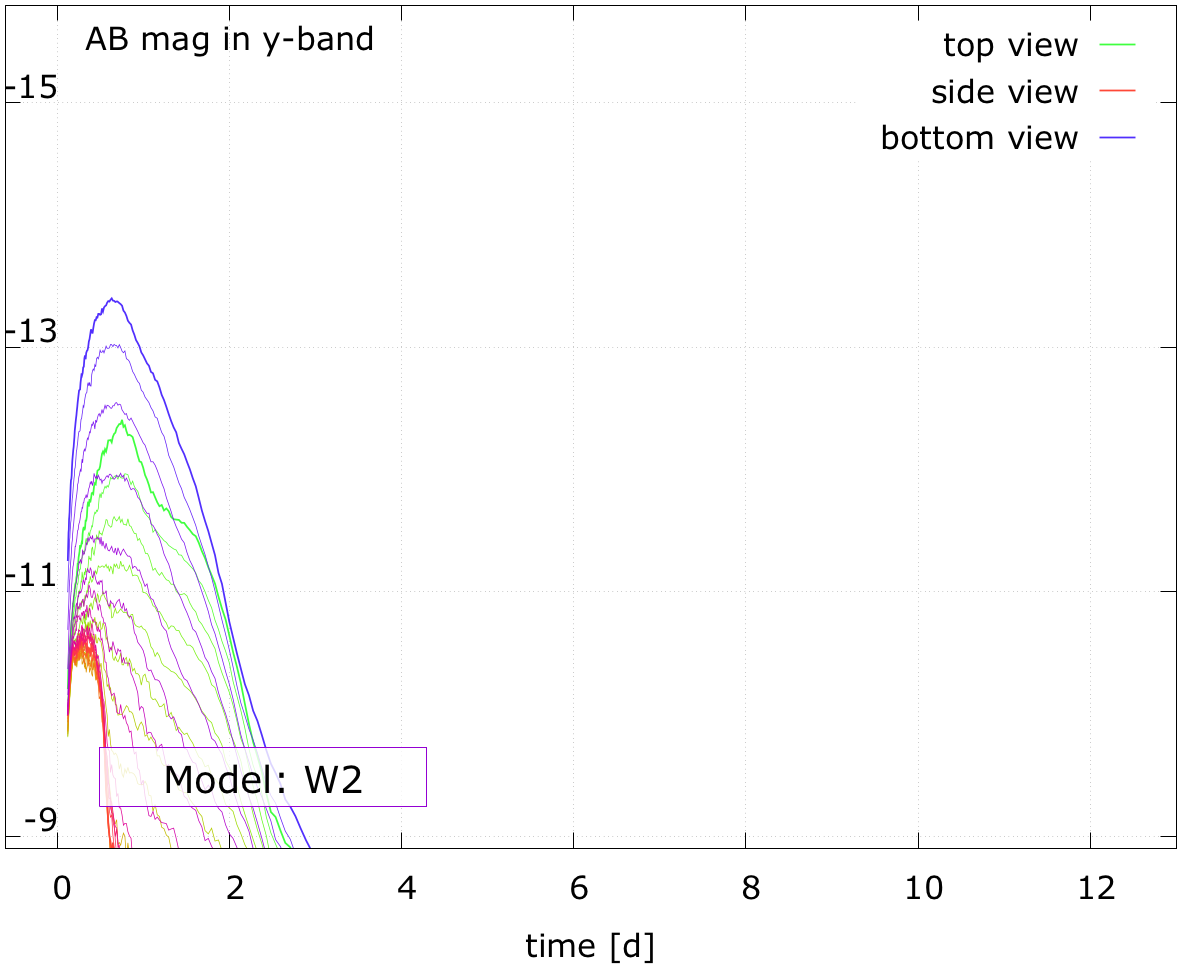} &
\includegraphics[height=0.24\textheight]{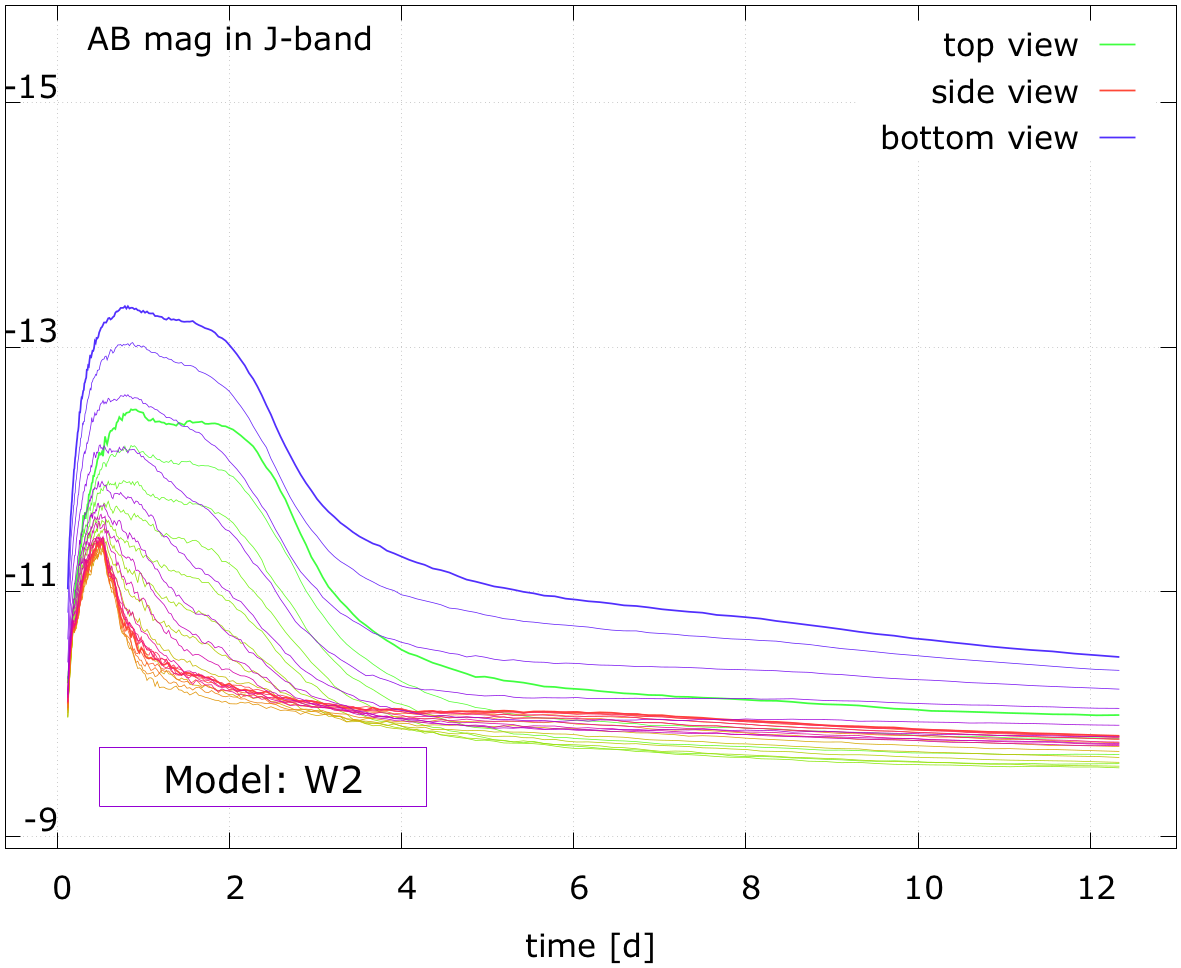}
\\
\includegraphics[height=0.24\textheight]{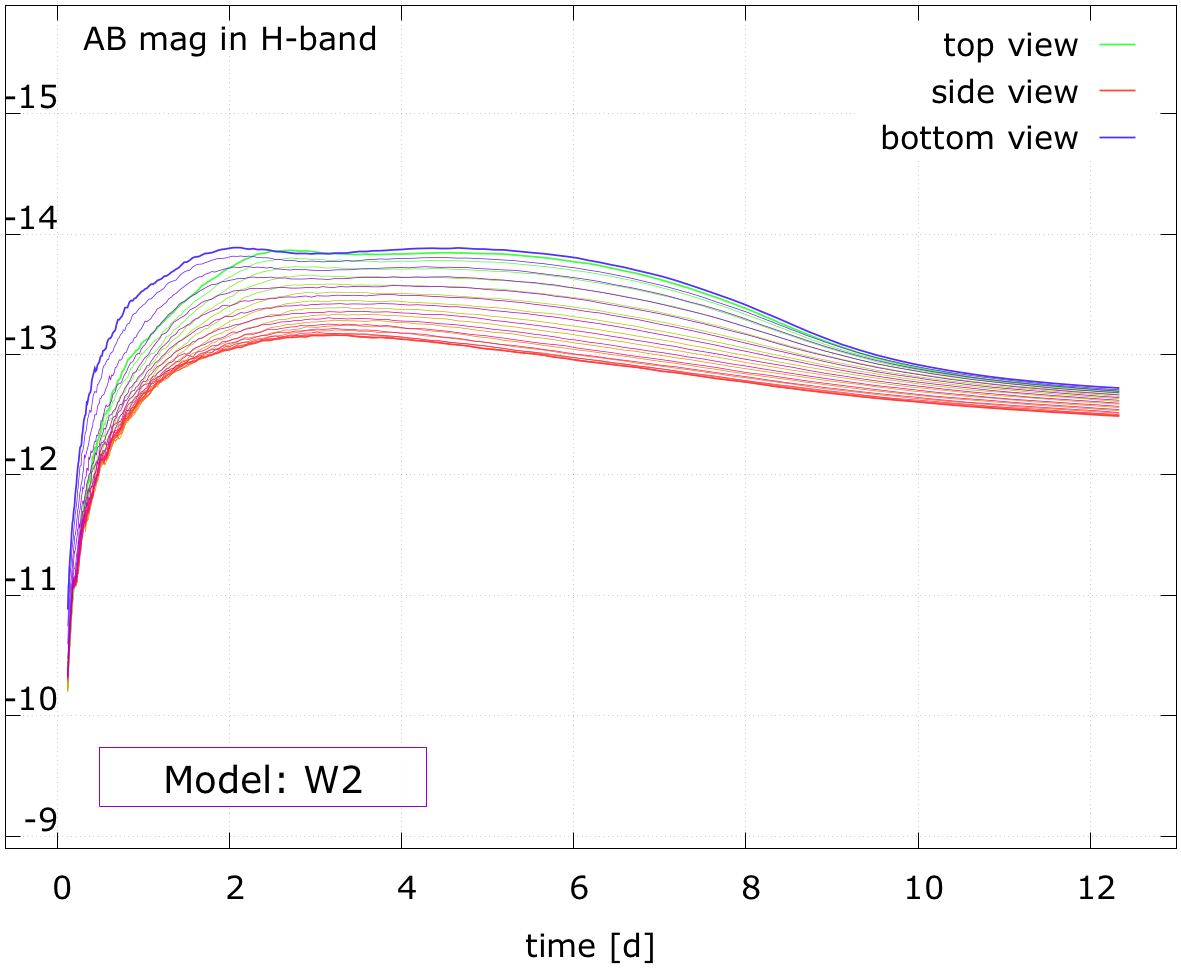} &
\includegraphics[height=0.24\textheight]{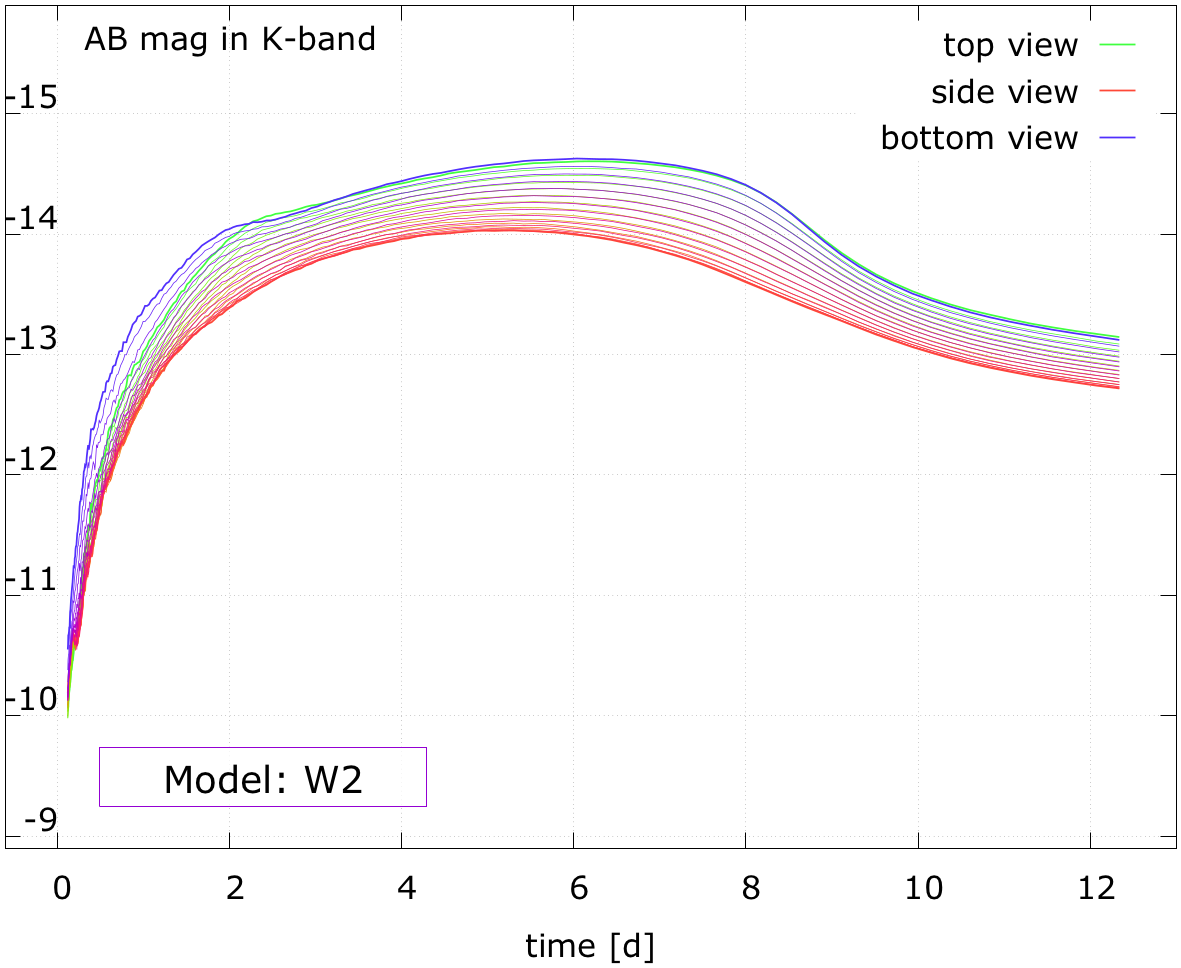}
\\
\end{tabular}
\caption{Broadband light curves for the optical $grizy$ and nIR $JHK$ bands
for model {\tt W2}. Color gradient is the same as in previous plot.}
\label{fig:mags_W2Zr}
\end{figure*}
This illustrates the presence of a blue transient, associated with the wind.
The transient is orientation-dependent, and clearly absent for "side"
orientations, showing that the wind is completely obscured for this view.
The on-axis configurations, on the other hand, display a peculiar double-peak
structure, with the first, early blue peak at $t\sim0.3^d$ generated by the
wind outflow, and the second nIR peak at $t\sim4^d$ generated by the dynamical
ejecta.

Table \ref{tab:models_2d} lists numerical values for the positions of both
peaks in all 2D models, for the "top" and "bottom" orientations of the
remnant. The "side" orientation is omitted, because it does not depend on the
wind and thus is the same as in dynamical ejecta-only models {\tt A2dSm}-{\tt
D2dSm}, considered in the previous section.  In more than half of the cases in
Table~\ref{tab:models_2d}, the blue peak is missing or substantially
suppressed in one orientation compared to the other.  This is due to the
irregular morphology of the ejecta around the axis and lack of symmetry with
respect to reflection in the equatorial plane. Configuration of the dynamical
ejecta density on the north and south poles of the remnant is different,
causing the wind transient to be partially or completely obscured.  In
general, this gives an idea of how sensitive the blue transient wind signal
is to the precise distribution of clumps in dynamical ejecta. Even though
the density of dynamical ejecta around the axis is rather small, it can still
interfere with the wind-generated transient.

Figure~\ref{fig:lums_W2} also displays the location of the blue peak (black
upward triangles) and its dependence on the wind mass and expansion velocity
(dashed thin lines).
Its behavior is qualitatively similar to expressions
(\ref{eq:tpeak_scaling}-\ref{eq:Lpeak_scaling}): more massive wind produces a
later and brighter transient, and faster wind produces an earlier transient.
Empirical fits for the peak times $t_{\rm p, wind}$ and luminosities
$L_{p, {\rm wind}}$ give the following formulae:
\begin{align}
  &t_{\rm p, wind} =
     0.32\ {\rm d}\;M_{{\rm w},2}^{0.14}\ v_{{\rm w},1}^{-0.83},
     \label{eq:tpeak_wind}
  \\
  &L_{\rm p, wind}=
     1.1\times10^{41}\ {\rm erg}\ {\rm s}^{-1}\ M_{{\rm w},2}^{0.63},
     \label{eq:Lpeak_wind}
\end{align}
where ${M_{{\rm w},2}:=M_{\rm wind}/10^{-2}M_{\odot}}$ and
      ${v_{{\rm w},1}:=\frac{v_{\rm wind, max}/2}{0.1\ c}}$ are
rescaled mass and expansion velocity parameters of the wind.
The power law indices in these expressions are different from the ones in
(\ref{eq:tpeak_scaling}-\ref{eq:Lpeak_scaling}) because of the presence of the
dynamical ejecta. In particular, the luminosity of the wind peak is almost
independent on the expansion velocity of the wind.

Models {\tt W2Se}-{\tt W2Cr} in Table~\ref{tab:models_2d} explore the effect
of light element composition on the blue transient. It turns out to be
surprisingly small: blue peak luminosities and peak epochs in
Table~\ref{tab:models_2d} are largely unaffected by which specific element
contributes to the wind. Fig.~\ref{fig:spectra_W2Zr} shows the time
evolution of angle-dependent spectra for our baseline combined model
{\tt W2}.

All models in this group show qualitatively similar spectral behaviour: the
spectra at early times show pronounced dependence on the remnant orientation.
The spectra in Fig.~\ref{fig:spectra_W2Zr} clearly resemble the spectrum of the
Zr models for on-axis orientations, and reduce to {\tt A2dSm}-like spectra for
side views. Several distinct spectral features of Zr can be clearly identified
in the early on-axis spectra.
At late times, the spectrum approaches that of
dynamical ejecta-only model {\tt A2dSm} for all orientations (which in turn is
close to the model {\tt SASm}).

Figure~\ref{fig:mags_W2Zr} shows corresponding angle-dependent AB
magnitudes in the optical $grizy$ and nIR $JHK$ bands. Optical magnitudes
reach values as high as $-13$ mag at $t\sim0.3^d$, while nIR magnitudes peak
on timescales of $t\sim4-6^d$ with peak magnitudes around $-14.5$.
This clearly demonstrates the double-peak nature of the transient, as well as
its dependence on orientation.

For references, Tables~\ref{tab:peak_mags_2d_top}, \ref{tab:peak_mags_2d_btm},
and \ref{tab:peak_mags_2d_side} contains detailed information on peak magnitudes,
epochs and transient durations for all optical and nIR bands, and for three
orientations: top, bottom and side respectively.

\section{Moving towards realistic models}
\label{sec:realistic_models}

\begin{figure*}
\begin{center}
 \begin{tabular}{cc}
 \includegraphics[width=0.49\textwidth]{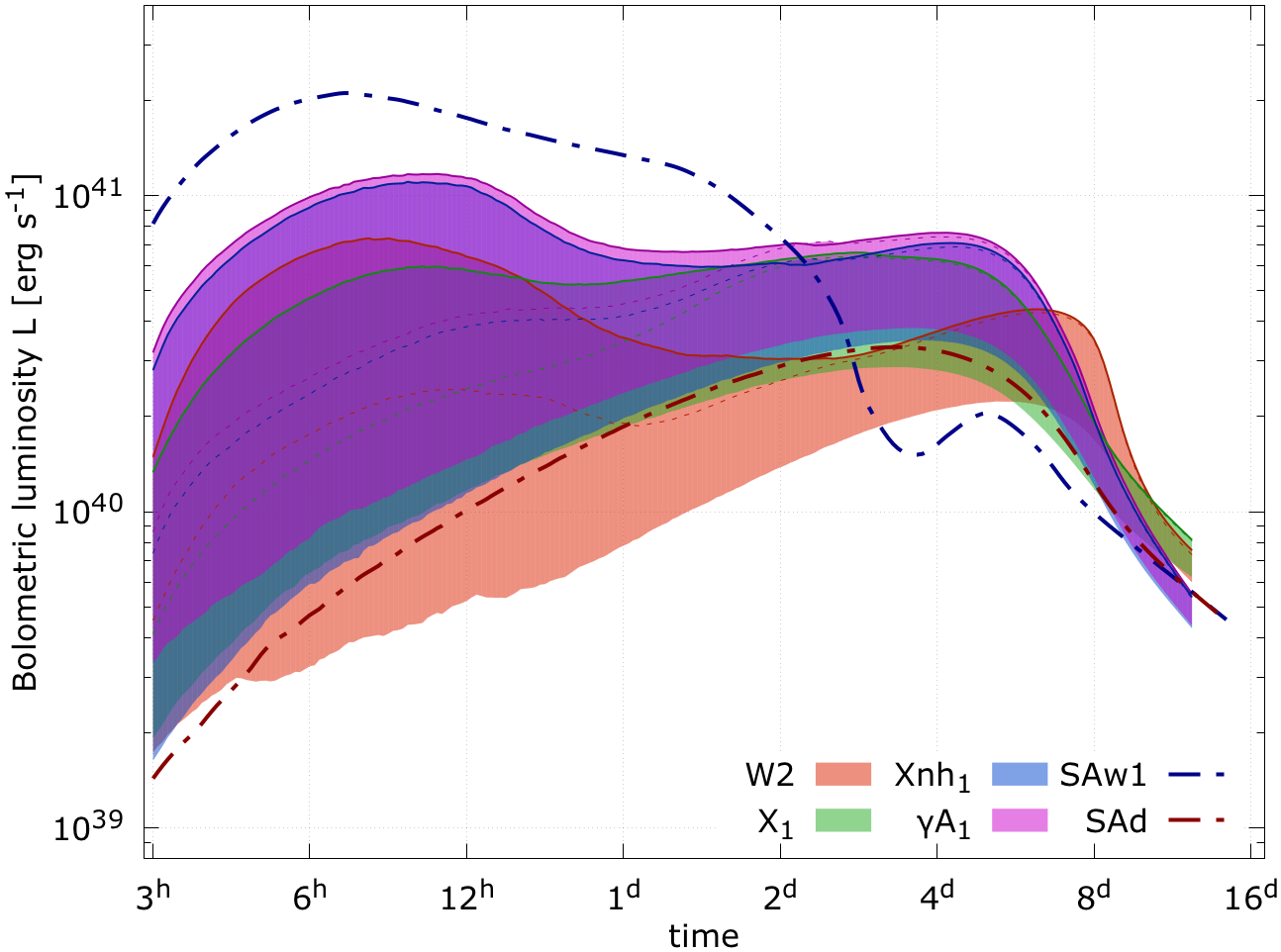}
 &
 \includegraphics[width=0.49\textwidth]{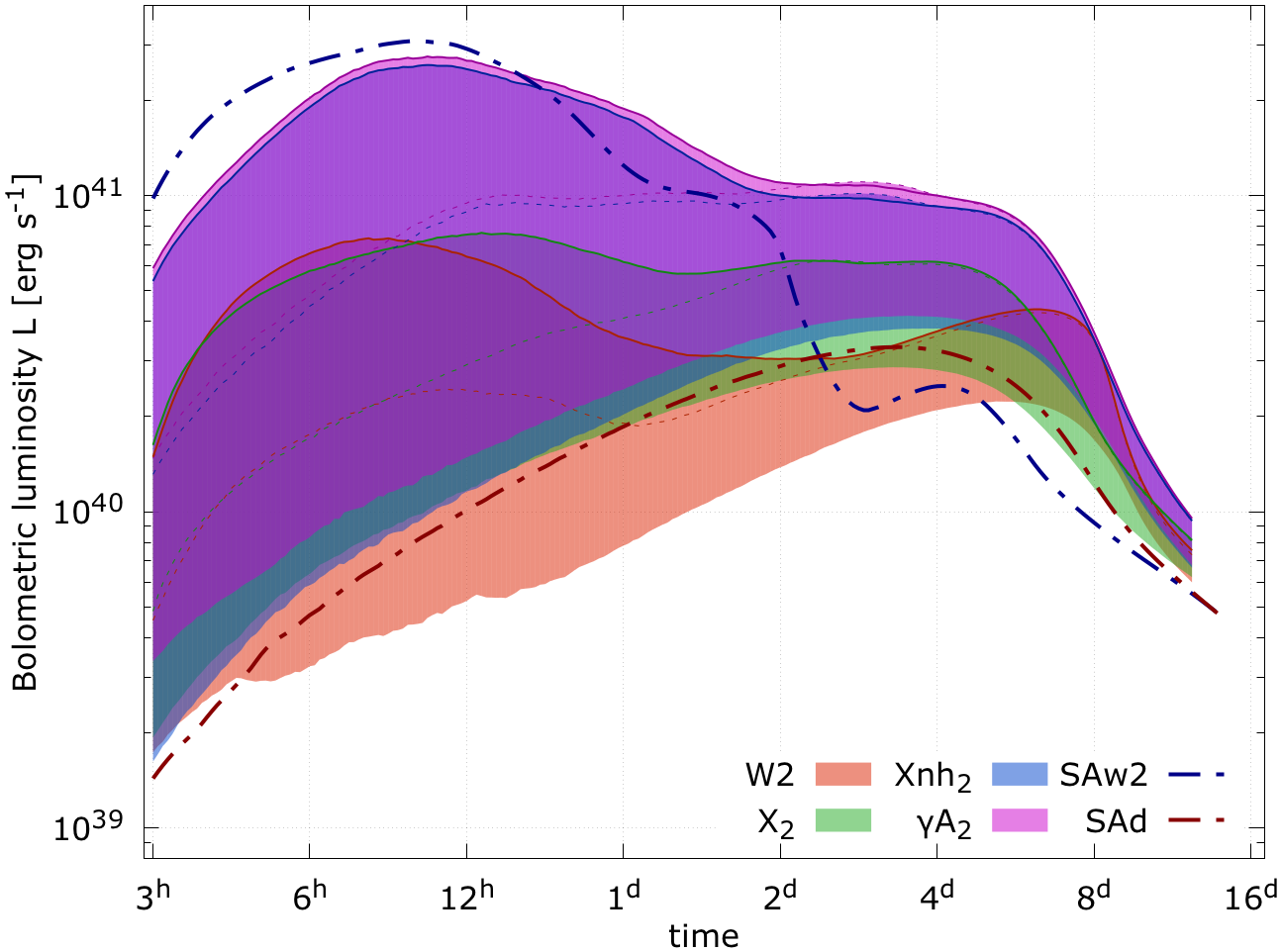}
 \end{tabular}
\end{center}
\caption{Viewing angle-dependent ranges of bolometric luminosities for the models
  with detailed heating and thermalization:
  {\tt Xnh}$_1$, $\gamma A_1$ (left), and {\tt Xnh}$_2$, $\gamma A_2$ (right).
  For comparison, also shown are models having the same composition but with
  analytic heating, {\tt X}$_1$ (left) and {\tt X}$_2$ (right), an axisymmetric
  Zr/Sm model {\tt W2}, and 1D models with dynamical ejecta mixture opacity
  {\tt SAd} and two types of wind mixed opacities {\tt SAw1} (left) and {\tt SAw2}
  (right).
  Solid lines on the upper edge of each range correspond to the "bottom"
  orientation of the remnant, and dashed lines of corresponding color indicate
  bolometric luminosities for the "top" orientation.
  }
\label{fig:lums_X12}
\end{figure*}

Previously, we approximated radioactive heating with a power law formula
(\ref{eq:analytic_nuclear_heating}), and used a constant thermalization
$\epsilon_{\rm th}=0.25$ to represent the fraction of this heating, which is
converted to thermal energy.  Here we lift these assumptions and exploit our
knowledge about composition and radioactive decays in the outflows to make our
models more realistic. By adding different ingredients one-by-one, we can
gauge their individual impact.

In models {\tt X}$_1$ and {\tt X}$_2$, we take model {\tt W2} and replace the
opacity of pure Sm in dynamical ejecta and Zr in the wind with the opacity
mixtures from Table~\ref{tab:mfs_repr}, similarly to the models {\tt SAd} and
{\tt SAw1}/{\tt SAw2} for two types of wind.
In {\tt Xnh}$_1$ and {\tt Xnh}$_2$ we replace analytic nuclear heating in {\tt
X}$_1$ and {\tt X}$_2$ with detailed nuclear heating generated by radioactive
decays, and add species-dependent thermalization, as explained in
Sect.~\ref{sec:composition}.

As pointed out in the end of Sect.~\ref{sec:composition}, much of the
nucleosynthesis occurs close to the neutron dripline where experimental
information is not available and one has to rely on theoretical models. The
available nuclear mass models agree overall reasonably well, but they make
different predictions for the amount of matter in the trans-lead region. Since
the corresponding nuclei undergo alpha-decay, their amount seriously impacts
the nuclear heating rate and "standard" mass formulae (e.g. FRDM vs DZ) can
differ by as much as an order of magnitude in the predicted heating rates.

To explore the impact of nuclear heating on macronovae, we introduce models
{\tt DZ}$_{1}$ and {\tt DZ}$_{2}$, which are identical to {\tt X}$_1$ and
{\tt X}$_2$ except for the heating rate, which is higher by a factor of 10 to
mimic the DZ heating rate.
These models produces substantially brighter transients, as can be seen in
Fig.~\ref{fig:GRB130603b} and the 2nd to last block of rows in
Tables~\ref{tab:models_2d}, \ref{tab:peak_mags_2d_top},~\ref{tab:peak_mags_2d_btm},
and~\ref{tab:peak_mags_2d_side}.

Our most sophisticated models, that we believe are best suited for
making claims about detectability, are the $\gamma$-models: $\gamma
A_1$-$\gamma D_1$ and $\gamma A_2$-$\gamma D_2$. Here we add grey $\gamma$-ray
transport sourced by the fraction of nuclear heating that is radiated in the
form of $\gamma$-radiation (see Sect.~\ref{sec:composition} for details).
The letters A-D in the model notation stand for the four different dynamical
ejecta morphologies used.
Shown in Fig.~\ref{fig:lums_X12}, the light curves for models {\tt Xnh}$_1$
and {\tt Xnh}$_2$ are nearly identical to those of the $\gamma A*$ models.
This indicates the ray-trace calculation of the $\gamma$-thermalization fractions
accurately estimate the energy depostion relative to the Monte Carlo model.
For this reason, the detection prospects of models {\tt Xnh}$_1$ and {\tt Xnh}$_2$
will be nearly identical to those of models $\gamma A_1$ and $\gamma A_2$,
respectively.

The second half of Table~\ref{tab:models_2d} contains parameters of these models
and their peak bolometric luminosities, for the blue and nIR peaks, for two
opposite on-axis remnant orientations, "top" and "bottom". Just as for the
{\tt W2*} models, visibility of the blue peak completely depends on the
dynamical ejecta configuration. In particular, models based on dynamical
ejecta configuration A ({\tt X*} and ${\gamma A*}$) feature the blue peak in
both orientations, models based on configurations B and D do not have any, whereas
${\gamma C_1}$ and ${\gamma C_2}$ have it in the "top" orientation but not in
the "bottom" one.

Figure \ref{fig:lums_X12} demonstrates bolometric luminosity ranges for the
models based on dynamical ejecta configuration A. The upper edge of each range
corresponds to the "bottom" orientation, the lower edge is the "side", and the
"top" orientation is shown with dashed lines. Thick dot-dashed blue and red lines
represent corresponding 1D analytic models {\tt SAw1}/{\tt SAw2} and {\tt
SAd}, which use wind or dynamical ejecta opacity mix. The light curves of the 2D
models lie roughly in between these two extreme cases: they are brighter than
the dynamical ejecta-only model {\tt SAd} and dimmer than the corresponding model
of the wind, {\tt SAw1} or {\tt SAw2}. Model {\tt W2} is also shown. It is based
on Sm, which has higher opacity than the dynamical ejecta mix that we use,
therefore the light curve of {\tt W2} is slightly dimmer and peaks later,
although the qualitative behaviour is the same.

While models {\tt X}$_1$ and {\tt X}$_2$ look very similar, once detailed
nuclear heating with thermalization is added, not only do they become
brighter, model {\tt Xnh}$_2$ significantly exceeds {\tt Xnh}$_1$ in peak
luminosity. This is simply a manifestation of the higher radioactive heating
rate for the "wind 2" than for the "wind 1" (shown in Fig.~\ref{fig:hrate_dyn},
bottom panel). Finally, adding grey $\gamma$-transport increases luminosity
only marginally.
This slight increase is due to the grey $\gamma$-transport accounting for
nonlocal deposition (from the fraction of energy that escapes each cell),
which is not accounted for in our implementation of the thermalization
fraction for $\gamma$-rays.

The evolution of spectra and light curves for the {\tt X*}-models and
$\gamma$-models is qualitatively similar to that of model {\tt W2}
(see Fig.~\ref{fig:spectra_W2Zr} and Fig.~\ref{fig:mags_W2Zr}).

For reference, Tables~\ref{tab:peak_mags_2d_top}, \ref{tab:peak_mags_2d_btm},
and \ref{tab:peak_mags_2d_side} contain detailed information on the
peak magnitudes, epochs and transient durations for all optical and nIR bands,
and for the three orientations: top, bottom and side respectively.

\section{Detection prospects}
\label{sec:detectability}

In this section, we will focus on our most sophisticated $\gamma$-models
(see Table~\ref{tab:allmodels}), and the most promising {\tt DZ}-models,
in which the nuclear heating rates from the FRDM mass model are --for the
dynamic ejecta-- multiplied by a factor of 10 to mimick the more optimistic
Duflo-Zuker heating rates (see Sect.~\ref{sec:realistic_models}).

We adopt methodology of estimating the number of potentially observable
macronovae similar to the one in \cite{rosswog17a}, by integrating the
expected NSM rate $\mathcal{R}_{\rm NSM}$ over the comoving volume in which
the macronova is observable. The only difference is that we also take into account
orientation with respect to a terrestrial observer, by weighing the integrand
with the probability of favorable orientation $P(z)$:
\begin{align}
 \mathcal{N}_{\rm MN}=\int \frac{d\mathcal{N}}{dz} dz
=\mathcal{R}_{\rm NSM}\int P(z) \frac{dV_c}{dz}
\frac{dz}{1+z},
\label{eq:nmn}
\end{align}
where $V_C$ is the comoving volume.
Following \cite{rosswog17a}, we calculate the limiting magnitudes for LSST and
VISTA surveys in $grizy$ and $JH$ bands, respectively, with two exposure
times: 60 and 180~seconds, using the same tools (ESO Infrared Exposure Time
Calculator for VISTA~\footnote{\url{https://www.eso.org/observing/etc/}} and a
Python exposure time calculator for
LSST~\footnote{\url{https://github.com/lsst-sims/exposure-time-calc}},
assuming a target signal-to-noise ratio of 5).
For the derivative $dV_C/dz$ of the comoving volume with respect to the
redshift, we adopt the flat cosmology parameters
$H_o=67\ {\rm km}\ {\rm s}^{-1}\ {\rm Mpc}^{-1}$,
$\Omega_m=0.307$ \citep{planck16m}.

\begin{figure*}
\begin{center}
 \begin{tabular}{cc}
 \includegraphics[width=0.49\textwidth]{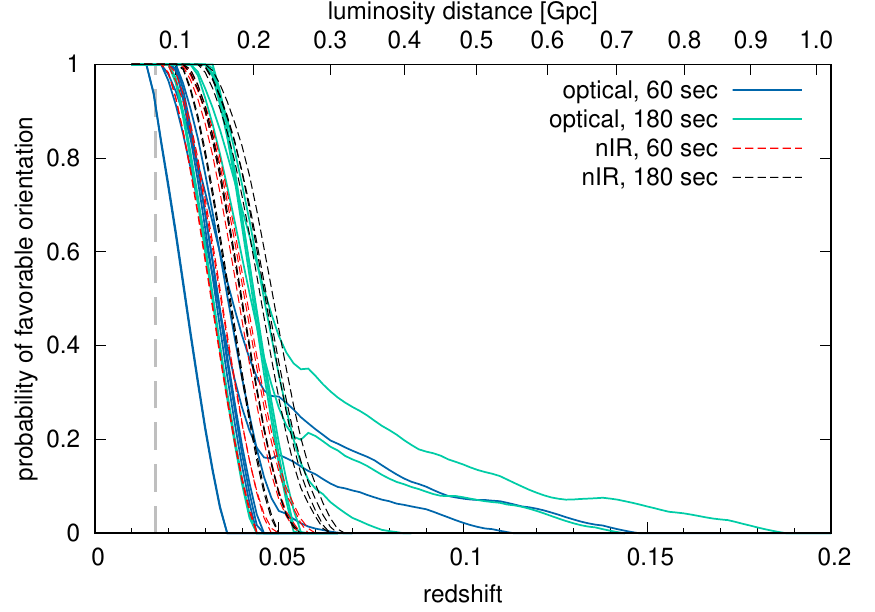}
 &
 \includegraphics[width=0.49\textwidth]{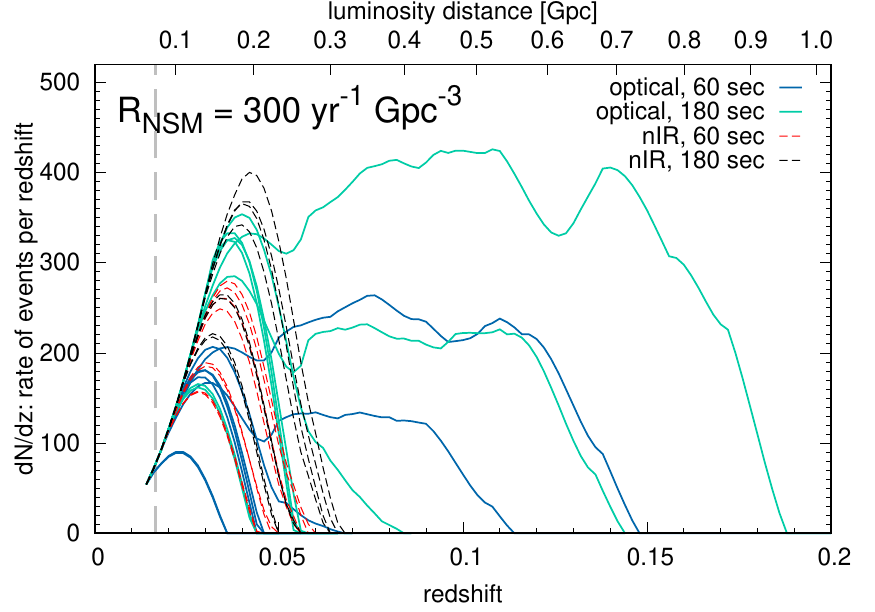} \\
 \includegraphics[width=0.49\textwidth]{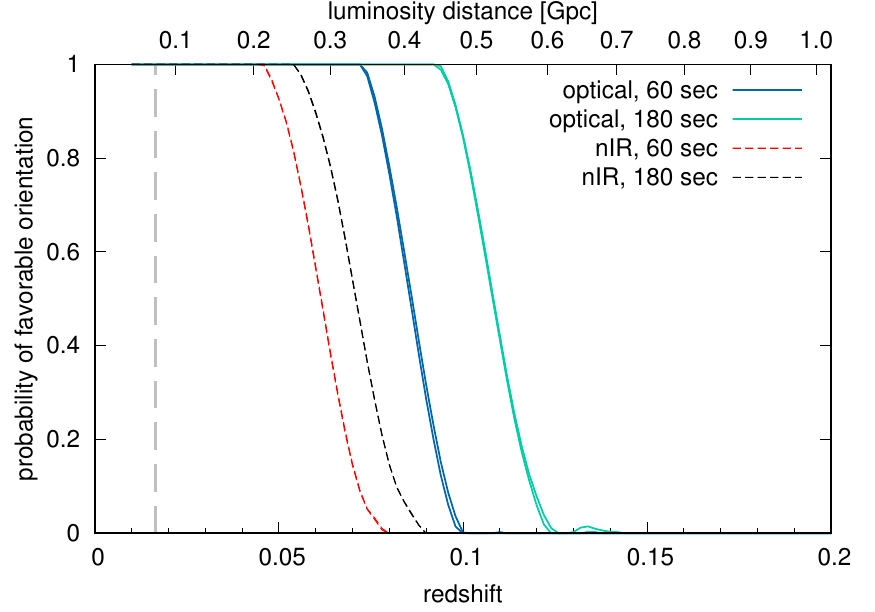}
 &
 \includegraphics[width=0.49\textwidth]{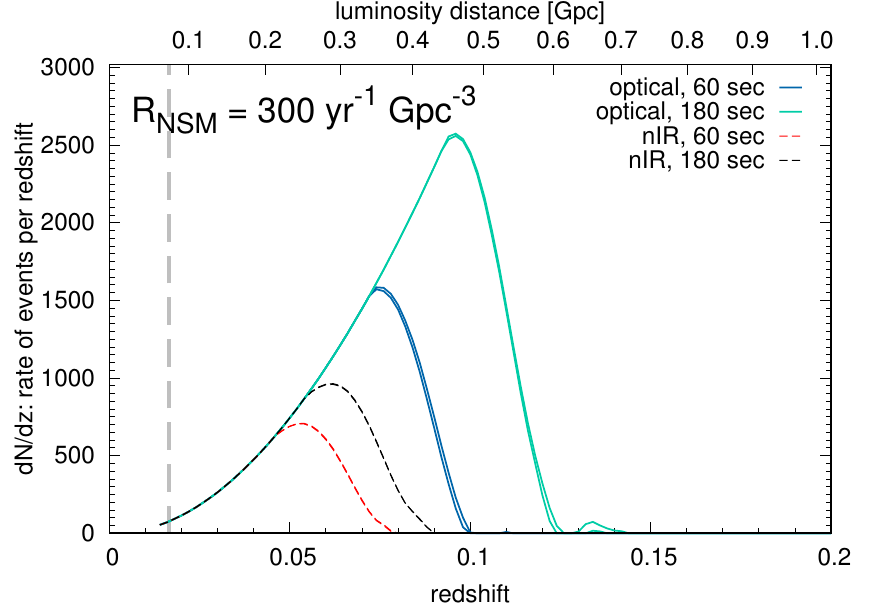}
 \end{tabular}
\end{center}
\caption{Left: probability of a macronova remnant being oriented such
that it can be observed at peak, in the optical or nIR bands, for exposure
times 60 and 180 seconds.
Right: rate of detectable events per redshift
$d\mathcal{N}/dz$, assuming constant volumetric NSM rate of $300\ {\rm
yr}^{-1}\ {\rm Gpc}^{-3}$, for optical or nIR bands, and for exposure times of
60 and 180 seconds.  The integral under each curve gives the total number of
detectable events.
Top row: our most realistic $\gamma$-models, bottom row: {\tt DZ}-models
with nuclear heating artificially enhanced by a factor of 10.
Among the $\gamma$-models in the top row, three models stand out: $\gamma
A_2$, $\gamma A_1$ and $\gamma C_2$. They have a long "tail" of nonzero
detection probability in the optical at $z>0.07$.
The grey dashed vertical line corresponds to the aLIGO
detection horizon for NSM events for the O1 observing run \citep[about 75 Mpc,
][]{martynov16}. }
\label{fig:dty_rates}
\end{figure*}

The NSM rate that we use to normalize our results \citep[same as in][]{rosswog17a}
is an "informed best guess" of
$\mathcal{R}_{\rm NSM}={300\ {\rm yr}^{-1}{\rm Gpc}^{-3}}$, or
${25.86\ {\rm Myr}^{-1}}$ per Milky Way size galaxy (assuming $11.6$ million
Milky Way-equivalent galaxies per ${\rm Gpc}^3$).
Note that since (\ref{eq:nmn}) depends linearly on $\mathcal{R}_{\rm NSM}$,
it is trivial to rescale our results to a different rate.
Our choice of this rate value is very conservatively above the expected upper
limit for the aLIGO future observation run O3 \citep{abbott16b}, to give us
room for discussing the prospects of follow-up observations after a future GW
trigger.
This rate value is also the median (in log scale) of the aLIGO compendium of
NSM rates \citep{abadie10}, as illustrated by Fig.~2 in \cite{rosswog17a}.
It is consistent with the revised NSM rate from the galactic binary pulsar
\citep[${7-49\ {\rm Myr}^{-1}}$,][]{kim15}, somewhat on the higher end of the
rates computed in \cite{deMink15} (${1.4-81}\ {\rm Myr}^{-1}$) and used in
\cite{belczynski16} ($5-15\ {\rm Myr}^{-1}$) for comparison with LIGO/Virgo
upper limits, ${0.06-77.4\ {\rm Myr}^{-1}}$ in \cite{dominik12} and
${0.01-80\ {\rm Myr}^{-1}}$ in \cite{fryer99}.
At the same time, these rates are on the lower side of the rates derived from
models of short GRBs: about ${8-96\ {\rm Myr}^{-1}}$ in \cite{fong12}, and about
${43-130\ {\rm Myr}^{-1}}$ in \cite{petrillo13}~\footnote{
But see \cite{guetta05}, where pre-SWIFT estimates of short GRBs statistics
were used to infer  rates as low as $\sim0.1\ {\rm Myr}^{-1}$}

We compute the probability of favorable orientation $P(z)$ as a ratio of the
number of angular bins for which the macronova magnitude is above detection
threshold, to the total number of angular bins. Fig.~\ref{fig:dty_rates}, left
column, shows $P(z)$ computed for each of the models at peak magnitude
using exposure thresholds for 60 and 180 seconds. Since the behaviour of the
macronova in the optical and nIR bands is different, we estimate optical and nIR
detection probabilities separately.
Optical detection probability is taken as the maximum probability over $grizy$
bands, and nIR detection probability is the maximum over the $JK$ bands.

As can be seen from Fig.~\ref{fig:dty_rates}, $P(z)$ for all our
$\gamma$-models in the infrared looks rather similar: it is $100\%$ up to about
$100\ {\rm Mpc}$, then steeply drops to zero, such that none of the models is
detectable at redshifts higher than $\sim0.07$.
The steep decrease in probability for the two {\tt DZ} models is qualitatively
similar, except they are detectable from farther out, about $300\ {\rm Mpc}$.
This behavior is due to the fact that even though the macronovae are brighter
in nIR than in optical, detection thresholds are also much higher.
Since macronovae light curves in nIR vary little with orientation
-- only within a factor of $\sim2-3$, corresponding to about 1~mag -- the
cutoff to zero is very steep.

In optical, $P(z)$ behaves rather differently, because the flux in optical
bands varies by several orders of magnitude depending on the orientation.
Three models in Fig.~\ref{fig:dty_rates} (top left) stand out as detectable to much
higher redshifts: with exposure of 180 seconds, $\gamma A_2$ is visible up to
$z=0.2$, $\gamma A_1$ up to $z=0.15$, and $\gamma C_2$ -- up to 0.08.
Visibility horizon for exposure of 60 seconds is not much smaller: $z=0.15$
for $\gamma A_2$ and $z=0.11$ for $\gamma A_1$.
For the two {\tt DZ} models with enhanced heating rates (bottom left), $P(z)$
is shifted to higher redshifts, because these models produce much brighter
optical transients.

The right column in Fig.~\ref{fig:dty_rates} displays the differential
quantity $d\mathcal{N}/dz$ from the first integral in (\ref{eq:nmn}), which
describes the rates of detectable macronovae per redshift. The integral under
each of the curves gives estimates of total number of detections, for an
assumed rate $\mathcal{R}_{\rm NSM}$. This plot illustrates that even though
the probability of favorable orientation at higher redshifts is small,
the majority of detections will be at higher redshifts due to rapidly increasing
comoving volume element.

The grey dashed vertical line in Fig.~\ref{fig:dty_rates} indicates the aLIGO
detection horizon for NSM events \citep[75 Mpc, ][]{martynov16}, computed for
the first aLIGO run O1. At such distance, all our models are above detection
threshold, both in optical and in nIR, independent of orientation.
For subsequent runs the detection horizon is pushed to $200\ {\rm Mpc}$, where
only a fraction of macronovae is observable. For more distant aLIGO horizon,
say $400\ {\rm Mpc}$, the majority of NSM GW signals will be unobservable in either
nIR or optical bands, at least in surveys with 60/180-second exposure
times.

\begin{table}
\centering
\caption{Total number of observable events across the whole sky, in nIR or
optical bands, with exposure times 60 and 180 s, sampled at peak $t_p$, and at
1~day and 2~days after the merger. Scaled to the NSM rate of
${300\ {\rm yr}^{-1}{\rm Gpc}^{-3}}$ ($\approx{25.86\ {\rm Myr}^{-1}}$
per Milky Way size galaxy).
High-latitude angles where afterglow may be detectable, are not excluded.}
\resizebox{\columnwidth}{!}{
\begin{tabular}{l|cc|ccc|ccc}
\hline\hline
      & \multicolumn{2}{|c}{nIR (VISTA)}
      & \multicolumn{3}{|c}{optical (LSST), 60 s}
      & \multicolumn{3}{|c}{optical (LSST), 180 s}\\
Model        & 60 s& 180 s &$t_p$ & 1d  & 2d   &$t_p$ & 1d  & 2d  \\
\hline
$\gamma A_1$ & 3.8 &  5.8  & 12.4 & 5.5 & 2.1  & 27.1 & 12.0 &  4.5 \\
$\gamma A_2$ & 4.9 &  7.5  & 27.8 &25.2 & 8.2  & 59.2 & 54.6 & 17.8 \\
$\gamma B_1$ & 3.8 &  5.9  &  1.8 & 0.9 & 0.1  &  3.9 &  1.8 &  0.5 \\
$\gamma B_2$ & 4.9 &  7.5  &  1.9 & 1.0 & 0.4  &  4.0 &  2.0 &  0.9 \\
$\gamma C_1$ & 7.1 & 10.9  &  4.5 & 3.9 & 1.5  &  9.5 &  8.4 &  3.1 \\
$\gamma C_2$ & 8.1 & 12.2  &  6.1 & 5.4 & 3.0  & 12.8 & 11.6 &  6.4 \\
$\gamma D_1$ & 7.5 & 11.7  &  4.2 & 3.6 & 1.2  &  9.0 &  7.5 &  2.5 \\
$\gamma D_2$ & 8.4 & 13.4  &  4.4 & 3.8 & 1.6  &  9.3 &  8.0 &  3.1 \\
\hline
{\tt DZ}$_1$ &24.8 & 38.2  & 65.3 &57.5 &49.4  &130.1 &118.6 & 98.2 \\
{\tt DZ}$_2$ &25.0 & 38.1  & 66.7 &60.6 &49.5  &131.9 &119.7 & 97.7 \\
\hline
\end{tabular}
}
\label{tab:num_detections}
\end{table}

Another factor which complicates the observability of macronovae is their
short duration in optical bands. In nIR, this is not such a big problem,
because peak times and durations are the order of a week.
Fig.~\ref{fig:dty_rates} shows only the most optimistic probabilities for
detections at peak times. If an observation is made several days after the GW
trigger, the macronova can be a few mags dimmer.
Fig.~\ref{fig:mags_W2Zr}, four top panels, shows a drop by 4 mags in $r$, $i$
and $z$ bands for model ${\tt W2}$ in its favorable on-axis orientation.

Table~\ref{tab:num_detections} displays the expected number of potentially
observable events for each model, either in nIR (for VISTA) or optical (for
LSST), depending on the exposure time and the epoch. The "$t_p$" columns for
optical observations correspond to the peak epochs, and "1d" and "2d"
correspond to the observation epochs of one and two days after the initial
trigger, respectively.  We can see from the table that observing at one (two)
day(s) after the trigger decimates the number of events by a factor of about
two (four).  One should also take into account that these numbers are for the
entire sky, but only a small patch of it will be accessible to generic surveys
with high enough cadence.

We could approach the LSST detection rates from another angle, where we
start from the supernova detection rate estimates \citep[e.g.][]{lien09,LSST09}.
Specifically, for core-collapse supernovae, LSST is estimated to detect as
many as $3.43\times10^5$ events per year \citep[][\ Table 2]{lien09}.
Our peaks are typically 3-5 magnitudes lower than core collapse supernovae,
which translates into rates of 0.4-6.3 detections per year, if we assume
that NSMs are $\sim1000$ times more rare and that all of them have a blue
transient. However, we also need to consider that only one detection epoch
will be possible with the LSST observing strategy. Thus, it would seem that
the identification of macronovae requires follow-up observations with other
telescopes. In this case, their reaction times can be as crucial as for the GW
triggers~\cite{abbott16a} due to the short duration of the blue transients.

One concern for the prospects of macronova detections is that it can be "drowned" in the
afterglow from a GRB. However, firstly, not all NSMs do necessarily produce GRBs,
because the wind from hypermassive neutron star can create a baryon-polluted
cloud around the engine, dense enough to prevent a GRB jet from getting
out~\citep{murguiaberthier14,just16} or reaching ultrarelativistic speeds.
Secondly, afterglows are only visible from very narrow opening angles,
\citep[$<10^\circ$,][]{fong12}, while optical macronovae for our models are
visible from much wider angles. In fact, in our numerical setting, where we
cover $4\pi$ with 27 axisymmetric angular bins equally spaced in
$\cos{\theta}$ (see Sect.~\ref{sec:radiative_transfer}), the $10^\circ$ solid
angle only covers $\sim1/5$ of one polar bin. Thus, our results for
optical detection are not affected by the invisibility due to the afterglow.

\begin{figure}
\begin{center}
 \includegraphics[width=0.49\textwidth]{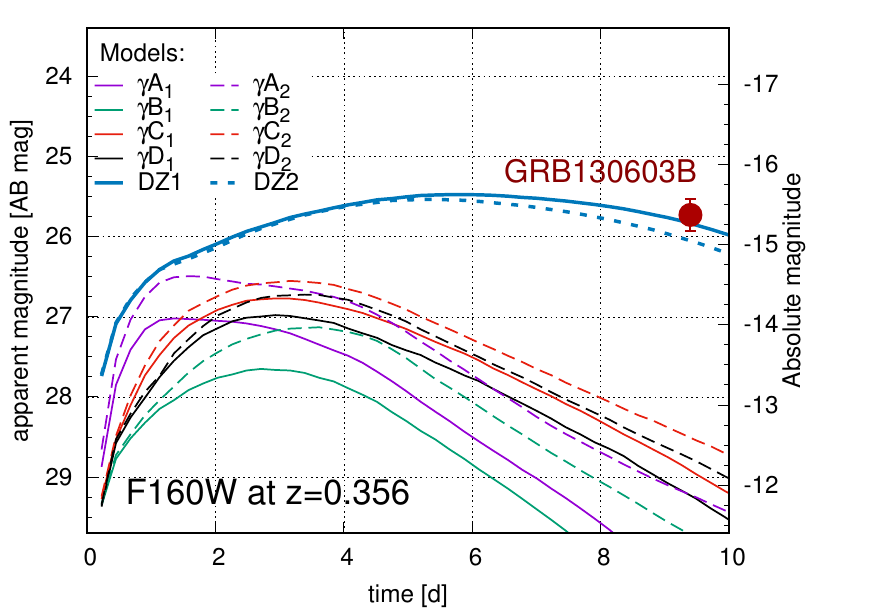}
\end{center}
\caption{Synthetic light curves from our models, calculated with the F160W
filter at redshift $z=0.356$, corresponding to the distance to GRB130603B. The
red circle with error bars indicates the measured nIR excess, interpreted as
a kilonova \citep{tanvir13}. }
\label{fig:GRB130603b}
\end{figure}

Can our models explain the observed nIR excess in GRB130603B, which was
reported in \cite{tanvir13} and \cite{deUgartePostigo13}
\citep[also][]{berger13a}?
Fig. \ref{fig:GRB130603b} shows the light curves in the F160W band for our most
sophisticated $\gamma$-models, and the two {\tt DZ}-models with artificially
enhanced nuclear heating rates, compared with the detection data point.
In all $\gamma$-models, the FRDM nuclear mass model is used, while {\tt DZ}
models are intended to mimick the Duflo-Zucker DZ31 model
\citep{wu16,rosswog17a}. Nuclear heating rates in the latter are higher than
in the former by almost one order of magnitude. While DZ31-based models
seem to have no problems reaching the observed brightness, none of the
$\gamma$-models is even close to the detection.

Moreover, some experimentation shows that no reasonable increase in the
ejecta mass or velocity, or wind mass or velocity, can possibly make FRDM-based
light curves agree with the GRB130603B observation. At the same time, DZ31-based
models explains it with ease. Thus, among other factors, the dominant impact of nuclear
heating rate is established. This is already hinted by the scaling expression
for bolometric luminosity (\ref{eq:Lpeak_scaling}): while other parameters
such as ejecta mass, velocity or even opacity enter this expression with
powers less than one, nuclear heating is directly proportional to the
luminosity. We conclude that future reliable macronova observations will
constrain nuclear heating rates in the first place, and so indirectly help
discriminating between nuclear mass models.

\section{Discussion and conclusion}
\label{sec:discussion}

We apply the multi-dimensional, multigroup Monte Carlo code
\SuperNu\ \citep{wollaeger14a} and detailed opacities from the LANL suite of atomic
physics codes~\citep{fontes15b} to simulate radiative transfer for a series of
spherically-symmetric and axisymmetric macronova models and produce light curves
and spectra. To demonstrate the accuracy and consistency of our code, we develop a
new analytic solution for macronovae (see Appendix \ref{app:analytic}) with
uniform homologously expanding background flow and grey opacity treatment.
This solution is then simulated with \SuperNu\ in spherical symmetry with full
multi-group opacity treatment (see Sect.~\ref{sec:radiative_transfer}).
For the nuclear heating, which supplies radiative power of the macronova,
we use \WinNET\ nuclear network~\citep{winteler12a,winteler12b} to calculate the
$r$-process nucleosynthesis and partitioning of released energy between
different decay products.

We systematically explore a series of models with varying level of detail in the
morphology, composition, and opacity to understand and disentangle various
factors affecting macronova light curves and spectra.
We consider two types of outflows: dynamical ejecta and winds, and combine
them in our models (see Fig.~\ref{fig:superimposed}).
Morphology is taken from 1D-spherically or 2D-axisymmetrically
averaged ejecta from sophisticated 3D SPH Newtonian simulations of neutron star
mergers~\citep[NSMs,][]{rosswog14a}. We also develop a new 1D analytic
spherically-symmetric hydrodynamic solution to model homologous expansion of the
ejecta (Sect.~\ref{sec:selfsim}).

For the UV/optical/IR absorption opacity, we explore a range of options:
effective grey, detailed for single elements, or detailed for multiple
elements with partial density weighted mixing.
Our final, most sophisticated models combine 2D dynamical ejecta from NSM
simulations with a spherical analytic wind solution, mixed composition opacity,
detailed $r$-process heating from nucleosynthesis network, partitioning of the
heating rates between decay products, and individual thermalization of different
decay products \citep[following the approach of][]{barnes16a}.
As in the work of~\cite{fontes17a}, we use a novel approach in which we depart
from the traditional Sobolev treatment of opacity during radiative transfer, and
instead use alternative multigroup with ``smeared'' lines.

For each simulation, Table~\ref{tab:allmodels} gives the relevant section and
model parameters.
Tables~\ref{tab:sa_models},~\ref{tab:dyn_models},~\ref{tab:models_2d}, give
model parameters and peak bolometric luminosity (including time of peak) for
the models with semi-analytic ejecta, averaged SPH ejecta, and dynamical
ejecta combined with wind, respectively.
Absolute peak magnitudes, peak times, and macronova durations
(defined as times to decrease one mag after peak) are provided
for $grizyJHK$ broadband filters in Table~\ref{tab:peak_mags_1d}
for all 1D models, and in Tables~\ref{tab:peak_mags_2d_top},~\ref{tab:peak_mags_2d_btm},
and~\ref{tab:peak_mags_2d_side} for ``top'', ``bottom'', and ``side'' views,
respectively, of each 2D model.
These tables show that opacity has a substantial impact on the luminosity,
with lanthanides and actinides contributing to broader, redder light curves
relative to the other elements tested.

We use the semi-analytic, grey opacity models to calculate power-law fits
of peak bolometric luminosity, peak epoch, and peak broadband
luminosity with respect to: ejecta mass, median velocity, and opacity.
These relationships are given in Sect.~\ref{sec:sa_grey}.
The scaling relations from our grey opacity models can be used for models
with detailed opacity in cases when the spectrum is close to Planckian.
For the set of semi-analytic ejecta models, we compare light curves
from simulations with grey opacity to light curves from simulations with detailed
opacity (see Fig.~\ref{fig:lum_SACrPdSm}).
These comparisons indicate that lanthanide opacities, representative of dynamical
ejecta, give peak \emph{bolometric} luminosities consistent with an effective grey
opacity of $\sim10$~$\cmg$, while lighter element wind-type opacities are consistent
with an effective grey opacity closer to $\sim1$~$\cmg$, justifying
the values that had been used in earlier work
\citep{kasen13,tanaka13,grossman14}.
However, in the optical and J bands, the effective grey lanthanide opacity
significantly overestimates macronova emission (see
Fig.~\ref{fig:spectra_A2dSm}), while for the H and K bands it gives a reasonable
approximation.

We summarize our main findings below, with sections divided according to
what features we examine in this study.

\subsection{Effect of geometry}

The impact of varying ejecta morphology on light curves and spectra, discussed in
Sect.~\ref{sec:dynamical}, is largely consistent with our expectations.
For the same compositions, the 1D semi-analytic models (presented in
Sect.~\ref{sec:sa_grey} and~\ref{sec:sa_multi}) produce comparable light curves to the
1D spherically symmetric averages of the SPH ejecta.
Moreover, these 1D results fall in between the brightest (top) and dimmest (side)
views of the equivalent 2D axisymmetric models.
In particular, the broadband data in Table~\ref{tab:peak_mags_1d} and the bolometric
luminosities shown in Fig.~\ref{fig:lum_SASmABCD} of models {\tt SASm}, {\tt A1dSm}
and {\tt B1dSm}, show similar transients.
The similarity arises from the comparable density profiles, mass, and velocity
of the ejecta (see  Fig.~\ref{fig:SKR_models_2d}).
{\tt SASm} has the analytic ansatz, and {\tt A1dSm} and {\tt B1dSm} have 1D
spherically averaged model A and model B SPH ejecta, respectively, as described
in Sect.~\ref{sec:exp_models}.
Also in Fig.~\ref{fig:lum_SASmABCD} is a plot of the light curve for the 2D
model {\tt A2dSm} at different viewing angles, showing {\tt SASm} falling between
the light curves in the top (or bottom) and side views.

Even without a wind component superimposed, the 2D axisymmetric models produce
brighter and bluer transients at top and bottom (or axial) views.
The brightness in the top and bottom views is a geometric effect, since a larger
projected area of the ejecta photosphere is visible relative to the side views.
We find that the transients appear bluer from the top and bottom views because the optical
depths from points on the equatorial (merger) plane to the surface are
generally lower.
Photons can reach the void above or below the merger plane more easily.
Consequently, early emission at higher temperature is suppressed less, relative
to side views.
Generally, morphologies that are non-spherical permit views with bluer and
brighter transients, relative to spherical morphologies.
For our models, the brightness varies by a factor of 2-3 between side and
top viewing angles, consistent with previous work
\citep[see, for instance,][]{grossman14}.

\subsection{Effect of composition}

In Section~\ref{sec:sa_multi}, we test the effect of opacity for pure
elements and mixtures of elements on the light curves and spectra, for a
fixed spherically-symmetric ejecta morphology.
We find that the brightness and timescale of the transients strongly depends
on the atomic electron configurations available to each element in the
macronova density-temperature regimes.
Specifically, for the set of elements we examine, we find the broadband
light curves can be categorized into three distinct types: bright blue transients
peaking in a few hours, intermediate red transients with double peaks over
1-2 days (in the $izy$ bands), and late nIR transients spanning a week
(in the $HK$ bands).
The blue, red, and nIR transients correspond to opacity from elements with
open $p$-shells, $d$-shells, and $f$-shells, respectively.
Figures~\ref{fig:spec_SACeNdU} and~\ref{fig:spec_SACrPdSm} show the spectra
at different times for each model discussed in Sect.~\ref{sec:sa_multi}.
The spectra from models with mixed composition are dominated by the features
from elements that have the most complex electronic configurations of their outer
atomic shell, because these tend to contribute the most opacity (as in
the findings of~\cite{kasen13}).
These broadband light curve features may be useful in characterizing the
composition of dominant sources of opacity in macronova ejecta.
However, the detectability of a transient from the wind relies substantially
on the composition; a time scale of several hours is not easily amenable to
detection in either follow-up or blind surveys~\citep{grossman14}.

The spectra of dynamical ejecta at late times most closely resemble those
of the $f$-shell elements --lanthanides and actinides-- and feature a peculiar
"spectral cliff", where emission is very strongly suppressed past certain
wavelength (for instance, about $1.6$ microns for Sm, right in the middle of the
H band). Compared to grey opacity models, this leads to much dimmer transients
in the optical and J bands, while light curves in the HK bands retain
comparable brightness. This can serve as a justification for using simple
models for estimating the brightness in the HK bands.

\subsection{Effect of ``lanthanide curtain''}

In Sections~\ref{sec:combined} and~\ref{sec:realistic_models}, we combine the
axisymmetric dynamical ejecta from NSM models A-D with various wind model ejecta
derived from the 1D analytic solution (presented in Sect.~\ref{sec:exp_models};
see Fig.~\ref{fig:superimposed}).
The combined models assess the visibility of potential blue transients for our
various types wind and dynamical ejecta.
These models are listed in Table~\ref{tab:models_2d}.

Figure~\ref{fig:lums_W2} shows the presence of a blue transient for several
of the combined models that employ model A dynamical ejecta.
The appearance of the blue transient is orientation-dependent (shown by the
shaded regions for the 2D models in Fig.~\ref{fig:lums_W2}).
This result is exhibited by the 2D combined mixed composition models as
well, seen in Fig.~\ref{fig:lums_X12}.

In more than half of the models in Table~\ref{tab:models_2d}, the irregular
morphology of the ejecta completely or substantially obscures the blue
transient in one on-axis view, relative to the other.
The dynamical ejecta is not completely symmetric when reflected through
the equatorial (merger) plane.
The sensitivity of the wind transient to small differences in dynamical
ejecta show the impact of lanthanide opacity in these regions.
This sensitivity propagates to the detection prospects, shown in
Fig.~\ref{fig:dty_rates}.
This ``lanthanide curtain'' has been found in other studies as well
\citep{barnes13,kasen15a}.

\subsection{Effect of nuclear heating rate}

Of the set of macronova properties we explore, the nuclear heating rate
has the largest impact on luminosity.
As noted in Sect.~\ref{sec:detectability}, this is implied by the
the exponents of ejecta mass, velocity, and opacity in the power-law
scaling relations discussed in Sect.~\ref{sec:sa_grey}, which are less
than one.
Luminosity scales directly proportional to the heating rate, which is
implied by the result in Appendix~\ref{app:analytic}.
Our {\tt DZ}-model light curves, when compared with the models that
use FRDM, indicate that reliable macronova observations will
constrain the nuclear heating rates and therefore nuclear physics far
from stability.
As a corollary, the thermalization efficiencies of the different
heating products should have a significant impact on brightness as well,
as originally found in the work of~\cite{barnes16a}.
However, the uncertainty in thermalization efficiency is subdominant
compared to the uncertainty in the nuclear heating rate due to the unknown
nuclear mass model \citep[also shown in][]{barnes16a}.

\subsection{Detection Prospects}

In Section~\ref{sec:detectability}, we discuss the detection prospects for
our most detailed models: $\gamma A_1$,  $\gamma A_2$,  $\gamma B_1$,
$\gamma B_2$, $\gamma C_1$, $\gamma C_2$, $\gamma D_1$, and $\gamma D_2$.
These models employ detailed r-process heating rates from \WinNET, and have
a grey multidimensional Monte Carlo energy deposition model for the $\gamma$-rays.
The models all have assumed the FRDM nuclear mass model for the r-process.
Consequently, in this section we also assess the detection prospects for models
{\tt DZ}$_{1}$ and {\tt DZ}$_{2}$, which apply the analytic power-law heating,
Eq.~\ref{eq:analytic_nuclear_heating}, but multiplied by a factor of 10 in the
dynamical ejecta.
The increase of the dynamical ejecta heating rate in the {\tt DZ}-models
substantially brightens the light curve in the nIR bands.
In particular, for the F160W filter, the light curves for the {\tt DZ}-models
come much closer to the GRB130603B data point than all other models (see
Fig.~\ref{fig:GRB130603b}).
Models with nuclear heating rates similar to what is delivered by the
DZ-mass models are consistent with the transient observed in the context of
GRB130603B being a macronova.

In Fig.~\ref{fig:dty_rates} we plot detection probabilities and
detection rates per redshift for two exposure times from the LSST
($grizy$ bands) and VISTA ($JK$ bands) surveys.
These values were calculated for each model, assuming an NSM rate
of 300 yr$^{-1}$Gpc$^{-3}$, perfect telescope coverage across the
whole sky, and that the model represents all macronovae.
The probability of detection is 100~\% up to about 100~Mpc for
$\gamma$-models, and up to 200~Mpc for {\tt DZ}-models, making them
detectable for all events within the LIGO horizon, both in the optical and in
the JK bands. A follow-up search is therefore possible, with the infrared
bands looking more promising, both due to the longer duration of the transients,
and low sensitivity to the orientation.

Only three of the $\gamma$-models, which have blue transients that
are not fully suppressed by lanthanide curtaining, are detectable
at $z>0.07$.
The detection probability decreases at higher redshift until only the
view {close to} the merger axis permits detection of the blue transient.
The non-monotonicity of the detection rate per redshift shows the
competing effects of increasing NSM sample volume while decreasing
apparent magnitude.
In the span of redshift where the models are visible in all orientations,
the rate of detections per redshift steadily increases.
However, once the redshift is sufficiently high, apparent magnitudes
are too dim, and the detection rates drop off unless there is a
sufficiently bright blue transient from the wind.

For the {\tt DZ}-models, the brighter transients increase the
detection prospects substantially past $z=0.07$.
Since the {\tt DZ}-models only increase the heating in the dynamical
ejecta, these models do not exhibit the same level of anisotropy
for the blue transient and hence do not have prominent tails in
the probability for optical detections.

In Table~\ref{tab:num_detections}, we have integrated the detection
rate per redshift with respect to redshift to get total (ideal)
detection rates for each model and each exposure.
Consequently, the assumed NSM rate of 300 yr$^{-1}$Gpc$^{-3}$ applies
to these numbers as well (they can be multiplicatively rescaled to
a different NSM rate).
Generally, for the $\gamma$-models, we find O(1-10) detections are
possible per year, assuming total coverage of the sky at all times.
For the {\tt DZ} models, we find O(10-100) possible detections,
again under the same assumptions.
The difference in the ideal detection rate between the
$\gamma$-models and the {\tt DZ}-models is consistent with the
difference in the heating rates for the models.

\subsection{Comparison with other macronova studies}

It is difficult to make precise comparisons with the existing
macronova literature, given the differences in ejecta morphologies,
r-process heating models, and assumed compositions.

First studies of macronovae \citep[e.g.][]{li98,roberts11a} esimated ejecta
opacities to be similar to the opacity of nickel ($0.2\ \cmg$) which was
proven to be overly optimistic~\citep{kasen13}. Such high opacities led to
the bolometric luminosities in the range of
$\sim 10^{42}-10^{44}\ \ergs$ (Fig.~2 of~\cite{li98}),
on par with supernovae and much brighter than all of our models.
\cite{roberts11a} performed full radiative transfer on multidimensional
ejecta, but applied a constant grey opacity of 0.1 cm$^2$/g.
Consequently, their peak bolometric luminosities are $\sim10^{42}\ \ergs$
as well
\footnote{If we apply equation (\ref{eq:Lpeak_scaling}) to rescale our
  1D peak bolometric luminosity to a grey opacity of 0.1 cm$^2$/g, the
  result is $\sim4.4\times10^{41}$.}.
The work of~\cite{barnes13},~\cite{tanaka13},~\cite{kasen13}, and
\cite{grossman14}, use opacity that should be more representative of
r-process ejecta.
These studies report peak bolometric luminosities in the range of
$\sim 10^{40}-10^{41}$ erg/s.
The recent work of~\cite{barnes16a} on thermalization fractions further
dims the transient, to a few times $10^{40}$ erg/s for their fiducial
model ($M_{\rm ej} = 5\times10^{-3}$ M$_{\odot}$, $v_{\rm ej} = 0.2c$).
In general, for emission from the dynamical ejecta, the bolometric
luminosities of our models are a few times $10^{40}$ erg/s.
Recently,~\cite{fontes15a} and~\cite{fontes17a} applied a line-smeared
multigroup approach in LTE light curve calculations, which is the method
we apply for opacity in this work (albeit, with a different treatment
of relativistic transformations).
Our luminosities in the mid-IR range are $\sim10^{40}$ erg/s for
dynamical ejecta without wind, consistent with the findings of
\cite{fontes15a,fontes17a}.
In comparison with \cite{rosswog17a}, our models are similar in
absolute brightness but differ by having much shorter durations in the
optical $grizy$ bands (as can be seen in Fig.~\ref{fig:mags_W2Zr})
making them much harder to detect for LSST.
This is due to the differences between the wind mass adopted in our models;
otherwise, the grey opacity of $\varkappa=1\ \cmg$ gives a reasonable
agreement with our multigroup study (see Sect.~\ref{sec:sa_multi}).
Moreover, \cite{rosswog17a} reports absolute brightness of $-15..-16$ in
the $K$ band, which is similar to the values obtained in this study (see
Tables~\ref{tab:peak_mags_2d_top},~\ref{tab:peak_mags_2d_btm},~\ref{tab:peak_mags_2d_side}).

\subsection{Caveats and Future work}

Our results have a number of approximations, both in the underlying
numerical methods, and in the problem configurations.
For the radiative transfer, \SuperNu\ assumes LTE, which limits
the reliability of the light curves and spectra in the late stages
of the expansion of the ejecta.
We also do not treat lines directly, but instead apply a multigroup
approach that we justify in Sect.~\ref{sec:radiative_transfer}.
For the opacity, we weight contributions from pure elements by their
partial density in the mixed compositions.
A more accurate approach would be to solve the Saha-Boltzmann equations
(for LTE) for each species, coupled through the free electron field.
This would give more accurate ion population densities for the subsequent
opacity calculation.

The explored matter configurations are based on essentially Newtonian SPH
simulations. Fully relativistic simulations, especially when coupled to a soft
nuclear matter equation of state that enhance the likelyhood of shocks, may
therefore lead to different matter configurations and possibly larger electron
fractions. While many of the quantities that are determined by the
'astrophysical engine' at work enter with powers smaller than unity into the
observables (such as peak times and luminosities, see Sect.~\ref{sec:sa_grey}),
the nuclear heating rate impacts the luminosity linearly. It is determined by
nuclear physics far from stability which may be decisive for whether
macronovae are detectable at interesting rates or not.

\section{Acknowledgments}
\label{app:ack}

Work at LANL was done under the auspices of the National Nuclear Security
Administration of the U.S. Department of Energy at Los Alamos National
Laboratory under Contract No. DE-AC52-06NA25396. All LANL calculations were
performed on LANL Institutional Computing resources.
SR has been supported by the Swedish Research Council (VR) under grant number
2016-03657\_3, by the Swedish National Space Board under grant number Dnr.
107/16 and by the research environment grant "Gravitational Radiation and
Electromagnetic Astrophysical Transients (GREAT)" funded by the Swedish
Research council (VR) under Dnr 2016-06012.  Some of the simulations for this
paper were performed on the facilities of the North-German Supercomputing
Alliance (HLRN).
We thank Mansi Kasliwal for useful input on detection prospects.

\appendix
\section{Analytic macronova solution}
\label{app:analytic}

The analytic solutions obtained for Fig.~\ref{fig:supernu_vs_analytic} follow the
prescription of~\cite{pinto00}.
For clarity, we outline the derivation here.
The semi-relativistic radiation diffusion equation is
\begin{equation}
  \label{eq1}
  \frac{D E}{D t}
  -\nabla\cdot\left(\frac{c}{3\kappa\rho}\nabla E\right)
  +\frac{4}{3}E\nabla\cdot\vec{v} = \rho\dot{\epsilon} \;\;,
\end{equation}
where $t$ is time, $\nabla$ is the gradient or divergence operator
($\nabla\cdot$) with respect to spatial coordinate $\vec{r}$, $E$
is comoving radiation energy density, $\vec{v}$ is velocity, $c$ is
the speed of light, $\kappa$ is a constant absorption opacity,
$\rho$ is gas density, and $\dot{\epsilon}$ is the radioactive heating
rate per unit mass.
In Eq.~\eqref{eq1}, it has been assumed that the thermal absorption and
emission rates cancel.
Restricting to 1D spherical geometry, the supporting equations are
\begin{subequations}
  \label{eq2}
  \begin{align}
    & \vec{v} = \frac{\vec{r}}{t} \;\;, \\
    & |\vec{v}| = v = v_{\max}x \;\;, \\
    & \rho = \rho_{0}\left(\frac{t_{0}}{t}\right)^{3} \;\;, \\
    & E = E_{0}\left(\frac{t_{0}}{t}\right)^{4}\psi(x)\phi(t) \;\;, \\
    & T = (E/a)^{1/4} \;\;, \\
    & \dot{\epsilon} = \epsilon_{0}t^{-\alpha} \;\;,
  \end{align}
\end{subequations}
where $v_{\max}$ is the maximum outflow speed, $x$ is a non-dimensional
radial coordinate, $t_{0}$ is an initial time, $\rho_{0}$ is density
at $t_{0}$, and $\psi(x)$ and $\phi(t)$ are the spatial and temporal profiles
of the radiation energy density.
Also, $T$ is gas or radiation temperature ($a$ is the radiation constant),
and $\alpha$ is a constant taken to be 1.3 for r-process heating.
In Eq.~\eqref{eq2}, it is assumed the outflow is homologous and the
radiation energy density solution is amenable to separation of variables.
Using Eq.~\eqref{eq2} to evaluate each term on the left side of
Eq.~\eqref{eq1},
\begin{subequations}
  \label{eq3}
  \begin{align}
    & \frac{D E}{D t} = E_{0}\left(\frac{t_{0}}{t}\right)^{4}
    \psi(x)\left(\phi'(t)-\frac{4}{t}\phi(t)\right) \;\;,\\
    & -\nabla\cdot\left(\frac{c}{3\kappa\rho}\nabla E\right) =
    -\frac{1}{(v_{\max}t)^{2}}\left(\frac{t_{0}}{t}\right)\phi(t)
    \left(\frac{cE_{0}}{3\kappa\rho_{0}}\right)\frac{1}{x^{2}}
    \left(x^{2}\psi'(x)\right)' \;\;,\\
    & \frac{4}{3}E\nabla\cdot\vec{v} = \frac{4}{t}
    E_{0}\left(\frac{t_{0}}{t}\right)^{4}\psi(x)\phi(t) \;\;.
  \end{align}
\end{subequations}
Summing Eqs.~\eqref{eq3} and cancelling $(t_{0}/t)^{3}$,
\begin{equation}
  \label{eq4}
  E_{0}\left(\frac{t_{0}}{t}\right)\psi(x)\phi'(t)
  -\frac{1}{(v_{\max}t_{0})^{2}}\phi(t)
  \left(\frac{cE_{0}}{3\kappa\rho_{0}}\right)\frac{1}{x^{2}}
  \left(x^{2}\psi'(x)\right)'
  = \rho_{0}\epsilon_{0}t^{-\alpha} \;\;.
\end{equation}

The homogeneous form of Eq.~\eqref{eq4} is solved first, allowing
separation of variables,
\begin{subequations}
  \label{eq5}
  \begin{align}
    & \frac{1}{x^{2}\psi(x)}
    \left(x^{2}\psi'(x)\right)'
    = -\lambda \;\;, \label{eq5a}\\
    & \tau_{0}\left(\frac{t_{0}}{t}\right)\frac{\phi'(t)}{\phi(t)}
    = -\lambda \;\;,
  \end{align}
\end{subequations}
where $\lambda$ is the separation constant and
\begin{equation}
  \label{eq6}
  \tau_{0} = \frac{3\kappa\rho_{0}}{c}(v_{\max}t_{0})^{2} \;\;.
\end{equation}
The boundary conditions of Eqs.~\eqref{eq5} are
\begin{subequations}
  \label{eq7}
  \begin{align}
    &\psi(0) = 0 \;\;, \\
    &\psi(x_{0}) = 0 \;\;, \\
    &\phi(t_{0}) = 1 \;\;, \\
    &\phi(\infty) = 0 \;\;.
  \end{align}
\end{subequations}
The solutions to Eqs.~\eqref{eq5} and~\eqref{eq7} are
\citep{pinto00}
\begin{subequations}
  \label{eq8}
  \begin{align}
    &\psi(x) = \frac{\sin(\sqrt{\lambda}x/x_{0})}{x} \;\;, \\
    &\phi_{h}(t) = \exp(-\lambda t^{2}/2\tau_{0}t_{0}) \;\;.
  \end{align}
\end{subequations}
For optically thick outflow, the radiative-zero boundary
condition can reasonably be set as $x_{0}=1$, $\psi(1)=0$
\citep{pinto00}.
Following~\cite{pinto00}, braket notation will be used for
non-dimensional spatial integrals of products of functions:
\begin{equation}
  \label{eq9}
  \braket{f|g} = \int_{0}^{1}f(x)\,g(x)\,x^{2}\,dx \;\;.
\end{equation}
Requiring
\begin{equation}
  \label{eq10}
  \braket{\psi|\psi} = 1 \;\;,
\end{equation}
the set of spatial eigenfunctions satisfying the boundary
conditions are
\begin{equation}
  \label{eq11}
  \psi_{m}(x) = \sqrt{2}\,\frac{\sin(m\pi x)}{x} \;\;.
\end{equation}
Multiplying Eq.~\eqref{eq4} by $\psi_{m}(x)x^{2}$ and integrating
over $x\in[0,1]$ yields
\begin{equation}
  \label{eq12}
  \phi_{m}'(t) + \left(\frac{t}{t_{0}\tau_{0}}\right)
  \left(m^{2}\pi^{2}\right)\phi_{m}(t)
  = \frac{\rho_{0}}{E_{0}t_{0}}\epsilon_{0}t^{1-\alpha}\braket{1|\psi_{m}} \;\;.
\end{equation}
It is convenient to rescale the time variables as well:
\begin{subequations}
  \label{eq13}
  \begin{align}
    &t\rightarrow t_{0}t \;\;,\\
    &\tau_{0}\rightarrow t_{0}\tau_{0} \;\;,\\
    &\frac{\rho_{0}}{E_{0}}\epsilon_{0}t_{0}^{1-\alpha}\rightarrow\epsilon_{0}
    \;\;.
  \end{align}
\end{subequations}
Substituting Eqs.~\eqref{eq13} into Eq.~\eqref{eq12} yields
\begin{equation}
  \label{eq14}
  \phi_{m}'(t) + \left(\frac{t}{\tau_{0}}\right)
  \left(m^{2}\pi^{2}\right)\phi_{m}(t)
  = \sqrt{2}\epsilon_{0}\frac{(-1)^{m+1}}{m\pi}t^{1-\alpha} \;\;,
\end{equation}
which provides an inhomogeneous temporal eigenfunction.
The bolometric luminosity solution is~\citep{pinto00},
\begin{equation}
  \label{eq15}
  L(t) = -\frac{4\pi cU_{\max}t_{0}E_{0}}{3\kappa\rho_{0}}
  \sum_{m=1}^{\infty}\phi_{m}(t)(x^{2}\psi_{m}'(x))|_{x=1} \;\;.
\end{equation}
Incorporating Eqs.~\eqref{eq5a} and~\eqref{eq11} into
Eq.~\eqref{eq15},
\begin{equation}
  \label{eq16}
  L(t) = \frac{4\pi cU_{\max}t_{0}E_{0}}{3\kappa\rho_{0}}
  \sum_{m=1}^{\infty}\phi_{m}(t)\lambda\braket{1|\psi_{m}} \;\;,
\end{equation}
or
\begin{equation}
  \label{eq17}
  L(t) = \frac{4\pi cU_{\max}t_{0}E_{0}}{3\kappa\rho_{0}}
  \sqrt{2}\sum_{m=1}^{\infty}(-1)^{m+1}m\pi\phi_{m}(t) \;\;.
\end{equation}
For the test examined in Sect.~\ref{sec:opacities}, we
find the solution is converged at $m\sim 500$.

\section{Macronovae: peak magnitudes, epochs and durations}
\label{app:tables}

\begin{table*}
\scriptsize
\caption{Properties of light curves for spherically symmetric (1D) models in LSST $grizy$ and VISTA $JHK$ bands.}
\begin{tabular}{l|cccccccc|cccccccc}
\hline\hline
      & \multicolumn{8}{|c}{Peak magnitude, $m$}
      & \multicolumn{8}{|c}{Peak epoch $t_p$ [d] and duration $\Delta t_{\rm 1 mag}$ [d]  }
\\
Model &$g$ &$r$ &$i$ &$z$ &$y$ &$J$ &$H$ &$K$
      &$g$ &$r$ &$i$ &$z$ &$y$ &$J$ &$H$ &$K$
\\
\hline
SAm1  & -10.8 &-11.3 &-11.6 &-11.8 &-11.9 &-12.1 &-12.2 &-12.3 &   0.25/0.65 & 0.31/0.85 & 0.37/1.05 & 0.43/1.22 & 0.52/1.37 & 0.67/1.74 & 0.9/2.2  & 1.2/2.8  \\
SAm2  & -11.9 &-12.4 &-12.7 &-12.9 &-13.1 &-13.3 &-13.5 &-13.6 &   0.33/1.05 & 0.47/1.49 & 0.60/1.94 & 0.73/2.35 & 0.94/2.71 & 1.25/3.63 & 1.8/4.9  & 2.5/6.3  \\
SAm3  & -12.9 &-13.4 &-13.7 &-13.9 &-14.1 &-14.3 &-14.6 &-14.8 &   0.51/1.63 & 0.73/2.45 & 1.00/3.34 & 1.32/4.17 & 1.42/4.93 & 2.14/7.04 & 3.6/10.2 & 5.0/13.7 \\
SAv1  & -11.4 &-11.9 &-12.2 &-12.3 &-12.5 &-12.7 &-12.9 &-13.0 &   0.55/1.86 & 0.87/2.73 & 1.16/3.62 & 1.41/4.41 & 1.60/5.12 & 2.44/6.95 & 3.5/9.5  & 4.7/12.1 \\
SAv2  & -11.9 &-12.4 &-12.7 &-12.9 &-13.0 &-13.3 &-13.5 &-13.6 &   0.41/1.27 & 0.56/1.82 & 0.75/2.41 & 0.91/2.93 & 1.09/3.39 & 1.55/4.61 & 2.3/6.3  & 3.1/8.1  \\
SAv3  & -12.3 &-12.8 &-13.1 &-13.3 &-13.4 &-13.6 &-13.8 &-13.9 &   0.33/1.01 & 0.44/1.45 & 0.61/1.90 & 0.69/2.30 & 0.87/2.65 & 1.25/3.59 & 1.8/4.9  & 2.4/6.3  \\
SAk0  & -13.8 &-14.1 &-14.2 &-14.3 &-14.3 &-14.3 &-14.3 &-14.2 &   0.48/1.33 & 0.63/1.74 & 0.79/2.09 & 0.95/2.37 & 1.03/2.62 & 1.32/3.23 & 1.7/4.2  & 2.1/5.6  \\
SAk1  & -12.1 &-12.6 &-12.9 &-13.1 &-13.2 &-13.5 &-13.7 &-13.8 &   0.36/1.12 & 0.51/1.61 & 0.65/2.11 & 0.79/2.56 & 0.98/2.97 & 1.37/4.02 & 2.0/5.5  & 2.7/7.1  \\
SAk2  &  -9.6 &-10.4 &-10.9 &-11.2 &-11.5 &-11.9 &-12.3 &-12.6 &   0.26/0.72 & 0.33/1.02 & 0.41/1.39 & 0.52/1.74 & 0.62/2.09 & 0.89/3.17 & 1.6/5.0  & 2.2/7.4  \\
SAk3  &  -6.3 & -7.5 & -8.3 & -8.8 & -9.1 & -9.7 &-10.3 &-10.8 &   0.17/0.44 & 0.18/0.61 & 0.38/0.82 & 0.36/0.94 & 0.42/1.15 & 0.59/1.78 & 0.7/3.0  & 1.3/4.8  \\
\hline
SASe  & -15.0 &-14.9 &-14.6 &-14.4 &-14.2 &-13.8 &-13.2 &-12.7 &   0.55/1.41 & 0.49/1.11 & 0.48/0.92 & 0.49/0.94 & 0.48/0.92 & 0.49/0.89 & 0.5/0.9  & 0.5/0.9  \\
SABr  & -15.0 &-14.9 &-14.7 &-14.6 &-14.6 &-14.1 &-13.5 &-13.0 &   0.51/1.06 & 0.51/1.24 & 0.54/1.60 & 0.59/1.97 & 0.59/1.77 & 0.58/2.39 & 0.6/3.1  & 0.6/3.4  \\
SATe  & -14.9 &-15.0 &-14.8 &-14.8 &-14.6 &-14.0 &-13.4 &-12.8 &   0.35/1.15 & 0.57/1.26 & 0.64/1.40 & 0.66/1.49 & 0.67/1.88 & 0.67/2.94 & 0.7/4.9  & 0.8/6.9  \\
SAPd  & -15.3 &-14.9 &-14.4 &-14.6 &-14.2 &-14.0 &-12.8 &-11.7 &   0.51/1.58 & 0.43/0.80 & 0.58/1.30 & 0.89/1.27 & 0.92/1.26 & 3.47/6.04 & 0.9/1.2  & 0.3/0.6  \\
SAZr  & -14.8 &-14.9 &-15.1 &-14.7 &-14.5 &-14.9 &-14.5 &-14.0 &   0.33/0.52 & 0.38/0.64 & 0.56/1.01 & 0.56/1.10 & 0.51/1.33 & 1.49/2.41 & 1.5/5.3  & 2.3/8.0  \\
SACr  & -13.8 &-14.1 &-14.8 &-15.1 &-15.0 &-14.8 &-14.0 &-13.0 &   0.23/0.60 & 0.28/2.49 & 1.02/3.68 & 1.13/2.80 & 0.99/2.05 & 1.07/1.85 & 1.2/1.7  & 1.2/1.7  \\
SASm  &  -5.5 & -7.5 & -8.5 & -9.4 &-10.2 &-11.0 &-13.2 &-14.2 &   0.15/0.37 & 0.25/0.46 & 0.25/0.54 & 0.25/0.48 & 0.25/0.47 & 0.35/0.81 & 3.3/14.4 & 4.8/9.4  \\
SACe  & -13.9 &-13.4 &-13.1 &-13.2 &-12.8 &-13.2 &-13.4 &-13.9 &   0.17/0.33 & 0.17/0.39 & 0.17/0.45 & 0.17/0.52 & 0.41/0.72 & 0.46/1.04 & 1.0/11.1 & 4.3/10.8 \\
SANd  &  -4.6 & -6.6 & -7.7 & -8.3 & -8.7 &-11.1 &-12.8 &-13.0 &   0.20/0.24 & 0.20/0.25 & 0.21/0.38 & 0.32/0.64 & 0.33/0.97 & 2.01/6.40 & 4.7/10.7 & 4.7/11.7 \\
SAU   &  -6.8 & -8.6 & -9.8 &-11.1 &-11.9 &-13.4 &-14.2 &-14.6 &   0.33/0.68 & 0.41/0.89 & 0.40/1.05 & 0.64/1.88 & 0.89/9.63 & 2.31/13.2 & 3.0/6.9  & 3.2/5.9  \\
\hline
SAw1  & -14.9 &-14.9 &-15.2 &-15.2 &-14.9 &-14.3 &-13.5 &-13.1 &   0.32/0.64 & 0.39/2.17 & 0.72/2.39 & 1.17/2.21 & 1.09/2.00 & 1.31/2.60 & 0.6/1.6  & 5.4/7.2  \\
SAw2  & -15.2 &-15.1 &-14.8 &-14.6 &-14.4 &-14.6 &-13.3 &-14.0 &   0.48/0.93 & 0.57/1.32 & 0.56/2.07 & 1.48/2.08 & 0.61/2.52 & 1.67/5.08 & 0.6/2.2  & 4.4/6.4  \\
SAd   &  -7.8 & -9.6 &-10.7 &-11.4 &-12.0 &-13.0 &-13.9 &-14.3 &   0.35/0.77 & 0.46/0.92 & 0.47/1.05 & 0.65/1.52 & 0.84/1.87 & 1.53/3.35 & 3.7/6.8  & 3.6/6.9  \\
\hline
A1dSm &  -6.0 & -7.9 & -8.9 & -9.4 &-10.4 &-11.2 &-13.2 &-14.3 &   0.21/0.28 & 0.15/0.28 & 0.15/0.29 & 0.25/0.37 & 0.27/0.38 & 0.35/0.84 & 3.4/13.9 & 4.7/8.4  \\
B1dSm &  -3.2 & -6.6 & -6.9 & -7.5 & -8.8 &-11.1 &-13.3 &-14.2 &   0.18/0.28 & 0.23/0.33 & 0.26/0.36 & 0.20/0.36 & 0.20/0.32 & 0.52/7.13 & 2.7/12.6 & 4.6/9.1  \\
C1dSm &  -6.7 & -9.0 & -9.7 &-10.1 &-10.6 &-11.8 &-13.9 &-14.8 &   0.20/0.25 & 0.00/0.24 & 0.00/0.25 & 0.24/0.29 & 0.25/0.33 & 1.11/10.8 & 3.0/12.9 & 5.6/11.7 \\
D1dSm &  -7.0 & -9.3 &-10.0 &-10.4 &-10.9 &-12.0 &-14.0 &-14.9 &   0.23/0.28 & 0.15/0.27 & 0.13/0.28 & 0.27/0.30 & 0.28/0.32 & 0.94/9.96 & 3.0/12.9 & 5.3/11.5 \\
\hline
A1dmSm&  -2.0 & -4.7 & -5.9 & -6.6 & -7.4 & -9.5 &-12.4 &-13.5 &   0.00/0.27 & 0.21/0.40 & 0.22/0.46 & 0.13/0.30 & 0.00/0.27 & 0.26/5.00 & 5.2/14.1 & 6.9/14.1 \\
B1dmSm&  -4.8 & -6.6 & -7.5 & -8.3 & -9.5 &-10.4 &-12.7 &-13.8 &   0.14/0.17 & 0.14/0.22 & 0.14/0.28 & 0.15/0.27 & 0.19/0.25 & 0.23/2.60 & 4.0/14.1 & 5.9/12.4 \\
C1dmSm&  -1.9 & -5.4 & -6.2 & -6.5 & -7.0 &-10.0 &-13.1 &-14.3 &   0.19/0.27 & 0.21/0.35 & 0.24/0.42 & 0.19/0.65 & 0.21/0.94 & 0.32/13.5 & 6.3/13.5 & 8.1/13.5 \\
D1dmSm&  -2.6 & -5.6 & -6.5 & -7.3 & -8.1 &-10.3 &-13.2 &-14.4 &   0.15/0.27 & 0.21/0.35 & 0.22/0.38 & 0.14/0.29 & 0.14/0.28 & 0.55/13.5 & 5.5/13.5 & 7.8/13.5 \\
\hline\hline
\end{tabular}
\label{tab:peak_mags_1d}
\end{table*}

\begin{table*}
\scriptsize
\caption{Properties of light curves for axisymmetric (2D) models, observed along the axis ("top view").}
\begin{tabular}{l|cccccccc|cccccccc}
\hline\hline
      & \multicolumn{8}{|c}{Peak magnitude, $m$}
      & \multicolumn{8}{|c}{Peak epoch $t_p$ [d] and duration $\Delta t_{\rm 1 mag}$ [d]  }
\\
Model &$g$ &$r$ &$i$ &$z$ &$y$ &$J$ &$H$ &$K$
      &$g$ &$r$ &$i$ &$z$ &$y$ &$J$ &$H$ &$K$
\\
\hline
A2dSm       &  -7.1 & -8.2 & -9.6 &-10.1 &-10.5 &-10.9 &-13.3 &-14.5 &   0.27/0.44 & 0.23/0.38 & 0.25/0.43 & 0.25/0.52 & 0.26/0.57 & 0.30/0.93 & 4.5/12.3 & 5.2/ 8.5 \\
B2dSm       &  -6.1 & -8.5 & -9.6 &-10.0 &-10.4 &-11.0 &-13.3 &-14.6 &   0.18/0.31 & 0.16/0.28 & 0.15/0.28 & 0.18/0.31 & 0.18/0.35 & 0.31/0.82 & 4.4/ 9.9 & 5.3/ 8.7 \\
C2dSm       &  -6.3 & -8.2 &-10.1 &-10.6 &-11.0 &-11.5 &-13.9 &-15.2 &   0.21/0.45 & 0.26/0.55 & 0.24/0.52 & 0.25/0.52 & 0.28/0.58 & 0.52/0.79 & 5.0/10.4 & 6.5/10.4 \\
D2dSm       &  -6.6 & -8.8 &-10.3 &-10.8 &-11.1 &-11.6 &-14.0 &-15.2 &   0.27/0.49 & 0.29/0.50 & 0.25/0.47 & 0.30/0.48 & 0.29/0.52 & 0.42/0.83 & 5.5/10.4 & 6.5/10.4 \\
\hline
W2A         & -12.1 &-12.0 &-12.3 &-12.4 &-12.4 &-12.5 &-13.9 &-14.6 &   0.43/0.78 & 0.64/0.91 & 0.71/1.39 & 0.76/1.69 & 0.75/1.64 & 0.87/2.81 & 2.7/10.2 & 6.2/ 9.7 \\
W2B         &  -6.1 & -8.5 & -9.6 &-10.0 &-10.4 &-11.0 &-13.8 &-14.6 &   0.20/0.31 & 0.16/0.28 & 0.15/0.28 & 0.17/0.31 & 0.18/0.35 & 0.31/2.31 & 4.6/ 9.9 & 5.7/ 9.9 \\
W2C         & -11.6 &-11.4 &-12.0 &-12.1 &-12.2 &-12.5 &-14.2 &-15.2 &   0.32/0.62 & 0.51/0.72 & 0.41/0.95 & 0.45/1.03 & 0.43/0.93 & 0.60/1.79 & 4.8/10.4 & 6.9/10.4 \\
W2D         &  -8.3 & -9.5 &-10.7 &-11.1 &-11.3 &-11.8 &-14.2 &-15.2 &   0.20/0.27 & 0.22/0.57 & 0.39/0.74 & 0.32/0.69 & 0.28/0.69 & 0.48/1.42 & 5.3/10.4 & 6.8/10.4 \\
\hline
W2Se        & -12.2 &-12.3 &-12.5 &-12.6 &-12.6 &-12.7 &-13.9 &-14.7 &   0.54/1.45 & 0.62/1.57 & 0.72/2.12 & 0.81/2.32 & 0.86/2.59 & 1.23/3.68 & 3.4/10.4 & 5.8/ 9.1 \\
W2Br        & -12.0 &-12.3 &-12.5 &-12.5 &-12.6 &-12.7 &-13.9 &-14.7 &   0.55/1.46 & 0.70/1.71 & 0.93/2.19 & 1.02/2.43 & 0.95/2.68 & 1.19/3.75 & 3.3/10.3 & 5.8/ 9.1 \\
W2Te        & -12.0 &-12.2 &-12.5 &-12.5 &-12.6 &-12.8 &-13.9 &-14.7 &   0.73/1.52 & 0.83/1.84 & 0.94/2.26 & 0.96/2.46 & 0.96/2.73 & 1.38/3.76 & 3.4/10.2 & 5.7/ 9.1 \\
W2Pd        & -12.3 &-12.4 &-12.5 &-12.5 &-12.6 &-12.7 &-13.9 &-14.7 &   0.61/1.46 & 0.66/1.52 & 0.79/2.21 & 0.89/2.42 & 0.95/2.66 & 1.39/3.92 & 3.3/10.2 & 5.7/ 9.1 \\
W2Zr        & -12.1 &-12.0 &-12.3 &-12.4 &-12.4 &-12.5 &-13.9 &-14.6 &   0.43/0.78 & 0.64/0.91 & 0.71/1.39 & 0.76/1.69 & 0.75/1.64 & 0.87/2.81 & 2.7/10.2 & 6.2/ 9.7 \\
W2Cr        & -11.3 &-11.3 &-11.7 &-12.2 &-12.3 &-12.7 &-13.9 &-14.7 &   0.34/0.60 & 0.45/2.41 & 0.54/3.12 & 1.60/3.05 & 1.78/3.21 & 1.96/3.93 & 3.3/10.0 & 5.8/ 9.1 \\
\hline
W2light     & -11.0 &-10.9 &-11.2 &-11.2 &-11.2 &-11.5 &-13.5 &-14.5 &   0.32/0.53 & 0.36/0.62 & 0.50/1.04 & 0.50/1.22 & 0.43/1.20 & 0.54/2.08 & 4.1/11.1 & 5.6/ 8.6 \\
W2heavy     & -12.9 &-12.9 &-13.5 &-13.6 &-13.6 &-13.8 &-14.6 &-14.7 &   0.65/1.11 & 0.90/1.30 & 1.00/1.68 & 1.05/2.05 & 1.08/1.94 & 1.19/3.68 & 3.2/11.2 & 6.6/12.3 \\
W2slow      & -12.7 &-12.5 &-12.6 &-12.7 &-12.7 &-12.8 &-14.0 &-14.7 &   0.74/1.32 & 1.07/1.60 & 1.11/1.83 & 1.04/2.59 & 1.22/2.68 & 1.48/3.90 & 3.4/10.0 & 5.7/10.0 \\
W2fast      & -13.4 &-13.7 &-14.0 &-13.7 &-13.7 &-13.8 &-13.9 &-14.6 &   0.18/0.29 & 0.21/0.43 & 0.25/0.50 & 0.25/0.58 & 0.26/0.78 & 0.68/1.80 & 3.4/ 9.2 & 5.7/ 8.9 \\
\hline
X$_1$       & -11.8 &-12.0 &-12.4 &-13.0 &-13.2 &-13.9 &-14.6 &-14.8 &   0.53/0.96 & 0.59/2.38 & 1.55/3.02 & 1.70/3.02 & 2.01/3.25 & 2.02/4.80 & 3.8/ 6.8 & 4.2/ 7.2 \\
X$_2$       & -12.1 &-12.3 &-12.6 &-13.0 &-13.1 &-13.9 &-14.6 &-14.8 &   0.71/1.27 & 0.82/1.73 & 1.19/2.66 & 1.82/2.81 & 1.99/2.94 & 2.10/4.56 & 4.0/ 6.8 & 4.1/ 7.3 \\
DZ$_1$      & -12.4 &-13.1 &-14.0 &-14.6 &-15.0 &-15.8 &-16.3 &-16.4 &   0.47/1.53 & 0.68/3.85 & 1.97/4.74 & 2.38/5.09 & 3.23/6.70 & 4.96/9.20 & 6.1/10.5 & 6.3/11.1 \\
DZ$_2$      & -12.6 &-13.3 &-14.0 &-14.6 &-14.9 &-15.8 &-16.3 &-16.4 &   0.63/1.51 & 0.96/3.11 & 1.46/4.00 & 2.34/4.44 & 2.88/6.01 & 4.70/9.02 & 6.2/10.6 & 6.3/11.2 \\
Xnh$_1$     & -12.5 &-12.6 &-12.8 &-13.3 &-13.4 &-14.1 &-14.9 &-15.1 &   0.55/0.95 & 0.64/1.57 & 0.82/2.40 & 1.57/2.54 & 1.73/2.77 & 1.89/4.59 & 4.2/ 6.7 & 4.5/ 7.0 \\
Xnh$_2$     & -13.3 &-13.4 &-13.7 &-13.9 &-13.9 &-14.5 &-15.1 &-15.2 &   0.73/1.68 & 1.05/2.21 & 1.44/3.16 & 1.95/3.49 & 2.19/3.65 & 2.27/5.35 & 4.4/ 7.6 & 4.8/ 8.1 \\
\hline
$\gamma A_1$& -12.5 &-12.6 &-13.0 &-13.4 &-13.5 &-14.3 &-15.0 &-15.2 &   0.56/0.98 & 0.66/1.74 & 0.85/2.58 & 1.62/2.72 & 1.81/2.98 & 2.06/4.68 & 4.1/ 6.7 & 4.2/ 7.1 \\
$\gamma B_1$&  -8.4 &-10.1 &-11.2 &-11.6 &-12.3 &-14.1 &-15.1 &-15.2 &   0.34/0.59 & 0.34/0.67 & 0.36/0.91 & 0.46/2.87 & 1.85/3.48 & 2.42/5.03 & 4.1/ 7.0 & 4.4/ 7.4 \\
$\gamma C_1$& -11.5 &-12.0 &-12.8 &-13.3 &-13.6 &-14.7 &-15.6 &-15.8 &   0.41/0.83 & 0.51/1.64 & 0.80/2.21 & 1.03/2.60 & 1.42/3.58 & 2.71/5.96 & 5.1/ 9.0 & 5.4/ 9.6 \\
$\gamma D_1$&  -9.4 &-10.9 &-12.0 &-12.7 &-13.3 &-14.6 &-15.7 &-15.9 &   0.00/1.30 & 0.56/1.88 & 0.97/2.20 & 1.15/2.70 & 1.63/3.70 & 2.77/5.97 & 5.0/ 8.8 & 5.2/ 9.4 \\
$\gamma A_2$& -13.4 &-13.5 &-13.9 &-14.1 &-14.2 &-14.8 &-15.3 &-15.3 &   0.75/1.68 & 1.10/2.25 & 1.60/3.12 & 1.98/3.38 & 2.21/3.51 & 2.35/5.25 & 4.3/ 7.3 & 4.4/ 7.9 \\
$\gamma B_2$&  -8.5 &-10.2 &-11.2 &-12.4 &-13.1 &-14.5 &-15.3 &-15.4 &   0.34/0.67 & 0.35/0.94 & 0.37/3.10 & 2.00/3.35 & 2.45/3.66 & 2.88/5.69 & 4.6/ 7.8 & 4.3/ 8.2 \\
$\gamma C_2$& -12.2 &-12.7 &-13.4 &-13.8 &-13.9 &-14.9 &-15.8 &-15.9 &   0.58/1.34 & 0.78/1.82 & 0.98/2.60 & 1.12/2.92 & 1.69/3.45 & 2.80/6.41 & 5.3/ 9.5 & 5.5/10.1 \\
$\gamma D_2$&  -9.7 &-11.2 &-12.4 &-13.1 &-13.6 &-14.8 &-15.8 &-15.9 &   0.39/1.84 & 0.54/2.51 & 1.04/2.85 & 1.49/3.11 & 1.88/3.74 & 2.82/6.38 & 5.1/ 9.3 & 5.5/ 9.9 \\
\hline
\end{tabular}
\label{tab:peak_mags_2d_top}
\end{table*}

\begin{table*}
\scriptsize
\caption{Properties of light curves for axisymmetric (2D) models, observed along the axis from the other side ("bottom view").}
\begin{tabular}{l|cccccccc|cccccccc}
\hline\hline
      & \multicolumn{8}{|c}{Peak magnitude, $m$}
      & \multicolumn{8}{|c}{Peak epoch $t_p$ [d] and duration $\Delta t_{\rm 1 mag}$ [d]  }
\\
Model &$g$ &$r$ &$i$ &$z$ &$y$ &$J$ &$H$ &$K$
      &$g$ &$r$ &$i$ &$z$ &$y$ &$J$ &$H$ &$K$
\\
\hline
A2dSm       &  -7.8 & -8.7 &-10.2 &-10.8 &-11.0 &-11.4 &-13.3 &-14.5 &   0.27/0.43 & 0.26/0.47 & 0.27/0.56 & 0.27/0.62 & 0.28/0.65 & 0.44/1.20 & 4.2/12.3 & 5.2/ 8.4 \\
B2dSm       &  -6.6 & -8.9 & -9.8 &-10.2 &-10.7 &-11.4 &-13.4 &-14.6 &   0.19/0.34 & 0.17/0.30 & 0.16/0.31 & 0.17/0.36 & 0.19/0.44 & 0.31/0.86 & 4.3/ 9.9 & 5.2/ 8.5 \\
C2dSm       &  -6.4 & -8.5 &-10.1 &-10.6 &-11.0 &-11.5 &-14.0 &-15.2 &   0.25/0.44 & 0.28/0.44 & 0.21/0.43 & 0.20/0.45 & 0.25/0.52 & 0.36/0.91 & 5.0/10.4 & 6.6/10.4 \\
D2dSm       &  -6.5 & -8.9 &-10.3 &-10.7 &-11.0 &-11.4 &-14.0 &-15.3 &   0.26/0.38 & 0.22/0.39 & 0.20/0.37 & 0.20/0.40 & 0.24/0.44 & 0.35/0.86 & 5.4/10.4 & 6.3/10.4 \\
\hline
W2A         & -13.5 &-13.3 &-13.5 &-13.5 &-13.4 &-13.3 &-13.9 &-14.6 &   0.40/0.70 & 0.57/0.82 & 0.63/1.19 & 0.74/1.38 & 0.63/1.32 & 0.79/2.55 & 2.1/10.2 & 6.1/ 9.6 \\
W2B         &  -9.3 & -9.3 & -9.9 &-10.3 &-10.7 &-11.5 &-13.8 &-14.6 &   0.33/0.77 & 0.38/0.83 & 0.18/0.82 & 0.18/0.77 & 0.18/0.61 & 0.37/2.31 & 4.7/ 9.9 & 5.7/ 9.9 \\
W2C         &  -8.5 & -9.2 &-10.4 &-10.8 &-11.1 &-11.8 &-14.2 &-15.2 &   0.25/0.38 & 0.29/0.56 & 0.35/0.77 & 0.31/0.73 & 0.37/0.87 & 0.40/1.54 & 5.0/10.4 & 6.9/10.4 \\
W2D         &  -6.5 & -8.9 &-10.3 &-10.7 &-11.0 &-11.4 &-14.2 &-15.3 &   0.27/0.38 & 0.20/0.39 & 0.20/0.37 & 0.21/0.39 & 0.23/0.45 & 0.37/0.89 & 5.5/10.4 & 6.7/10.4 \\
\hline
W2Se        & -13.4 &-13.4 &-13.5 &-13.5 &-13.4 &-13.3 &-13.8 &-14.7 &   0.35/1.22 & 0.51/1.17 & 0.57/1.69 & 0.66/2.00 & 0.79/2.27 & 1.07/3.56 & 4.2/10.8 & 5.8/ 9.0 \\
W2Br        & -13.2 &-13.4 &-13.5 &-13.4 &-13.4 &-13.4 &-13.9 &-14.7 &   0.50/1.19 & 0.51/1.28 & 0.62/1.84 & 0.79/2.15 & 0.90/2.40 & 0.98/3.66 & 4.1/10.7 & 5.8/ 9.0 \\
W2Te        & -13.3 &-13.4 &-13.5 &-13.5 &-13.5 &-13.4 &-13.9 &-14.7 &   0.52/1.18 & 0.53/1.39 & 0.56/1.84 & 0.75/2.13 & 0.84/2.33 & 1.04/3.60 & 4.1/10.6 & 5.9/ 9.0 \\
W2Pd        & -13.7 &-13.6 &-13.6 &-13.6 &-13.5 &-13.4 &-13.9 &-14.7 &   0.51/1.19 & 0.56/1.10 & 0.60/1.64 & 0.65/1.96 & 0.72/2.21 & 0.99/3.91 & 3.9/10.6 & 5.9/ 9.0 \\
W2Zr        & -13.5 &-13.3 &-13.5 &-13.5 &-13.4 &-13.3 &-13.9 &-14.6 &   0.40/0.70 & 0.57/0.82 & 0.63/1.19 & 0.74/1.38 & 0.63/1.32 & 0.79/2.55 & 2.1/10.2 & 6.1/ 9.6 \\
W2Cr        & -12.6 &-12.7 &-13.2 &-13.5 &-13.4 &-13.5 &-13.9 &-14.7 &   0.28/0.61 & 0.35/1.89 & 0.55/2.46 & 0.71/2.58 & 0.72/2.74 & 1.42/3.59 & 4.0/10.4 & 5.7/ 9.0 \\
\hline
W2light     & -12.4 &-12.2 &-12.4 &-12.4 &-12.4 &-12.4 &-13.5 &-14.6 &   0.27/0.47 & 0.33/0.54 & 0.38/0.84 & 0.53/1.00 & 0.41/0.96 & 0.51/1.77 & 4.0/10.7 & 5.5/ 8.5 \\
W2heavy     & -14.4 &-14.1 &-14.5 &-14.5 &-14.3 &-14.3 &-14.7 &-14.7 &   0.50/0.98 & 0.75/1.18 & 0.92/1.72 & 1.04/1.92 & 0.96/1.85 & 1.27/3.58 & 2.7/11.0 & 6.9/12.3 \\
W2slow      & -13.5 &-13.1 &-13.3 &-13.4 &-13.4 &-13.3 &-14.0 &-14.7 &   0.72/1.17 & 1.04/1.51 & 1.05/1.73 & 1.22/2.42 & 1.21/2.40 & 1.38/3.87 & 3.3/10.2 & 5.8/ 9.9 \\
W2fast      & -13.1 &-13.5 &-13.8 &-13.5 &-13.5 &-13.8 &-14.0 &-14.6 &   0.17/0.33 & 0.22/0.55 & 0.30/0.61 & 0.29/0.68 & 0.30/0.90 & 0.62/1.93 & 3.5/ 9.1 & 5.7/ 8.8 \\
\hline
X$_1$       & -13.2 &-13.3 &-13.4 &-13.6 &-13.6 &-13.9 &-14.7 &-14.9 &   0.40/0.72 & 0.45/1.32 & 0.60/2.33 & 0.79/2.78 & 0.83/2.98 & 1.64/4.82 & 3.6/ 6.7 & 4.1/ 7.2 \\
X$_2$       & -13.3 &-13.4 &-13.5 &-13.6 &-13.6 &-13.9 &-14.7 &-14.9 &   0.44/1.03 & 0.60/1.27 & 0.72/1.99 & 0.77/2.44 & 0.69/2.53 & 1.73/4.66 & 3.7/ 6.8 & 4.2/ 7.3 \\
DZ$_1$      & -13.5 &-13.8 &-14.2 &-14.8 &-15.1 &-15.9 &-16.4 &-16.4 &   0.42/0.95 & 0.49/2.74 & 1.18/4.50 & 2.03/5.21 & 3.41/6.76 & 5.16/9.20 & 6.0/10.4 & 6.3/11.0 \\
DZ$_2$      & -13.6 &-14.0 &-14.4 &-14.7 &-15.0 &-15.9 &-16.4 &-16.4 &   0.51/1.16 & 0.71/1.92 & 0.99/3.62 & 2.08/4.57 & 2.61/6.19 & 4.66/9.06 & 5.9/10.5 & 6.3/11.1 \\
Xnh$_1$     & -13.8 &-13.8 &-13.8 &-14.0 &-14.0 &-14.1 &-15.0 &-15.1 &   0.44/0.75 & 0.51/0.99 & 0.61/1.61 & 0.73/2.09 & 0.81/2.24 & 1.61/4.76 & 4.1/ 6.6 & 4.3/ 7.0 \\
Xnh$_2$     & -14.4 &-14.5 &-14.5 &-14.5 &-14.5 &-14.5 &-15.1 &-15.2 &   0.62/1.35 & 0.74/1.65 & 0.90/2.14 & 0.88/2.76 & 0.94/2.97 & 1.38/5.43 & 4.4/ 7.6 & 4.8/ 8.1 \\
\hline
$\gamma A_1$& -13.9 &-13.9 &-13.9 &-14.0 &-14.0 &-14.3 &-15.0 &-15.2 &   0.46/0.78 & 0.52/1.05 & 0.63/1.70 & 0.75/2.28 & 0.83/2.49 & 1.85/4.82 & 4.0/ 6.7 & 4.4/ 7.0 \\
$\gamma B_1$&  -9.6 &-10.7 &-11.4 &-11.8 &-12.2 &-14.1 &-15.1 &-15.2 &   0.36/0.80 & 0.34/0.80 & 0.36/1.34 & 0.44/2.58 & 0.58/3.42 & 2.47/5.09 & 4.2/ 7.0 & 4.3/ 7.3 \\
$\gamma C_1$&  -9.4 &-11.1 &-12.4 &-13.1 &-13.6 &-14.7 &-15.7 &-15.8 &   0.81/1.69 & 0.88/2.08 & 1.02/2.33 & 1.28/2.68 & 1.69/3.73 & 2.65/5.98 & 5.1/ 9.0 & 5.5/ 9.5 \\
$\gamma D_1$&  -8.9 &-10.7 &-11.7 &-12.3 &-13.1 &-14.6 &-15.7 &-15.9 &   0.00/0.85 & 0.47/1.27 & 0.48/1.89 & 0.98/3.06 & 1.60/3.98 & 2.75/6.03 & 5.0/ 8.8 & 5.4/ 9.3 \\
$\gamma A_2$& -14.5 &-14.5 &-14.6 &-14.6 &-14.7 &-14.7 &-15.3 &-15.3 &   0.61/1.37 & 0.78/1.65 & 0.88/2.21 & 0.94/2.81 & 1.01/2.93 & 2.07/5.41 & 4.2/ 7.3 & 4.5/ 7.8 \\
$\gamma B_2$&  -9.9 &-10.8 &-11.6 &-12.3 &-12.9 &-14.5 &-15.3 &-15.4 &   0.40/1.41 & 0.37/1.83 & 0.37/2.80 & 2.11/3.37 & 2.43/3.70 & 2.86/5.75 & 4.5/ 7.8 & 4.4/ 8.2 \\
$\gamma C_2$&  -9.9 &-11.5 &-12.8 &-13.4 &-13.8 &-14.9 &-15.8 &-15.9 &   0.88/2.02 & 1.03/2.54 & 1.23/2.77 & 1.40/3.02 & 1.95/3.62 & 2.78/6.42 & 5.2/ 9.5 & 5.4/10.0 \\
$\gamma D_2$&  -9.0 &-10.7 &-11.7 &-12.5 &-13.4 &-14.8 &-15.8 &-16.0 &   0.30/0.97 & 0.49/2.03 & 0.52/2.85 & 1.77/3.47 & 2.16/3.96 & 2.96/6.41 & 5.1/ 9.3 & 5.3/ 9.8 \\
\hline
\end{tabular}
\label{tab:peak_mags_2d_btm}
\end{table*}

\begin{table*}
\scriptsize
\caption{Properties of light curves for axisymmetric (2D) models, observed in equatorial plane of dynamical ejecta ("side view").}
\begin{tabular}{l|cccccccc|cccccccc}
\hline\hline
      & \multicolumn{8}{|c}{Peak magnitude, $m$}
      & \multicolumn{8}{|c}{Peak epoch $t_p$ [d] and duration $\Delta t_{\rm 1mag}$ [d]  }
\\
Model &$g$ &$r$ &$i$ &$z$ &$y$ &$J$ &$H$ &$K$
      &$g$ &$r$ &$i$ &$z$ &$y$ &$J$ &$H$ &$K$
\\
\hline
A2dSm       & -6.4 & -8.8 & -9.9 &-10.3 &-10.6 &-11.4 &-13.1 &-13.9 &   0.30/0.47 & 0.18/0.42 & 0.18/0.40 & 0.18/0.52 & 0.27/0.56 & 0.53/ 1.1 & 3.4/12.3 & 4.8/ 8.5 \\
B2dSm       & -6.8 & -9.0 & -9.8 &-10.2 &-10.5 &-11.5 &-13.2 &-13.9 &   0.21/0.31 & 0.15/0.28 & 0.16/0.31 & 0.17/0.40 & 0.17/0.47 & 0.49/ 4.8 & 2.5/ 9.9 & 4.6/ 9.0 \\
C2dSm       & -7.4 & -9.7 &-10.5 &-10.8 &-10.9 &-11.8 &-13.6 &-14.3 &   0.19/0.28 & 0.20/0.31 & 0.20/0.31 & 0.21/0.40 & 0.21/0.49 & 1.42/10.4 & 3.4/10.4 & 6.4/10.4 \\
D2dSm       & -7.7 &-10.1 &-10.6 &-11.0 &-11.2 &-12.0 &-13.8 &-14.4 &   0.22/0.38 & 0.26/0.34 & 0.23/0.39 & 0.23/0.47 & 0.27/0.56 & 1.29/10.4 & 3.8/10.4 & 6.0/10.4 \\
\hline
W2A         & -6.4 & -8.8 &-10.0 &-10.3 &-10.6 &-11.4 &-13.2 &-14.0 &   0.28/0.49 & 0.18/0.41 & 0.18/0.40 & 0.18/0.50 & 0.28/0.56 & 0.53/ 1.2 & 3.2/12.3 & 5.3/10.1 \\
W2B         & -6.8 & -9.0 & -9.8 &-10.2 &-10.6 &-11.5 &-13.2 &-14.0 &   0.17/0.30 & 0.16/0.29 & 0.15/0.29 & 0.18/0.41 & 0.21/0.48 & 0.43/ 4.6 & 2.8/ 9.9 & 5.5/ 9.9 \\
W2C         & -7.4 & -9.8 &-10.4 &-10.7 &-10.9 &-11.8 &-13.6 &-14.4 &   0.18/0.30 & 0.23/0.28 & 0.24/0.34 & 0.19/0.44 & 0.21/0.49 & 1.37/10.4 & 3.3/10.4 & 6.9/10.4 \\
W2D         & -7.6 &-10.0 &-10.6 &-11.0 &-11.3 &-12.0 &-13.8 &-14.5 &   0.21/0.37 & 0.27/0.34 & 0.27/0.41 & 0.24/0.46 & 0.28/0.52 & 1.24/ 9.8 & 3.5/10.4 & 6.0/10.4 \\
\hline
W2Se        & -6.6 & -8.8 &-10.0 &-10.3 &-10.6 &-11.4 &-13.2 &-14.0 &   0.30/0.43 & 0.18/0.42 & 0.18/0.41 & 0.18/0.50 & 0.28/0.56 & 0.51/ 1.2 & 3.6/12.3 & 5.2/10.1 \\
W2Br        & -6.4 & -8.8 &-10.0 &-10.3 &-10.7 &-11.4 &-13.2 &-14.0 &   0.26/0.50 & 0.18/0.43 & 0.18/0.39 & 0.18/0.51 & 0.26/0.56 & 0.52/ 1.3 & 3.4/12.3 & 5.3/10.1 \\
W2Te        & -6.4 & -8.8 & -9.9 &-10.3 &-10.6 &-11.4 &-13.2 &-14.0 &   0.31/0.49 & 0.18/0.43 & 0.18/0.41 & 0.18/0.51 & 0.27/0.56 & 0.51/ 1.2 & 3.3/12.3 & 5.1/10.1 \\
W2Pd        & -6.5 & -8.8 &-10.0 &-10.3 &-10.6 &-11.4 &-13.2 &-14.0 &   0.28/0.45 & 0.18/0.42 & 0.18/0.41 & 0.18/0.51 & 0.31/0.56 & 0.51/ 1.2 & 3.5/12.3 & 5.0/10.1 \\
W2Zr        & -6.4 & -8.8 &-10.0 &-10.3 &-10.6 &-11.4 &-13.2 &-14.0 &   0.28/0.49 & 0.18/0.41 & 0.18/0.40 & 0.18/0.50 & 0.28/0.56 & 0.53/ 1.2 & 3.2/12.3 & 5.3/10.1 \\
W2Cr        & -6.5 & -8.8 & -9.9 &-10.3 &-10.6 &-11.4 &-13.2 &-14.0 &   0.29/0.48 & 0.18/0.41 & 0.18/0.40 & 0.18/0.51 & 0.26/0.56 & 0.52/ 1.0 & 3.5/12.3 & 5.0/10.1 \\
\hline
W2light     & -6.4 & -8.9 & -9.9 &-10.3 &-10.6 &-11.4 &-13.1 &-13.9 &   0.30/0.49 & 0.18/0.42 & 0.18/0.41 & 0.18/0.51 & 0.28/0.55 & 0.47/ 1.2 & 3.0/12.3 & 5.0/ 8.8 \\
W2heavy     & -6.5 & -8.9 & -9.9 &-10.3 &-10.6 &-11.4 &-13.4 &-14.3 &   0.24/0.47 & 0.27/0.41 & 0.18/0.41 & 0.18/0.50 & 0.28/0.56 & 0.51/ 1.5 & 3.9/12.3 & 6.3/12.3 \\
W2slow      & -6.4 & -8.8 &-10.0 &-10.3 &-10.6 &-11.4 &-13.2 &-13.9 &   0.28/0.48 & 0.18/0.43 & 0.18/0.41 & 0.18/0.50 & 0.30/0.55 & 0.52/ 1.2 & 3.6/12.3 & 4.8/11.7 \\
W2fast      &-12.3 &-13.0 &-13.3 &-13.0 &-13.0 &-13.3 &-13.6 &-14.1 &   0.15/0.25 & 0.15/0.35 & 0.21/0.38 & 0.16/0.43 & 0.18/0.59 & 0.45/ 1.4 & 2.3/10.6 & 4.4/ 8.7 \\
\hline
X$_1$       & -8.2 &-10.1 &-11.1 &-11.7 &-12.2 &-13.0 &-13.7 &-14.1 &   0.40/0.86 & 0.41/0.97 & 0.42/1.12 & 0.62/1.63 & 0.88/1.90 & 1.55/ 3.7 & 3.9/ 7.2 & 3.7/ 7.2 \\
X$_2$       & -8.1 &-10.0 &-11.1 &-11.7 &-12.2 &-13.0 &-13.7 &-14.2 &   0.47/0.90 & 0.44/0.97 & 0.44/1.14 & 0.68/1.62 & 0.85/1.89 & 1.47/ 3.7 & 3.9/ 7.3 & 3.8/ 7.2 \\
DZ$_1$      & -12.3 &-13.4 &-14.1 &-14.4 &-14.7 &-15.3 &-15.4 &-15.7 &   0.39/1.77 & 0.55/2.76 & 0.60/3.53 & 1.98/5.13 & 2.66/6.34 & 4.10/9.11 & 5.7/12.1 & 5.7/11.3 \\
DZ$_2$      & -12.3 &-13.4 &-14.1 &-14.4 &-14.7 &-15.3 &-15.4 &-15.7 &   0.38/1.80 & 0.56/2.79 & 0.56/3.54 & 2.01/5.12 & 2.81/6.36 & 4.07/9.11 & 5.7/12.1 & 6.0/11.4 \\
Xnh$_1$     & -8.1 &-10.0 &-11.0 &-11.7 &-12.2 &-13.2 &-13.9 &-14.4 &   0.48/0.95 & 0.42/1.05 & 0.59/1.26 & 0.68/1.81 & 0.96/2.11 & 1.72/ 4.0 & 4.2/ 7.1 & 4.1/ 7.1 \\
Xnh$_2$     & -8.0 &-10.0 &-11.0 &-11.7 &-12.3 &-13.3 &-14.0 &-14.5 &   0.47/0.96 & 0.44/1.14 & 0.45/1.50 & 0.82/2.01 & 0.95/2.39 & 1.76/ 4.2 & 4.5/ 8.3 & 4.5/ 8.2 \\
\hline
$\gamma A_1$& -9.2 &-10.8 &-11.8 &-12.3 &-12.8 &-13.6 &-14.1 &-14.5 &   0.51/1.05 & 0.49/1.23 & 0.64/1.50 & 0.92/2.08 & 1.08/2.42 & 1.97/4.30 & 4.3/ 7.0 & 3.9/ 6.9 \\
$\gamma B_1$& -9.2 &-10.7 &-11.5 &-12.1 &-12.6 &-13.7 &-14.1 &-14.5 &   0.36/0.69 & 0.37/0.84 & 0.44/1.14 & 0.68/1.66 & 0.84/1.92 & 1.73/4.19 & 4.1/ 7.4 & 3.9/ 7.3 \\
$\gamma C_1$& -9.2 &-11.1 &-11.9 &-12.5 &-12.9 &-13.9 &-14.5 &-15.0 &   0.49/0.92 & 0.49/0.95 & 0.57/1.37 & 0.96/1.83 & 1.05/2.14 & 2.03/4.97 & 5.6/ 9.6 & 5.4/ 9.5 \\
$\gamma D_1$& -9.7 &-11.3 &-12.1 &-12.7 &-13.2 &-14.1 &-14.6 &-15.0 &   0.46/0.93 & 0.50/1.03 & 0.59/1.47 & 0.92/1.90 & 1.07/2.32 & 2.18/5.05 & 4.8/ 9.4 & 5.1/ 9.4 \\
$\gamma A_2$& -9.2 &-10.8 &-11.8 &-12.4 &-12.8 &-13.7 &-14.2 &-14.6 &   0.53/1.12 & 0.60/1.33 & 0.64/1.63 & 0.96/2.20 & 1.18/2.67 & 1.94/4.51 & 4.8/ 8.1 & 4.5/ 8.0 \\
$\gamma B_2$& -9.3 &-10.8 &-11.6 &-12.2 &-12.6 &-13.7 &-14.3 &-14.7 &   0.36/0.68 & 0.36/0.81 & 0.44/1.10 & 0.58/1.69 & 0.88/1.98 & 1.93/4.54 & 4.9/ 8.6 & 4.5/ 8.5 \\
$\gamma C_2$& -9.2 &-11.1 &-11.9 &-12.5 &-13.0 &-14.0 &-14.5 &-15.0 &   0.46/0.94 & 0.48/1.00 & 0.58/1.51 & 0.88/1.94 & 1.04/2.18 & 2.20/5.04 & 5.8/10.5 & 5.8/10.2 \\
$\gamma D_2$& -9.6 &-11.3 &-12.1 &-12.7 &-13.2 &-14.2 &-14.6 &-15.1 &   0.00/1.02 & 0.48/1.04 & 0.60/1.57 & 0.75/1.97 & 1.16/2.42 & 2.45/5.12 & 5.0/10.2 & 5.5/10.1 \\
\hline
\end{tabular}
\label{tab:peak_mags_2d_side}
\end{table*}
\clearpage

\bibliographystyle{mnras}
\bibliography{refs}

\end{document}